\documentclass{aastex61}
\usepackage{amsmath,amsthm,amssymb}
\usepackage{titlesec}
\usepackage{graphicx}
\usepackage{tabularx}
\usepackage{float}
\usepackage{color}
\usepackage{scalefnt}
\usepackage{hyperref}
\usepackage[numbib]{tocbibind}
\usepackage{enumitem}
\usepackage{multirow}
\usepackage{wasysym}
\usepackage{mathrsfs}

\begin{document}

\title{Exo-Milankovitch Cycles II: Climates of G-dwarf Planets in Dynamically Hot Systems}
\shorttitle{Exo-Milankovitch II}
\author{Russell Deitrick}
\affil{Department of Astronomy, University of Washington, Seattle, WA 98195-1580, USA}
\affil{Center for Space and Habitability, University of Bern, Gesellschaftsstrasse 6, CH-3012, Bern, Switzerland}
\affil{Virtual Planetary Laboratory, University of Washington, Seattle, WA 98195-1580, USA}
\email{russell.deitrick@csh.unibe.ch}
\author{Rory Barnes}
\affil{Department of Astronomy, University of Washington, Seattle, WA 98195-1580, USA}
\affil{Virtual Planetary Laboratory, University of Washington, Seattle, WA 98195-1580, USA}
\author{Cecilia Bitz}
\affil{Department of Atmospheric Sciences, University of Washington, Seattle, WA 98195-1580, USA}
\affil{Virtual Planetary Laboratory, University of Washington, Seattle, WA 98195-1580, USA}
\author{David Fleming}
\affil{Department of Astronomy, University of Washington, Seattle, WA 98195-1580, USA}
\affil{Virtual Planetary Laboratory, University of Washington, Seattle, WA 98195-1580, USA}
\author{Benjamin Charnay}
\affil{LESIA, Observatoire de Paris, PSL Research University, CNRS, Sorbonne Universit\'{e}s, UPMC Univ. Paris 06, Univ. Paris Diderot, Sorbonne, Paris Cit\'{e}, 5 Place Jules Janssen, 92195 Meudon, France}
\affil{Virtual Planetary Laboratory, University of Washington, Seattle, WA 98195-1580, USA}
\author{Victoria Meadows}
\affil{Department of Astronomy, University of Washington, Seattle, WA 98195-1580, USA}
\affil{Virtual Planetary Laboratory, University of Washington, Seattle, WA 98195-1580, USA}
\author{Caitlyn Wilhelm}
\affil{Department of Astronomy, University of Washington, Seattle, WA 98195-1580, USA}
\affil{Virtual Planetary Laboratory, University of Washington, Seattle, WA 98195-1580, USA}
\author{John Armstrong}
\affil{Department of Physics, Weber State University, Ogden, UT 84408-2508, USA}
\affil{Virtual Planetary Laboratory, University of Washington, Seattle, WA 98195-1580, USA}
\author{Thomas R. Quinn}
\affil{Department of Astronomy, University of Washington, Seattle, WA 98195-1580, USA}
\affil{Virtual Planetary Laboratory, University of Washington, Seattle, WA 98195-1580, USA}
\date{}

\begin{abstract}
Using an energy balance model with ice sheets, we examine the climate response of an Earth-like planet
orbiting a G dwarf star and experiencing
large orbital and obliquity variations. We find that ice caps 
 couple strongly to the orbital forcing, leading to 
extreme ice ages. In contrast with previous studies, we find that 
such exo-Milankovitch cycles tend to impair habitability by inducing 
snowball states within the habitable zone. The large 
amplitude changes in obliquity and eccentricity cause the ice edge,
the lowest 
latitude extent of the ice caps, to become unstable and grow to the 
equator. We apply an 
analytical theory of the ice edge latitude to show that obliquity is the 
primary driver of the instability. The thermal 
inertia of the ice sheets and the spectral energy distribution of 
the G dwarf star increase the sensitivity of the model to triggering 
runaway glaciation. Finally, we apply a machine learning algorithm to
demonstrate how this technique can be used to extend the power of 
climate models. This work illustrates the importance of orbital 
evolution for habitability in dynamically rich planetary systems. We emphasize that as potentially habitable planets are discovered around G dwarfs, we need to consider orbital dynamics.
\end{abstract}

\maketitle

\section{Introduction}
Milankovitch cycles, or orbitally-induced climate variations, are 
thought to influence, if not control, Earth's ice ages 
\citep{hays1976,imbrie1980,raymo1997,lisiecki2007}. This mechanism 
has also been proposed as an important player in the habitability of 
exoplanets, which may have orbital evolution very different from that
of Earth \citep{spiegel2010,brasser2014,armstrong2014}. In \cite{deitrick2018} (hereafter, Paper I), 
we discussed much of the work 
that has been done to understand Milankovitch cycles, both for Earth 
and for exoplanets. Briefly, we review the subset of the literature most 
concerned with the modeling of climate.

Milutin Milankovi\'{c} and Wladimir K\"oppen supplied a plausible 
explanation for the orbital forcing of Earth's ice ages: small 
variations in summer-time insolation at high latitudes controls 
whether ice sheets on the continent grow or retreat. This idea is
generally accepted as at least part of the story 
\citep{hays1976, roe2006, huybers2008, lisiecki2010}, though the 
reality is somewhat more complicated because of geography, ice shelf 
calving, atmospheric circulation, and changes in greenhouse gases 
\citep{clark1998,abeouchi2013}, and some studies have 
challenged the role of orbital forcing entirely 
\citep{wunsch2004,maslin2016}. 

Much of the controversy surrounding Milankovitch theory stems from 
the fact that Earth's orbital and obliquity variations are rather 
small---Earth's obliquity varies by $\sim 2.5^{\circ}$ and its 
eccentricity by $\sim0.05$ \citep{laskar1993}. For exoplanets, the role of orbital 
forcing may be more compelling---many exoplanets have variations that 
are much larger than Earth's, and there is evidence that 
primordial obliquities (\emph{i.e.}, the obliquity after the
formation stage) can be very different from Earth's present value 
\citep{miguel2010}. 

In this study, we are interested in how planetary habitability 
is affected by obliquity, eccentricity, and variations of these parameters.
For example, it was proposed that, at zero 
obliquity, the lack of insolation at the poles of an
Earth-like planet would cause the 
ice caps to grow uncontrollably and trigger a snowball state 
\citep{laskar1993}, however, climate models demonstrated that this 
is not the case \citep{williamskasting97}. In fact, the models indicate
that Earth's climate can
remain stable (and warm) at any obliquity 
\citep{williamskasting97,williams2003,spiegel2009} at its current
solar flux. 
 
 For obliquities larger
than Earth's, the seasonality of the planet is intensified 
\citep{williamskasting97,williams2003,spiegel2009}, \emph{i.e.},
mid- and high-latitudes experience extremely warm summers and 
extremely cold, dark winters. At obliquity $\gtrsim55^{\circ}$, 
the poles begin to receive more insolation over an orbit than the 
equator \citep{vanwoerkom1953,williams1975,williams1993,lissauer2012,rose2017}. In such conditions, it is
possible that ice sheets form at the equator (``ice-belts''), rather
than at the poles \citep{williams2003,rose2017}, but this 
phenomenon appears to be sensitive to the atmospheric properties
and the details of the model \citep{ferreira2014,rose2017}. The other
important development is that high obliquity ($\gtrsim55^{\circ}$) 
tends to increase
the distance (from the host star) to the outer edge of the habitable zone (HZ), because 
the insolation distribution is more even across the surface than
at low obliquity \citep{spiegel2009,rose2017}. The habitable zone, as we discuss it here,
is the range of stellar flux at which a planet with an Earth-like atmosphere
can maintain liquid water on its surface
\citep[see][]{kasting1993,selsis2007,kopparapu2013}.

The effect of planet's eccentricity, $e$, on the orbitally-averaged 
stellar flux, $\langle S \rangle$, can be directly calculated \citep{laskar1993},
which results in a dependence of the form:
\begin{equation}
    \langle S \rangle \propto (1-e^2)^{-1/2}. \label{eqn:annualinsol}
\end{equation}
Thus, the insolation increases as the eccentricity 
increases, and some studies have indeed shown that the outer-edge 
of the habitable zone can increase as a result
\citep{williams2002,dressing2010}. This relationship is complicated
by the fact that eccentricity can introduce a global 
``seasonality''---a result of the varying distance between the planet
and host star over an orbit. Because of Kepler's second law, the planet
spends much of its orbit near apoastron, and if
the orbit is sufficiently long period, snowball states can be triggered
at these times \citep{bolmont2016}. Thus an increase in eccentricity 
does not warm an Earth-like planet in all cases.

How orbital and obliquity \emph{variations} (exo-Milankovitch cycles) 
affect habitability
is only beginning to be understood. Some studies have found that 
increases in eccentricity can rescue a planet from a snowball state
\citep{dressing2010,spiegel2010}. Others have shown that strong
variations can affect the boundaries of the habitable zone 
\citep{armstrong2014,way2017}. There may be some threat to the planet
in the form of water loss if the planet is near the inner edge 
because of periastron's proximity to the host star during high 
eccentricity times \citep{way2017}. Exo-Milankovitch cycles may also
increase or decrease the outer edge of the habitable zone, as suggested in 
\cite{armstrong2014}. \cite{forgan2016} showed that Milankovitch
cycles can be very rapid for circumbinary planets, though that study
did not find them a threat to planetary habitability in the cases 
considered.

Though the effects of different eccentricity and obliquity values and 
their variations have been studied by the previously discussed 
works, their remains no complete synthesis of orbital evolution, 
obliquity evolution, and climate, including the effects of ice 
sheets and oceans. The majority of the aforementioned works 
examined only static orbits and obliquities 
\citep{williamskasting97,williams2002,williams2003,spiegel2009,dressing2010,ferreira2014,bolmont2016,rose2017}. 
The studies that did model climate under varying orbital 
conditions were limited in various ways. 
\cite{spiegel2010} and \cite{way2017} allowed eccentricity to vary, but did 
not include obliquity variations. \cite{armstrong2014} included 
obliquity variations in addition to orbital variations. Unfortunately, that paper
contained a sign error in the obliquity equations (though the 
code was correct) that was
propagated to \cite{forgan2016}. The climate models used by 
\cite{spiegel2010} and \cite{forgan2016} did not include ice 
sheets and the  thermal inertia associated with them, and so 
produced climates that are potentially too warm and too stable against the 
snowball instability. The climate model used in 
\cite{armstrong2014} included ice sheets, but the outgoing 
longwave radiation prescription and the lack of latitudinal heat 
diffusion makes that model excessively stable against snowball 
states, and that model did not include oceans (see Section
\ref{sec:armstrongcomp}). \cite{spiegel2010} and 
\cite{forgan2016} included oceans only in a limited capacity: the 
albedo and heat capacities used are the average of land and ocean 
properties. This mutes the seasonal response of land and the 
thermal inertia of water. \cite{way2017} used a 3D GCM, easily 
the most robust model of the lot, but because that model is so 
computationally expensive, only a handful of simulations were 
run. 


Here, we present the first fully coupled model of orbits, 
obliquities, and climates of Earth-like exoplanets. This model 
treats land and ocean as separate components and includes ice 
sheet growth and decay on land. Because the model is 
computationally inexpensive, thousands of coupled 
orbit-obliquity-climate simulations can be run in a reasonable 
time frame. This facilitates the exploration of broad regions of 
parameter space and will help in the prioritization of planet 
targets for characterization studies. 

The purpose of this study is to examine the effect of obliquity 
and orbital evolution on potentially habitable planets. In Paper I,
we modeled 
the orbit and obliquity of an Earth-mass planet, in the habitable 
zone of a G dwarf star, with an eccentric gas giant companion.
This ``dynamically hot'' scenario represents
an end-member case, in which the orbital evolution has
a large impact on the climate of the planet, without 
catastrophic destruction of the planetary system.
In this paper, we couple the climate model 
described in Section \ref{sec:climatemodel} to the orbit and 
obliquity model and analyze the ultimate climate state of the 
planet. In a number of interesting scenarios, we apply a 
fully-analytic climate model \citep{rose2017} to gain some deeper 
understanding of the results. Finally, we revisit the G dwarf
systems from \cite{armstrong2014} with this new climate model
to update the results in that paper.

\section{Methods}
We use a combination of a secular orbital model
(\texttt{DISTORB}), an N-Body model (\texttt{HNBody} 
\citep{rauch2002}), a 
secular obliquity model (\texttt{DISTROT}), and a one-dimensional (1D) latitudinal 
energy balance model (EBM) with ice-sheets. For a more detailed 
description of \texttt{DISTORB} and 
\texttt{DISTROT}, and a description of how we employ the N-Body 
model, see Paper I. We describe the EBM and ice-sheet model below.

\subsection{Climate model}
\label{sec:climatemodel}
The climate model, \texttt{POISE} (Planetary Orbit-Influenced 
Simple EBM), is a one-dimensional EBM \citep{budyko1969, sellers1969}
based on \cite{northcoakley1979}, with a number of modifications, 
foremost of which is the inclusion of a model of ice sheet 
growth, melting, and flow. The model is one-dimensional in $x = 
\sin{\phi}$, where $\phi$ is the latitude. In this fashion, 
latitude cells of size $dx$ will not have equal width in 
latitude, but will be equal in area. The general energy balance 
equation is:
\begin{align}
&\begin{aligned}
C(x) \frac{\partial T}{\partial t}(x,t) -& D(x,t) \nabla^2 T(x,t) + I(x,T,t) = S(x,t) (1-\alpha(x,T,t)), \label{eqn:ebmeq} \\
\end{aligned}
\end{align}
where $C(x)$ is the heat capacity of the surface at location $x$, 
$T$ is the surface temperature, $t$ is time, $D$ is the 
coefficient of heat diffusion between latitudes (due to 
atmospheric circulation), $I(x,t)$ is the outgoing long-wave 
radiation (OLR) to space (i.e., the thermal infrared 
flux), $S(x,t)$ is the incident insolation (stellar flux), and 
$\alpha$ is the planetary albedo and represents the 
percent of the insolation that is reflected back into space.

Though the model lacks a true longitudinal dimension, each 
latitude is divided into a land portion and a water portion. The 
land and water have distinct heat capacities and albedos, and 
heat is allowed to flow between the two regions. The energy 
balance equation can then be separated into two equations, one equation for the water component and one for the land component:
\begin{align}
&\begin{aligned}
C_L \frac{\partial T_L}{\partial t} - D \frac{\partial}{\partial x} (1-x^2) \frac{\partial T_L}{\partial x} +& \frac{\nu}{f_L} (T_L-T_W) + I(x,T_L,t) \\& = S(x,t) (1-\alpha(x,T_L,t)),\label{eqn:ebland}\\
\end{aligned}\\
&\begin{aligned}
C_W^{eff} \frac{\partial T_W}{\partial t} - D \frac{\partial}{\partial x} (1-x^2) \frac{\partial T_W}{\partial x} + &\frac{\nu}{f_W} (T_W-T_L) + I(x,T_W,t)\\& = S(x,t) (1-\alpha(x,T_W,t)),\label{eqn:ebwater}\\
\end{aligned}
\end{align}
where we have employed the co-latitudinal component of the 
spherical Laplacian, $\nabla^2$ (the radial and 
longitudinal/azimuthal components vanish). The effective heat 
capacity of the ocean is $C_W^{eff} = m_d C_W$, where $m_d$ is 
an adjustable parameter representing the 
mixing depth of the ocean.
The parameter $\nu$ is used to adjust the land-ocean heat 
transfer to reasonable values, and $f_L$ and $f_W$ are the 
fractions of each latitude cell that are land and ocean, 
respectively.

The insolation (or solar/stellar flux) received as a function of 
latitude, $\phi$, and declination of the host star, $\delta$, is 
calculated using the formulae of \cite{berger1978}. Declination, 
$\delta$, varies over the course of the planet's orbit for 
nonzero obliquity. For Earth, for example, $\delta \approx 
23.5^{\circ}$ at the northern summer solstice, $\delta = 
0^{\circ}$ at the equinoxes, and $\delta \approx -23.5^{\circ}$ 
at the northern winter solstice. Because $\delta$ is a function 
of time (or, equivalently, orbital position), the insolation varies, and gives 
rise to the seasons (again, assuming the obliquity is nonzero). 
For latitudes and times where there is no sunrise (e.g., polar 
darkness during winter):
\begin{equation}
S(\phi,\delta) = 0,
\end{equation}
while for latitudes and times where there is no sunset:
\begin{equation}
S(\phi,\delta) = \frac{S_{\star}}{\rho^2} \sin{\phi} \sin{\delta},
\end{equation}
and for latitudes with a normal day/night cycle:
\begin{equation}
S(\phi,\delta) = \frac{S_{\star}}{\pi \rho^2} (H_0 \sin{\phi} \sin{\delta} + \cos{\phi} \cos{\delta} \sin{H_0}).
\end{equation}
Here, $S_{\star}$ is the solar/stellar constant (in W m$^{-2}$), 
$\rho$ is the distance between the planet and host star 
normalized by the semi-major axis (\emph{i.e.} $\rho = r/a$), and 
$H_0$ is the hour angle of the of the star at sunrise and sunset, 
and is defined as:
\begin{equation}
\cos{H_0} = - \tan{\phi}\tan{\delta}.
\end{equation}
The declination of the star with respect to the planet's 
celestial equator is a simple function of its obliquity $\varepsilon$ 
and its true longitude $\theta$:
\begin{equation}
\sin{\delta} = \sin{\varepsilon} \sin{\theta}.
\label{eqn:decl}
\end{equation}
See also \cite{laskar1993} for a comprehensive derivation. For 
these formulas to apply, the true longitude should be defined as 
$\theta = f + \Delta^*$, where $f$ is the true anomaly (the 
angular position of the planet with respect to its periastron) and 
$\Delta^*$ is the angle between periastron and the planet's position at
its northern spring equinox, given by 
\begin{equation}
    \Delta^* = \varpi + \psi + 180^{\circ}.
\end{equation}
Above, $\varpi$ is the longitude of periastron, and $\psi$ is
the precession angle. 
Note that we add $180^{\circ}$ because of the convention of 
defining $\psi$ based on the vernal point, $\vernal$, which is
the position of the \emph{sun} at the time of the northern
spring equinox. For exoplanets, there is likely a more sensible
definition, however, we adhere to the Earth conventions for the sake
of consistency with past literature.

A point of clarification is in order: EBMs (at least, the models 
employed in this study) can be either \emph{seasonal} or 
\emph{annual}. The EBM component of \texttt{POISE} is a seasonal 
model---the variations in the insolation throughout the 
year/orbit are resolved and the temperature of the surface at 
each latitude varies in response, according to the leading terms 
in Equations (\ref{eqn:ebland}) and (\ref{eqn:ebwater}). In an annual 
model (we utilize one in this study to understand ice sheet 
stability; see Section \ref{sec:rosemodel}), the insolation at 
each latitude is averaged over the year, and the energy balance 
equation (Eq. \ref{eqn:ebmeq}) is forced into ``steady state'' by 
setting $\partial T/\partial t$ equal to zero (this can be done 
numerically or analytically). By ``steady state'', we mean that 
the surface conditions (temperature and albedo) come to final 
values and remain there. Seasonal EBMs, on the other 
hand, can be in ``equilibrium'', in that the orbitally averaged 
surface conditions remain the same from year to year, but the 
surface conditions vary \emph{throughout} the year.

The planetary albedo is a function of surface type (land or 
water), temperature, and zenith angle. For land grid cells, the 
albedo is:

\begin{equation}
\alpha = \left\{ \begin{array}{cc}
			\alpha_L + 0.08 P_2(\sin{Z}) & \begin{array}{c}\hspace{1mm} \text{if } M_{\text{ice}} = 0 \text{ and } T > -2^{\circ} \text{ C} \end{array}\\
			\alpha_i & \begin{array}{c} \hspace{1mm}  \text{if }M_{\text{ice}} > 0\text{ or } T <= -2^{\circ} \text{ C},
			\end{array}\\
		     \end{array} \right.
\label{eqn:albland}
\end{equation}
while for water grid cells it is:
\begin{equation}
\alpha = \left\{ \begin{array}{cc}
			\alpha_W + 0.08 P_2(\sin{Z}) & \hspace{1mm} \text{if } T > -2^{\circ} \text{ C}\\
			\alpha_i &\hspace{4mm}  \text{if } T <= -2^{\circ} \text{ C},\\
		     \end{array} \right.
		     \label{eqn:albwater}
\end{equation}
where $Z$ is the zenith angle of the sun at noon and $P_2(x) = 
1/2 (3 x^2-1)$ (the second Legendre polynomial). This last 
quantity is used to approximate the additional reflectivity seen 
at shallow incidence angles, \emph{e.g.} at high latitudes on 
Earth. The zenith angle at each latitude is given by 
\begin{equation}
Z = | \phi - \delta |.
\end{equation}

The albedos, $\alpha_L$, $\alpha_W$ (see Table 
\ref{tab:ebm_tab}), not accounting for zenith angle effects, are 
chosen to match Earth data \citep{northcoakley1979} and account, 
over the large scale, for clouds, various surface types, and 
water waves. Additionally, the factor of $0.08$ in Equations 
(\ref{eqn:albland}) and (\ref{eqn:albwater}) is chosen to reproduce the 
albedo distribution in \cite{northcoakley1979}. The functional form of Equations \ref{eqn:albland} and \ref{eqn:albwater} is also given by \cite{northcoakley1979}---those authors fit Earth measurements using Fourier-Legendre series, finding that the dominant albedo term is the second order Legendre polynomial.
The ice albedo, $\alpha_i$, is a single value that
does not depend on zenith angle due to the fact that ice
tends to occur at high zenith angle, so that the zenith angle
is essentially already accounted for in the choice of $\alpha_i$.
Equation (\ref{eqn:albland}) indicates that when there is 
ice on land ($M_{\text{ice}}>0$), or the temperature is below freezing, 
the land takes on the albedo of ice. Though there are multiple 
conditionals governing the albedo of the land, in practice the 
temperature condition is only used when ice sheets are turned off 
in the model, since ice begins to accumulate at $T = 0^{\circ}$ 
C, and so is always present when $T < -2^{\circ}$ C. Equation 
(\ref{eqn:albwater}) indicates a simpler relationship for the albedo 
over the oceans: when it is above freezing, the albedo is that of 
water (accounting also for zenith angle effects); when it is 
below freezing, the albedo is that of ice.

We take the land fraction and water fraction to be constant 
across all latitudes. This is roughly like having a single 
continent that extends from pole to pole. The effect of geography 
on the climate is beyond the scope of this work, which is to 
isolate the orbitally-induced climate variations.

Like \cite{budyko1969} and subsequent studies, including \cite{northcoakley1979}, we utilize a linearization of the OLR with 
temperature:
\begin{equation}
I = A + BT, \label{eqn:olrnc}.\\
\end{equation}
We adopt the values for Earth as determined by \cite{northcoakley1979}: $A = 203.3$ W m$^{-2}$ and $B = 2.09$ W 
m$^{-2}$ $^{\circ}$C$^{-1}$, and $T$ is the surface temperature 
in $^{\circ}$C. The purpose of this linearization is that it 
allows the coupled set of equations to be formulated as a matrix 
problem that can be solved using an implicit Euler scheme \citep{press1987}
with 
the following form:
\begin{equation}
\mathscr{M}\cdot T_{n+1} = \frac{C T_n}{\Delta t} - A + S (1-\alpha),\label{eqn:matrixform}\\
\end{equation}
where $T_n$ is a vector containing the current surface 
temperatures, $T_{n+1}$ is a vector representing the temperatures 
to be calculated, and $C$, $A$, $S$, and $\alpha$ are vectors 
containing the heat capacities, OLR offsets (Equation 
\ref{eqn:olrnc}), insolation at each latitude, and albedos, 
respectively. The matrix $\mathscr{M}$ contains all of the 
information on the left-hand sides of Equations \ref{eqn:ebland} and 
\ref{eqn:ebwater} related to temperature. The time-step, $\Delta t$, 
is chosen so that conditions do not change significantly between 
steps, resulting in typically 60 to 80 time-steps per orbit. The 
new temperature values can then be calculated by taking the 
dot-product of $\mathscr{M}^{-1}$ with the right-hand side of 
Equation \ref{eqn:matrixform}. The large time step allowed by this 
integration scheme greatly speeds the climate model, allowing us 
to run thousands of simulation for millions of years.

The ice sheet model consists of three components: mass balance (that is, local ice accumulation and ablation), longitudinal flow across the surface, and isostatic rebound of the bedrock. Longitudinal flow ensures that the ice sheets maintain a realistic size and shape, for example, they do not grow to unrealistic heights at the poles, while bedrock rebound is necessary to accurately model ice flow.

We model ice accumulation and ablation in a similar fashion to 
\cite{armstrong2014}. Ice accumulates on land at a constant rate, 
$r_{\text{snow}}$, when temperatures are below 0$^{\circ}$ C. 
Melting/ablation occurs when ice is present and temperatures are 
above 0$^{\circ}$ C, according to the formula:
\begin{equation}
\frac{dM_{\text{ice}}}{dt} = \frac{2.3 \sigma (T_{\text{freeze}}^4 - (T+T_{\text{freeze}})^4)}{L_h},
\label{eqn:ablation}
\end{equation}
where $M_{\text{ice}}$ is the surface mass density of ice, $\sigma = 
5.67 \times 10^{-8}$ W m$^{-2}$ K$^{-4}$ is the Stefan-Boltzmann 
constant, $L_h$ is latent heat of fusion of ice, $3.34 \times 
10^5$ J kg$^{-1}$ and $T_{\text{freeze}} = 273.15$ K. The factor of 2.3 
that appears here, though not in \cite{armstrong2014}, is added 
to scale the melt rate to roughly Earth values of 3 mm 
$^{\circ}$C$^{-1}$ day$^{-1}$ 
\citep[see][]{braithwaitezhang2000,Lefebre2002,huybers2008}. 

The ice sheets flow across the surface via 
deformation and sliding at the base. We use the formulation from 
\cite{huybers2008} to model the changes in ice height due to 
these effects. Bedrock depression is moderately important in this model 
(despite the fact that we have only one atmospheric layer and 
thus do not resolve elevation-based effects), because the flow 
rate is affected. This ultimately affects the ice sheet height---without the bedrock component, the ice sheets grow to be $\sim10\%$ taller, but less massive (see Section \ref{sec:repmilank}). The ice flow \citep[via][]{huybers2008} is:
\begin{align}
&\begin{aligned}
\frac{\partial h}{\partial t} = \frac{\partial}{\partial y} &\left[ \frac{2A_{\text{ice}}(\rho_i g)^n}{n+2} \left | \left ( \frac{\partial (h+H)}{\partial y} \right )^{n-1} \right | \right. \left. \cdot \frac{\partial (h+H)}{\partial y}~ (h+H)^{n+2} + u_b h \right], \label{eqn:iceflow1}\\
\end{aligned}
\end{align}
where $h$ is the height of the ice, $H$ is the height of the 
bedrock (always negative or zero, in this case), $A_{\text{ice}}$ 
represents the deformability of the ice, $\rho_i$ is the density 
of ice, $g$ is the acceleration due to gravity, and $n$ is the 
exponent in Glen's flow law \citep{glen1958}, where $n=3$. The ice height and ice 
surface mass density, $M_{\text{ice}}$ are simply related via $M_{\text{ice}} = 
\rho_i h$. The first term inside the derivative represents the 
ice deformation; the second term is the sliding of the ice at the base. 
The latitudinal coordinate, $y$, is related to the radius of the 
planet and the latitude, $y = R \phi$, thus $\Delta y = R \Delta 
x (1-x^2)^{-1/2}$. Finally, $u_b$, the ice velocity across the 
sediment, is:
\begin{align}
&\begin{aligned}
u_b = &\frac{2 D_0 a_{\text{sed}}}{(m+1)b_{\text{sed}}} \left ( \frac{ |a_{\text{sed}}|}{2D_0 \mu_0} \right )^m \cdot  \left ( 1 - \left [ 1- \frac{b_{\text{sed}}}{|a_{\text{sed}}|} \min \left ( h_s,\frac{|a_{\text{sed}}|}{b_{\text{sed}}} \right ) \right ]^{m+1} \right ), \label{eqn:basalvel}
\end{aligned}
\end{align}
as described by \cite{jenson1996}. The constant $D_0$ represents 
a reference deformation rate for the sediment, $\mu_0$ is the 
reference viscosity of the sediment, $h_s$ is the depth of the 
sediment, and $m=1.25$. The shear stress from the ice on the 
sediment is:
\begin{equation}
a_{\text{sed}} = \rho_i g h \frac{\partial (h+H)}{\partial y}, \label{eqn:shear}
\end{equation}
and the rate of increase of shear strength with depth is:
\begin{equation}
b_{\text{sed}} = (\rho_s-\rho_w)g \tan{\phi_s},
\end{equation}
where $\rho_s$ and $\rho_w$ are the density of the sediment and 
water, respectively, and $\phi_s$ is the internal deformation 
angle of the sediment. We adopt the same numerical values as 
\cite{huybers2008} for all parameters related to ice and sediment 
(see Table \ref{tab:ice_tab}), with a few exceptions. We use a value 
of $A_{\text{ice}}$ (ice deformability) that is consistent with ice at 
270 K \citep{paterson1994}, and a value of $r_{\text{snow}}$ (the 
precipitation rate) that best allows us to reproduce Milankovitch 
cycles on Earth (see Section \ref{sec:valid}). Note also that the 
value of $D_0$ in Table A2 of \cite{huybers2008} appears to be 
improperly converted for the units listed (the correct value, 
from \cite{jenson1996}, is listed in the text, however). With 
Equations (\ref{eqn:basalvel}) and (\ref{eqn:shear}), Equation (\ref{eqn:iceflow1})
can be treated numerically as a diffusion equation, with the 
form:
\begin{equation}
\frac{\partial h}{\partial t} = D_{\text{ice}} \frac{\partial^2 (h+H)}{\partial y^2},\\
\end{equation}
where,
\begin{align}
&\begin{aligned}
D_{\text{ice}} & = \frac{2A_{\text{ice}}(\rho_i g)^n}{n+2} \left | \left ( \frac{\partial (h+H)}{\partial y} \right )^{n-1} \right | ~ (h+H)^{n+2}\\
&+\frac{2 D_0 \rho_i g h^2}{(m+1)b_{\text{sed}}} \left ( \frac{ |a_{\text{sed}}|}{2D_0 \mu_0} \right )^m \cdot  \left ( 1 - \left [ 1- \frac{b_{\text{sed}}}{|a_{\text{sed}}|} \min \left ( h_s,\frac{|a_{\text{sed}}|}{b_{\text{sed}}} \right ) \right ]^{m+1} \right ),
\end{aligned}
\end{align}
and $D_{\text{ice}}$ is evaluated at each time-step, at every boundary 
to provide mass continuity. We solve the diffusion equation numerically using
a Crank-Nicolson scheme \citep{crank1947}.

The bedrock depresses and rebounds locally in response to the changing weight of ice above, always seeking isostatic equilibrium. The equation governing the bedrock height, $H$, is \citep{clark1998,huybers2008}:
\begin{equation}
    \frac{\partial H}{\partial t} = \frac{1}{T_b}\left( H_{eq} - H - \frac{\rho_i h}{\rho_b} \right),
    \label{eqn:brock}
\end{equation}
where $T_b$ is a characteristic relaxation time scale, $H_{eq} = 0$ is the ice-free equilibrium height, and $\rho_b$ is the bedrock density. We again adopt the values used by \cite{huybers2008} (see Table \ref{tab:ice_tab}).

Because of the longer time-scales (years) associated with the ice 
sheets, the growth/melting and ice-flow equations are run asynchronously in \texttt{POISE}. First, the 
EBM (Equation \ref{eqn:ebmeq}) is run for 4-5 
orbital periods, and ice accumulation and ablation is tracked 
over this time frame, but ice-flow (Equation \ref{eqn:iceflow1}) is 
ignored. The annually-averaged ice accumulation/ablation is then 
calculated from this time-frame and passed to the ice-flow 
time-step, which can be much longer (years). The EBM is then 
re-run periodically to update accumulation and ablation and ensure that conditions vary smoothly and 
continuously.

To clarify, the hierarchy of models and their time-steps is as 
follows:
\begin{enumerate}
\item The EBM (shortest time-step): run for a duration of several 
orbital periods with time-steps on the order of days. The model 
is then rerun at the end of every orbital/obliquity time-step and 
at user-set intervals throughout the ice-flow model.
\item The ice-flow model (middle time-step): run at the end of 
every orbital time-step (with time-steps of a few orbital 
periods), immediately after the EBM finishes. The duration of the 
model will follow one of two scenarios:
\begin{enumerate}[label=\alph*]

\item If the orbital/obliquity time-step is sufficiently long, 
the EBM is rerun at user-set intervals, then the ice-flow model 
continues. The ice-flow model and the EBM thus alternate 
back-and-forth until the end of the orbit/obliquity time-step. 
\item If the orbital/obliquity time-step is shorter than the 
user-set interval, the ice-flow model simply runs until the end 
of the orbital time-step. 
\end{enumerate}
\item The orbital/obliquity model (longest time-step). The 
time-steps are set by the fastest changing variable (see Paper I) amongst those parameters.
\end{enumerate}

\begin{figure}
\includegraphics[width=\textwidth]{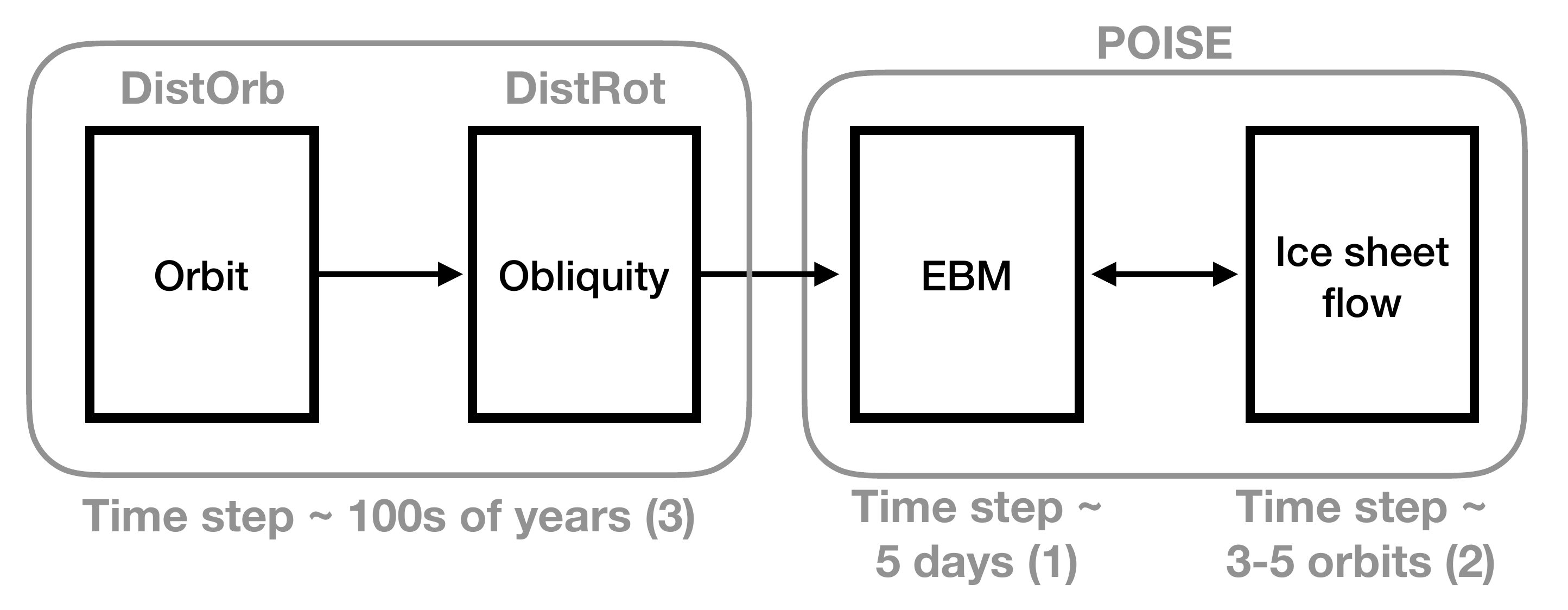}
\caption{\label{fig:poisestruct} Hierarchy of \texttt{POISE} and 
the orbit and obliquity models. The orbit and obliquity models 
(\texttt{DISTORB} and \texttt{DISTROT}) are run for $\sim$ 
hundreds of years (with an adaptive time step determined by the 
rates of change of the orbital/obliquity parameters). 
\texttt{POISE} is run at the end of each orbit/obliquity time 
step. First, the EBM is run for several orbits, with time steps 
of $\sim$ 5 days. Then the ice flow model is run with time steps 
of $\sim 3-5$ orbits. The ice flow model runs until the next 
orbit/obliquity time step, or until a user-set time, at which 
point the EBM is rerun for several orbits.}
\end{figure}

This approach is shown schematically in 
Figure \ref{fig:poisestruct}. The user-set interval 
discussed above must be considered carefully. The assumption is 
that annually-averaged climate conditions like surface 
temperature and albedo do not change much during the time span 
over which the ice-flow model runs. Hence, we choose a value that 
ensures that the ice-flow does not run so long that it 
dramatically changes the albedo without updating the temperature 
and ice balance (growth/ablation) via the EBM.

The initial conditions for the EBM are as follows. The first time the 
EBM is run, the planet has zero ice mass on land, the temperature 
on both land and water is set by the function
\begin{equation}
T_0 = 7.5^{\circ}\text{C} + (20^{\circ}\text{C})(1-2\sin^2{\phi}),\label{eqn:inittemp}
\end{equation}
where $\phi$ is the latitude. This gives the planet a mean 
temperature of $\sim14^{\circ}$ C, ranging from 
$\sim28^{\circ}$ C in the tropics to $\sim-13^{\circ}$ at the 
poles. This is thus a ``warm start'' condition. The initial albedo 
of the surface is calculated from the initial temperatures. We 
then perform a ``spin-up'' phase, running the EBM iteratively 
until the mean temperature between iterations changes by 
$<0.1^{\circ}$ C, \emph{without} running the orbit, obliquity, or 
ice-flow models, to bring the seasonal EBM into equilibrium at 
the actual stellar flux the planet receives and its actual 
initial obliquity. Then, every time the EBM is rerun (at the 
user-set interval or the end of the orbit/obliquity time-step), 
the initial conditions are taken from the previous EBM run 
(temperature distribution) and the end of the ice-flow run 
(albedo, ice mass).

\begin{table}
\caption{\textbf{Parameters used in the EBM}}
\centering
\begin{tabular}{lllp{0.5\linewidth}}
\hline\hline \\ [-1.5ex]
Variable & Value & Units & Physical description \\ [0.5ex]
\hline \\ [-1.5ex]
$C_L$ & $1.55 \times 10^7$ & J m$^{-2}$ K$^{-1}$ & land heat capacity \\
$C_W$ & $4.428 \times 10^6$ & J m$^{-2}$ K$^{-1}$ m$^{-1}$ & ocean heat capacity per meter of depth \\
$m_d$ & 70 & m & ocean mixing depth \\
$D$ & 0.58 & W m$^{-2}$ K$^{-1}$ & meridional heat diffusion coefficient\\
$\nu$ & 0.8 &  & coefficient of land-ocean heat flux\\
$A$ & 203.3 & W m$^{-2}$ & OLR parameter\\
$B$ & 2.09 & W m$^{-2}$ K$^{-1}$& OLR parameter\\
$\alpha_L$ & 0.363 &  & albedo of land \\
$\alpha_W$& 0.263 &  & albedo of water \\
$\alpha_i$& 0.6  & & albedo of ice\\
$f_L$ & 0.34 & & fraction of latitude cell occupied by land\\
$f_W$ & 0.66 & & fraction of latitude cell occupied by water\\
\hline 
\end{tabular}
\label{tab:ebm_tab}
\end{table}

\begin{table}
\caption{\textbf{Parameters used in the ice sheet model}}
\centering
\begin{tabular}{lllp{0.5\linewidth}}
\hline\hline \\ [-1.5ex]
Variable & Value & Units & Physical description \\ [0.5ex]
\hline \\ [-1.5ex]
$T_{freeze}$ & 273.15 & K & freezing point of water \\
$L_h$ & $3.34 \times 10^5$ & J kg$^{-1}$ & latent heat of fusion of water\\
$r_{\text{snow}}$ & $2.25 \times 10^{-5}$ & kg m$^{-2}$ s$^{-1}$ & snow/ice deposition rate \\
$A_{\text{ice}}$ & $2.3 \times 10^{-24}$ & Pa$^{-3}$ s$^{-1}$ & deformability of ice \\
$n$ & 3 & & exponent of Glen's flow law \\
$\rho_i$ & 916.7 & kg m$^{-3}$ & density of ice \\
$\rho_s$ & 2390 & kg m$^{-3}$ & density of saturated sediment \\
$\rho_w$ & 1000 & kg m$^{-3}$ & density of liquid water\\
$D_0$ & $7.9 \times 10^{-7}$ & s$^{-1}$ & reference sediment deformation rate\\
$\mu_0$ & $3 \times 10^9$ & Pa s & reference sediment viscosity \\
$m$ & 1.25 & & exponent in sediment stress-strain relation\\
$h_s$ & 10 & m & sediment depth \\
$\phi_s$ & 22 & degrees & internal deformation angle of sediment \\
$T_b$ & 5000 & years & bedrock depression/ rebound timescale\\
$\rho_b$ & 3370 & kg m$^{-3}$ & bedrock density\\
\hline 
\end{tabular}
\label{tab:ice_tab}
\end{table}

\subsection{Analytical solution for ice stability}
\label{sec:rosemodel}

To better understand the snowball instability, we compare our 
results to the analytical EBM from \cite{rose2017}. Their model is 
an \emph{annual} EBM and is analytic in that the solution is algebraic, rather than numerical. While this model does 
not capture seasonal variations or the thermal inertia associated 
with ice sheets, it is nonetheless instructive for understanding 
how the snowball state is triggered. We utilize the Python 
code\footnote{Available at https://github.com/brian-rose/ebm-analytical} 
developed
by those authors for our results in Section \ref{sec:icestab}.

According to the ``slope-stability theorem'' \citep{cahalan1979}, 
the ice edge is stable as long as
\begin{equation}
\frac{dq}{dx_s} > 0,
\label{eqn:slopestable1}
\end{equation}
where $x_s = \sin{\phi_s}$, $\phi_s$ is the latitude of the ice 
edge (land and ocean are not separate component in the analytic 
model), and $q$ is the non-dimensional quantity
\begin{equation}
q = \frac{a_0 Q}{A+B T_{\text{ref}}}.
\end{equation}
The quantity $q$ represents the absorbed solar/stellar radiation, 
divided by the planet's cooling function (or outgoing longwave 
radiation) at some temperature. Thus, it is analogous to the total 
heating that the planet receives, both from the host star and its 
own greenhouse effect. Here, $Q$ is the global average incoming flux ($4Q$ is the 
solar/stellar constant, $S_{\star}$) and $T_{\text{ref}}$ is the 
temperature threshold at which the planetary albedo switches from 
a value appropriate for ice free to ice covered ($T_{\text{ref}}$ is the 
freezing point, in other words). For ice free latitudes, the 
co-albedo, $a_0$, is a single value in the annual model. In our 
comparison using our seasonal model, we take this to be the 
average co-albedo of the unfrozen surfaces, $a_0 = f_L 
(1-\alpha_L)+f_W (1-\alpha_W)$, and we set $T_{\text{ref}} = -2^{\circ}$ 
C.

Equation (\ref{eqn:slopestable1}) applies to low obliquity planets. 
If the planet has high obliquity, ice will tend to form at the 
equator, and the stability condition is
\begin{equation}
\frac{dq}{dx_s} > 0. 
\label{eqn:slopestable2}
\end{equation}
In the annual model, there is a distinct boundary between ``low'' 
and ``high'' obliquity, and the transition occurs at
\begin{equation}
\varepsilon = \sin^{-1}\left (\sqrt{\frac{2}{3}} \right) \approx 54.74^{\circ}.
\end{equation}
See Equation (3b) of \cite{rose2017}. This angle is the obliquity 
at which the average annual insolation is the same at all 
latitudes.

At a single value of 
$q$, there can be multiple equilibrium locations for the ice 
edge---but only some of these ``branches'' are stable (those with 
positive or zero slopes) according to the slope-stability theorem.
At Earth's obliquity, the slope (Equation \ref{eqn:slopestable1}) is negative
at high latitudes, which gives rise to the ``small ice cap instability''
(SICI), and near the equator, giving rise to the ``large ice cap
instability'' (LICI). The slope is positive between $\sim 35^{\circ}$ and 
$\sim 80^{\circ}$---in other words, an ice cap extending to this range of
latitudes is stable.

As we will show, this stability concept is useful in 
understanding how the snowball states occur in many of our 
simulations. However, because the seasonal EBM (\texttt{POISE}) 
is not an equilibrium model, it does deviate from the annual 
model at times. Hence, the ice stability diagrams that we analyze in 
Section \ref{sec:icestab} do not always 
accurately predict the occurrence of snowball states.

\subsection{Statistics and machine learning}
\label{davealgo}

To extend the predictive power and utility of the model, we calculate correlations between orbital parameters and snowball states and area of ice coverage. We then employ a machine learning algorithm to determine how often we can correctly predict the climate state of the planet considered here, given a set of orbital properties. The properties that go into this analysis are shown in Table \ref{tab:davebot}. There are 10 model inputs (orbit/spin parameters) and 2 model outputs ($\delta_{\text{snow}}$ and $f_{\text{ice}}$).The fractional ice cover, $f_{\text{ice}}$, is the fractional area of the globe that is covered in ice year-round at the end of the simulation (the last orbital time-step). The other output parameter, $\delta_{\text{snow}}$, is 1 if the planet is in a snowball state at the end of the simulation and 0 if it is not. Note that $\delta_{\text{snow}} = 1$ when the \emph{oceans} are frozen year-round; this means that there exist circumstances in which $\delta_{\text{snow}}=1$ but $f_{\text{ice}}\neq1$ (the land component can warm above freezing seasonally, even if the oceans are frozen). In practice, this only occurs when the ice sheet model is not used, as the ice significantly alters the thermal inertia of the land. It is usually the case that $\delta_{\text{snow}}=1$ when $f_{\text{ice}}=1$ and $\delta_{\text{snow}}=0$ when $f_{\text{ice}}<1$.

\begin{table}
\caption{\textbf{Parameters used in statistical analysis and machine-learning algorithm}}
\centering
\begin{tabular}{lp{0.45\linewidth}}
\hline\hline \\ [-1.5ex]
\multicolumn{1}{c}{Parameter} & \multicolumn{1}{c}{Description} \\ [0.5ex]
\hline \\ [-1.5ex]
$S$ & Incident stellar flux (stellar constant) \\
$e_0$ & Initial eccentricity\\
$\Delta e$ & Maximum change in eccentricity\\
$\langle e \rangle$ & Mean eccentricity\\
$i_0$ & Initial inclination\\
$\Delta i$ & Maximum change in inclination\\
$\langle i \rangle$ & Mean inclination\\
$\varepsilon_0$ & Initial obliquity\\
$\Delta \varepsilon$ & Maximum change in obliquity\\
$\langle \varepsilon \rangle$ & Mean obliquity\\
\hline 
$\delta_{\text{snow}}$ & Equal to 1 in snowball state, 0 otherwise\\
$f_{\text{ice}}$ & Fractional area permanently (year-round) covered in ice\\
\hline 
\end{tabular}
\label{tab:davebot}
\end{table}

We examine how the input features of our model (Table \ref{tab:davebot}) correlate with the final climate state ($\delta_{\text{snow}}$ and $f_{\text{ice}}$) to gain insight into how the underlying physical processes influence the outcomes of our simulations.  For example, if the mean eccentricity correlates with likelihood that the planet enters a snowball state, we can infer that orbital dynamical processes could influence the climate evolution.  Note that we cannot and do not seek to show causal relationships in the correlation analysis, but rather identify features that may impact the climate evolution.

The relationship between any feature of our model and the final state of the simulated planet climate likely has a non-linear correlation given the inherent complexities of our coupled orbital dynamics and climate model. To characterize these correlations, we compute the simple Pearson correlation coefficient ($R$) and the maximal information coefficient \citep[MIC;][]{reshef2011}.  Pearson's $R$ measures the linear relationship between two variables and ranges from [-1,1] with 0 representing no linear correlation and 1 and -1 represent a perfect positive and negative linear correlation, respectively. We also compute the $p-$values associated with each correlation, which are measure of statistical significance: the $p-$value indicates that there is a $p$-percent chance that the null hypothesis produces the observed correlation $R$. A $p<0.05$ is the traditional definition of significance for when testing a single hypothesis, however, since we are testing multiple hypotheses (10 in total for each climate parameter), we set the threshold for significance to $p<0.05/10$ or $p<0.005$ \citep[a Bonferroni correction;][]{dunn1959}.

The MIC characterizes non-linear relationships between variables by estimating the maximum mutual information between two variables. Mutual information between two variables characterizes the reduction in uncertainty of one variable after observing the other \citep[see][]{reshef2011}.  For independent variables, their mutual information is 0 as observing one does not provide any insight into the other.  The MIC ranges from [0,1] where MIC $= 0$ represents no relationship while  MIC $= 1$ represents some noiseless functional relationship of any form. The MIC depends on the estimate of the joint distribution of the two variables when computing the maximum mutual information and hence is sensitive to how the variables are binned.  Following the suggestion of \cite{reshef2011}, we set the number of bins to be $N^{0.6}$ for $N$ simulations. We computed the MIC using the Python package \texttt{minepy} \citep{albanese2013} for each feature versus the final surface area of ice ($f_{\text{ice}}$) and the final climate state ($\delta_{\text{snow}}$). We also define a measurement of the non-linearity associated with each parameter:
\begin{equation}
\zeta_{NL} = \text{MIC} - R^2.
\end{equation}
By subtracting out a measure of the linearity of the relationship ($R^2$, in this case), $\zeta_{NL}$ captures the degree to which the measured correlation is non-linear. This quantity allows us to probe how the coupling between our models impact a planet's final climate state as opposed to direct climate scalings. 

As an alternative method to estimate the correlation between various features and simulation results, we turn to a machine learning (ML) approach akin to that of \cite{tamayo2016}. The purpose of this method is to look for correlations not found by either of the previous methods. Following the procedure of that study, we use an ML algorithm to predict the results of our simulations as a function of the features of our model (Table \ref{tab:davebot}).  We use the \texttt{scikit-learn} \citep{pedregosa2011} implementation of the random forest algorithm \citep{brieman2001}.  The random forest algorithm is a particularly powerful and flexible algorithm that fits an ensemble of decision trees on numerous randomized sub-samples of the data set and averages the predictions of the decision trees to produce an accurate, low-variance prediction.  The random forest algorithm has a particular advantage for our purposes in that it can compute ``feature importances" as a means to estimate how the algorithm weights various inputs when producing an output.  An input with a high feature importance implies that the algorithm weights that feature more heavily when making a prediction.  Feature importances, $\xi_i$, can hence be considered as a proxy for how much that feature correlates with the predicted variable (the simulation output).  The feature importances are all normalized such that they sum to 1, \emph{i.e.}, $\sum \xi_i = 1$.

We cast our ML problem in two forms.  First, we consider the binary classification problem in which we use a random forest classifier (RFC) to predict whether or not the simulation results in a snowball planet state, $\delta_{\text{snow}}$.  Second, we consider the regression problem in which we use a random forest regressor (RFR) to predict the area fraction of the planet covered in ice, $f_{\text{ice}}$, a continuous quantity that ranges from 0 to 1.  In both cases, we fit the ML algorithms with the following procedure.  We divide our data set using 75\% of the data for our training set in which we fit and calibrate our algorithms and the remaining 25\% as the testing set used to estimate the performance of our fitted algorithms on unseen data. We fit each algorithm, a process commonly referred to as ``training", and use $k-$folds cross-validation with $k=5$ to tune the hyperparameters of our model using only the training set.  After training the algorithms, we find that both the RFC and RFR algorithms generalize exceptionally well. For example, the RFC's predictions of $\delta_{\text{snow}}$ achieve a classification accuracy of $\sim97\%$ on the testing set.  After training the models and verifying their accuracy, we extract the feature importances ($\xi_i$) for each algorithm as shown in Tables \ref{tab:daveresults1} and \ref{tab:daveresults2}. Note that in order prevent the random forest regressor (RFR) from predicting negative values for $f_{\text{ice}}$, we instead use the value $\log_{10}{(f_{\text{ice}}+1)}$ as the model output. 

\subsection{Initial orbital and obliquity conditions}

We model the climate of planet 2 in the dynamically evolving 
system, TSYS, from the previous study (Paper I), 
over a narrower range of rotational periods. This hypothetical system, 
which consists of a warm Neptune, an Earth-mass planet (planet 2), and 
a Jovian exterior to the HZ, allows us to test the effects on 
habitability of exo-Milankovitch cycles. This test system is chosen as an
end-member scenario, i.e., the effect of orbital evolution on climate is
maximized (without destabilizing the system). The initial orbital 
and spin properties are shown in Table \ref{tab:sys1table2nd}. As mentioned in Paper I,
the warm Neptune has almost no dynamical effect on the rest of the system. To 
understand the effects of orbital evolution over a range of 
stellar fluxes, we leave the semi-major axis fixed at $a = 
1.0031$ au and instead vary the luminosity of the star over the 
range $L_{\star} = 3.6 \times 10^{26}$ W to $L_{\star} = 
3.95\times10^{26}$ W. This corresponds to an incident stellar 
flux range of $S = 1304.00$ W m$^{-2}$ to $S = 1395.88$ W 
m$^{-2}$. 

\begin{table}[h]
\centering
\caption{{\bf Initial conditions for TSYS}}

\begin{tabular}{lccc}
\hline\hline \\ [-1.5ex]
Planet & 1 & 2 & 3 \\ [0.5ex]
\hline \\ [-1.5ex]

$m$ ($M_{\oplus}$) & 18.75 & 1 & 487.81 \\
$a$ (au) & 0.1292 & 1.0031 & 3.973 \\
$e$ & 0.237 & 0.001-0.4 & 0.313 \\
$i$ ($^{\circ}$) & 1.9894 & 0.001-35 & 0.02126 \\
$\varpi$ ($^{\circ}$) & 353.23 & 100.22 & 181.13 \\
$\Omega$ ($^{\circ}$) & 347.70 & 88.22 & 227.95 \\
$P_{rot}$ (days) & & 0.65,1,1.62 & \\
$\varepsilon$ ($^{\circ}$) & & 0-90 & \\
$\psi$ & & 281.78 & \\

\hline
\end{tabular}
\label{tab:sys1table2nd}
\end{table}

The planet Kepler-62 f, discussed in the previous study, requires 
some additional adjustments to the climate model because of its 
(cooler) location in the HZ and the different stellar spectrum. 
It is also interesting enough to warrant its own study and so we 
will reserve a climate analysis of this planet for a future work.

\section{Model Validation}
\label{sec:valid}
To validate the climate model, we adjust our input parameters to 
reproduce Earth-like values. We use the OLR parameters, $A$ and 
$B$, and heat diffusion coefficient $D$ from 
\cite{northcoakley1979} and surface albedos for land, water, and 
ice that give us good agreement to the data used in that paper, 
see Table \ref{tab:ebm_tab}. 

\subsection{Comparison with Earth and LMDG}
\label{sec:validcomp}
\begin{figure*}
\begin{centering}
\includegraphics[width=0.8\textwidth]{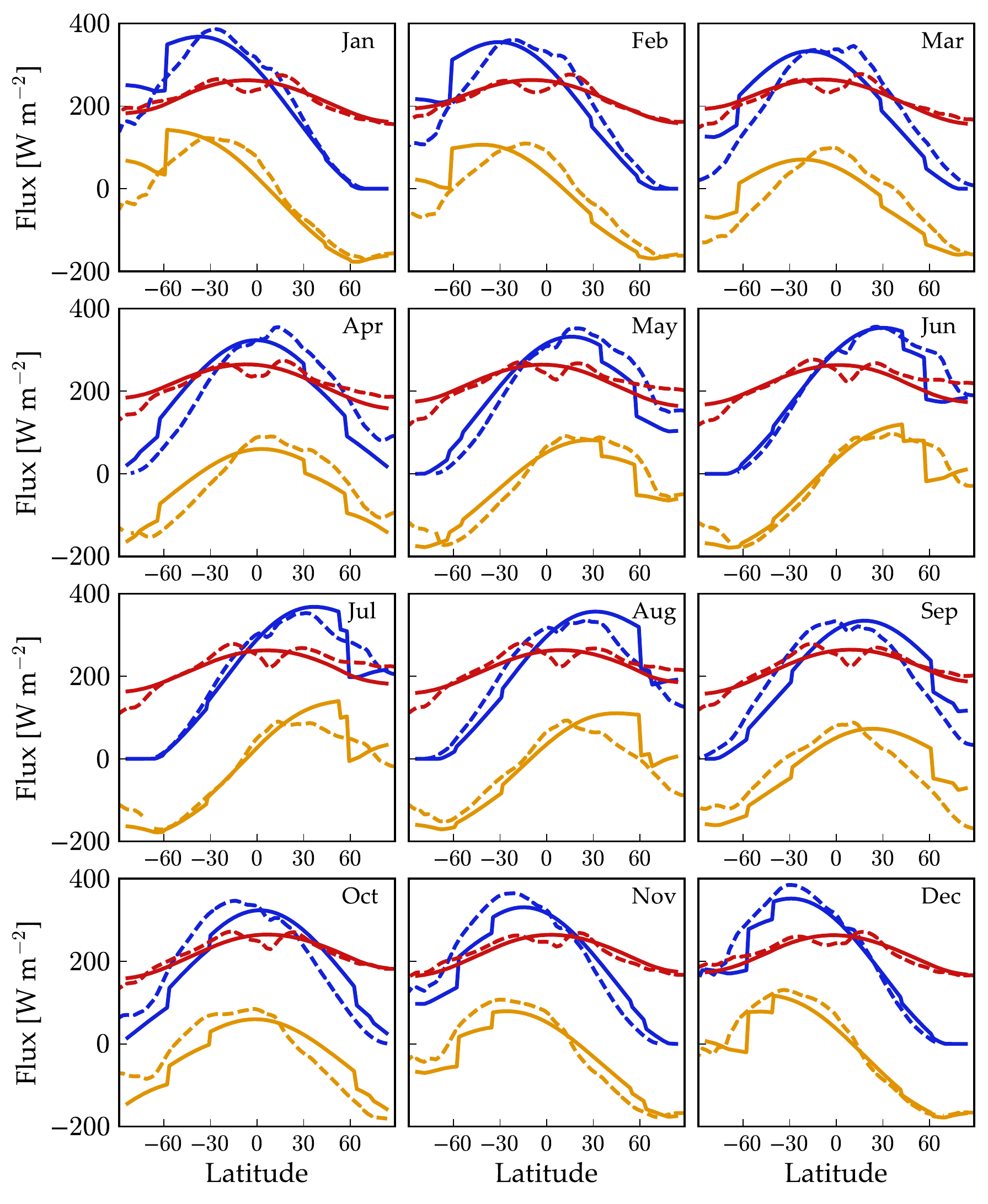}
\caption{\label{fig:monthf} Monthly averaged vertical fluxes for the 
EBM (solid lines) and satellite data for Earth (dashed lines). 
Blue corresponds to incoming flux (equal to $(1-\alpha)S(\phi)$), 
red is the outgoing long-wave radiation (OLR), and orange is the 
difference (net heating). }
\end{centering}
\end{figure*}

Like \cite{spiegel2009}, we compare our vertical heat fluxes to 
the Earth Radiation Budget Experiment satellite data 
\citep{barkstrom1990}. In Figure \ref{fig:monthf} we show the values 
for the flux in (blue), flux out (red), and the difference, or 
net heating (orange), as a function of latitude, for the Earth, 
using our climate model \texttt{POISE}. Our model compares well 
with the zonally- and monthly-averaged satellite data, though it 
is too simple to capture all of the variations. Our model also 
produces sharp jumps at high latitudes because of the sudden 
change in albedo at freezing temperatures. For the Earth, this 
sudden change is not seen because of a combination of 
geographic variations, darkening of snow and ice, 
clouds, etc., which are not captured in our model. 

\begin{figure}
\includegraphics[width=0.5\textwidth]{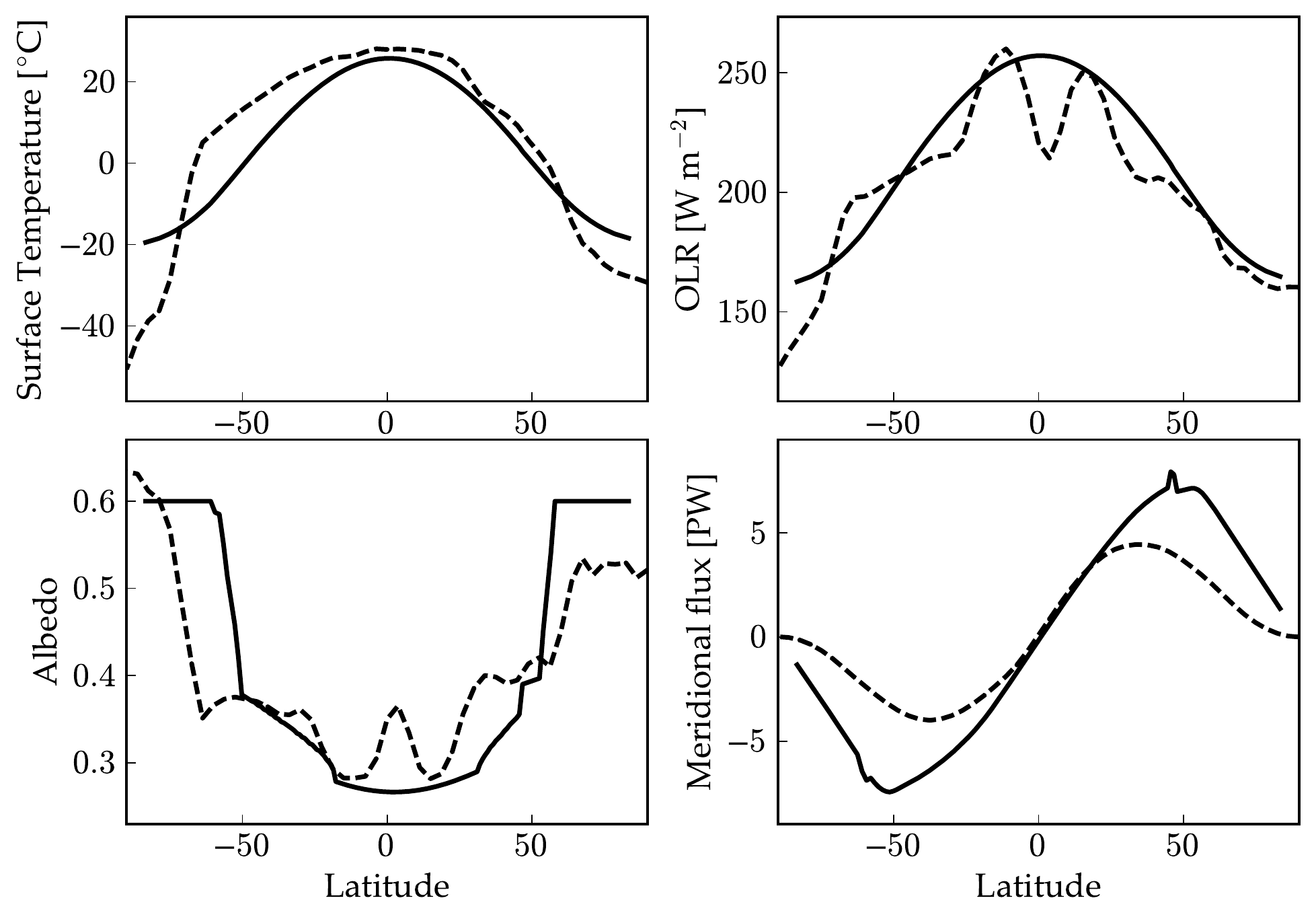}
\includegraphics[width=0.5\textwidth]{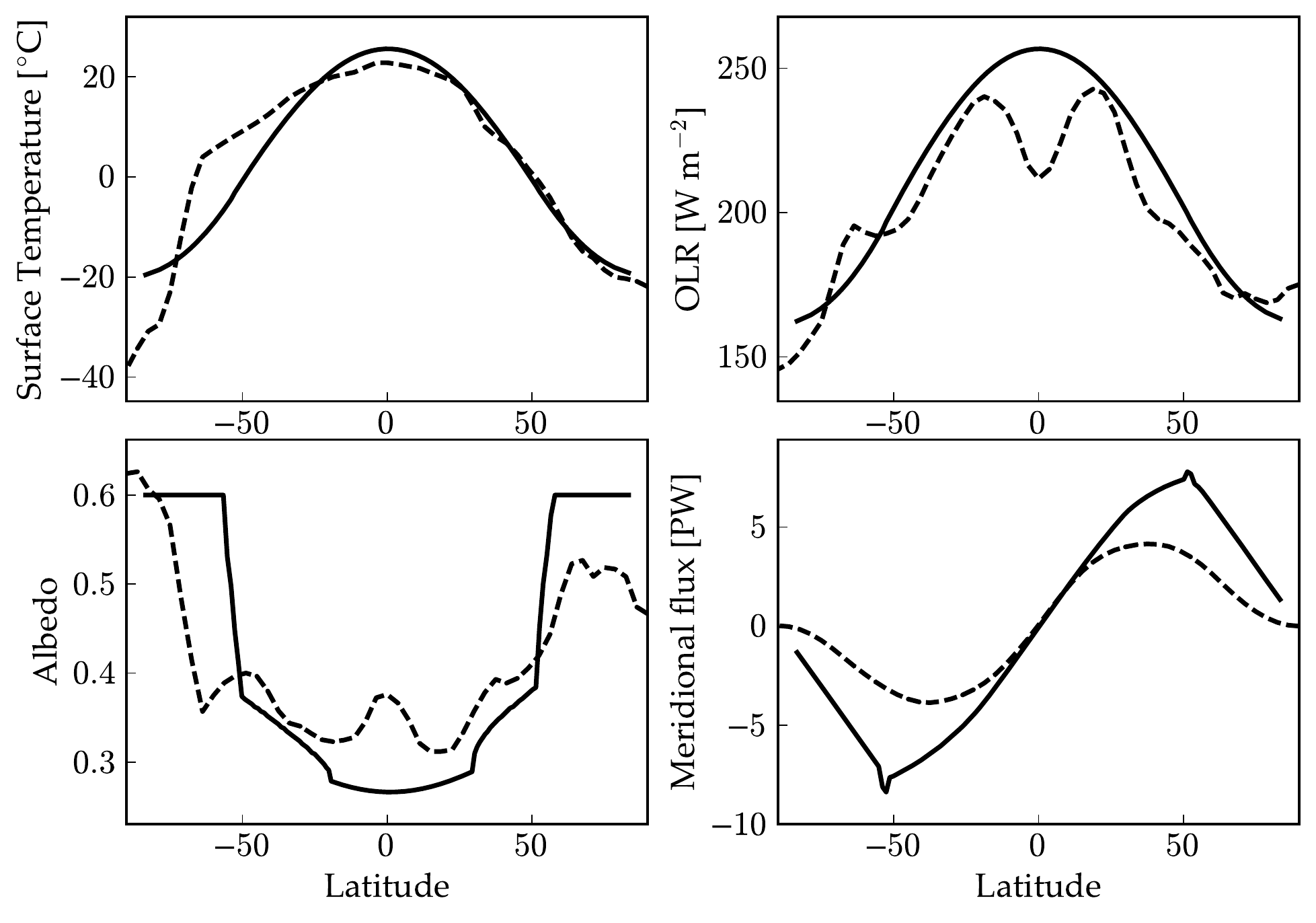}
\caption{\label{fig:lmdcompe1o1r1} Comparison between our EBM (solid 
lines) and the LMDG 3D GCM (dashed lines), for $\varepsilon = 
23.5^{\circ}$, $P_{rot} = 0.65$ day, and $e = 0.1$ (left two columns)
and $\varepsilon = 23.5^{\circ}$, $P_{rot} = 1.62$ day, and $e = 0.1$
(right two columns). 
The surface temperature, OLR, and albedo compare reasonably well to the 
zonally-averaged quantities from LMDG considering the differences 
in geography and missing physics (\emph{e.g.} clouds and Hadley 
cells). The meridional flux in the EBM peaks at $\sim 7$ PW, a 
bit higher than Earth's $\sim 6$ PW, while LMDG's peak is a tad 
low at $\sim 5$ PW. For $P_{rot} = 1.62$ day, despite the slower rotation 
the meridional flux is very similar to that of the $P_{rot} = 0.65$ 
day rotator, which suggests that parameterizations of the heat 
flux with rotation rate $\Omega$ 
\citep[$D \propto \Omega^{-2}$; see][]{williamskasting97,spiegel2008}
probably overestimate the latitudinal heat flow.}
\end{figure}

\begin{figure}
\includegraphics[width=0.5\textwidth]{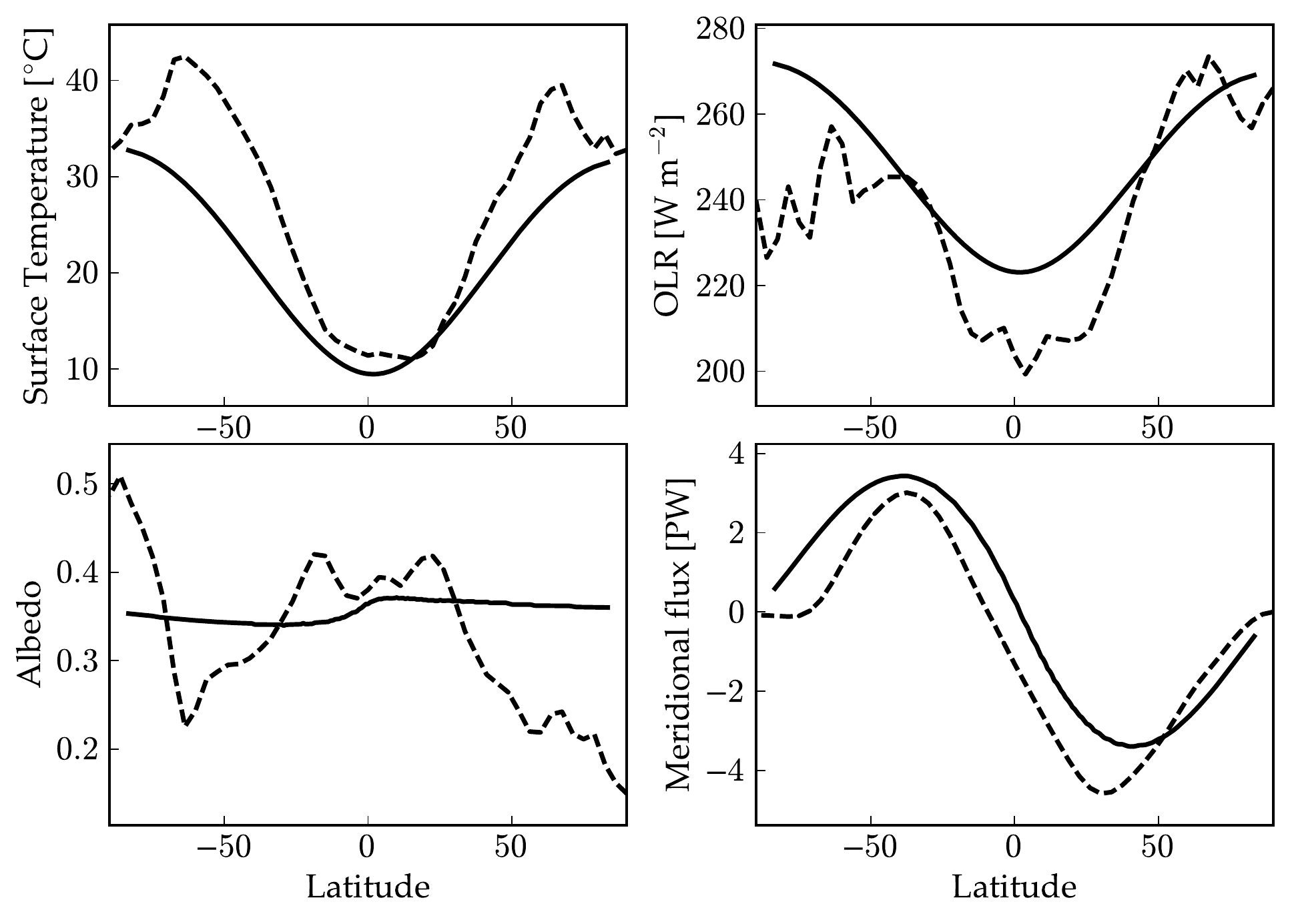}
\includegraphics[width=0.5\textwidth]{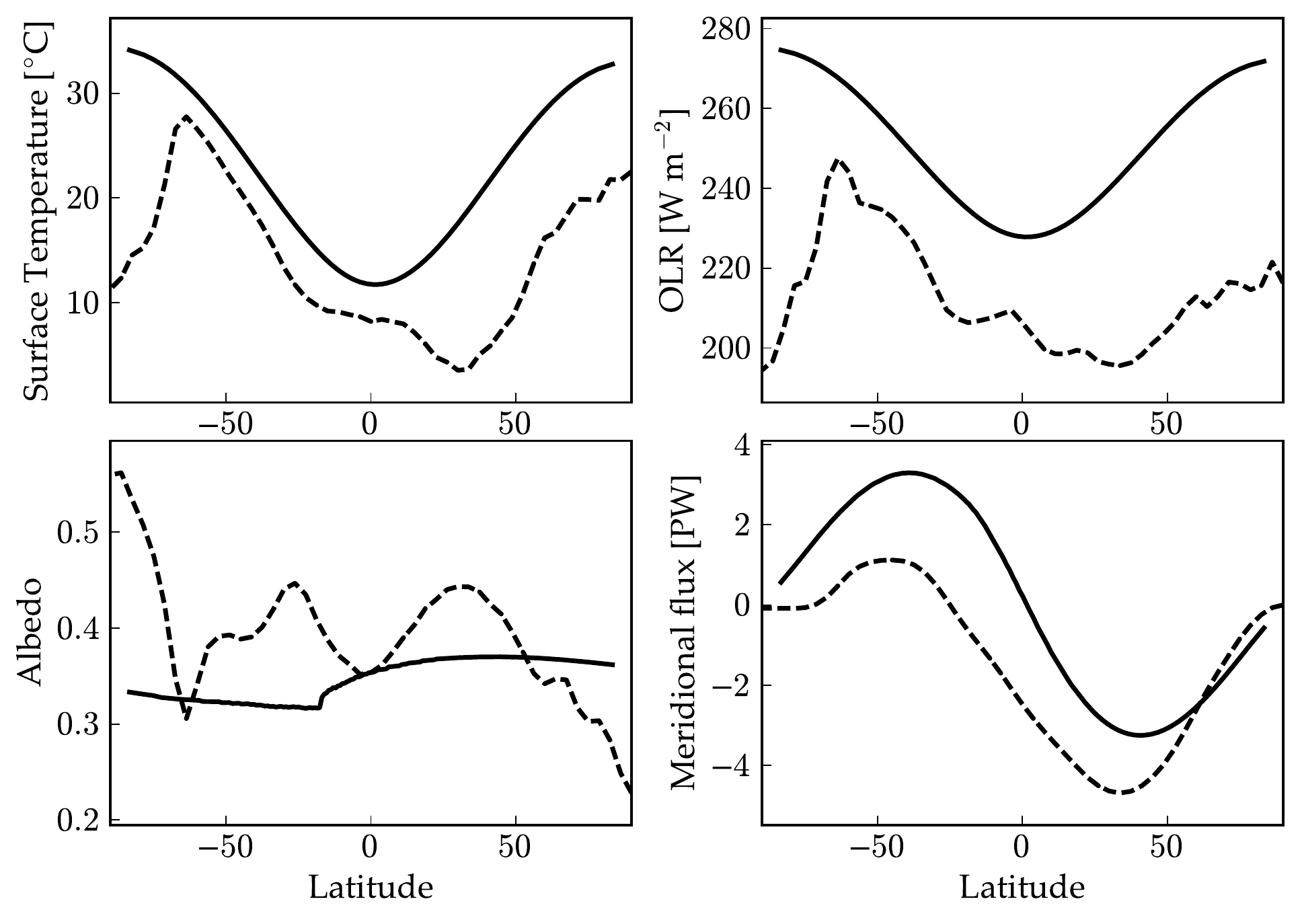}
\caption{\label{fig:lmdcompe1o1r2} Same as Figure \ref{fig:lmdcompe1o1r1}, 
but for $\varepsilon = 85^{\circ}$. The left two columns again 
correspond to $P_{rot} = 0.65$ days, the right two to $P_{rot}=1.62$ 
days. \texttt{POISE} compares worse with LMDG in these high obliquity
cases. \texttt{POISE} captures the general patterns but underestimates the 
surface temperature at mid-latitudes and overestimates the OLR at 
the equator and south pole. At high obliquity, the geography may 
play a larger role than at low obliquity, due to the extreme 
seasonality---land and ocean have different heat capacities and 
so will heat on different time-scales, possibly explaining  
the discrepancy between the models.}
\end{figure}

Further, in Figures \ref{fig:lmdcompe1o1r1}-\ref{fig:lmdcompe1o1r2}, we 
compare \texttt{POISE} to the Generic LMD 3D Global Climate 
Model (LMDG) \citep{wordsworth2011,leconte2013,leconte2013b,charnay2013},
for rotation periods of 0.65 and 1.62 days, obliquities of 
23.5$^{\circ}$ and 85$^{\circ}$, and eccentricities of $0.1$ and 
$0.3$ (eight GCM simulations in total). These initial orbital and 
rotational conditions sample a broad range of the conditions we 
explore further with the EBM. We use present Earth geography in 
the LMDG simulations, though in the EBM there is a fixed quantity 
of land at each latitude, so some difference in the models is 
attributable to geography. All LMDG simulations are started
from an initial state corresponding to present-day Earth, with present-day
topography, and run for 30 years (the typical timescale required for
convergence). 

In Figures \ref{fig:lmdcompe1o1r1}-\ref{fig:lmdcompe1o1r2} we plot the 
annually-averaged surface temperature, OLR, albedo, and meridional flux as a function of latitude 
for the \texttt{POISE} and LMDG simulations. With a climate model 
as simple as an EBM, we cannot replicate all of the 
variations with latitude in these quantities found by LMDG. 
Still, \texttt{POISE} captures LMDG's general patterns in 
surface temperature and heat fluxes. It captures the surface 
temperature better in the low obliquity cases than in the high 
obliquity cases, though, oddly, the meridional flux in 
\texttt{POISE} matches LMDG more closely in the high obliquity 
cases. 

A primary source of error in the high obliquity cases is that 
the EBM simply does not capture all of the physical processes that 
occur during the planet's extreme summers. During 
the summer, nearly an entire hemisphere experiences sunlight for 
months on end, leading to extremely high temperatures and 
strong circulation. Ultimately, the simple parameterization of 
the OLR ($I =A+B T$) probably breaks down under such conditions, 
and convection should lead to cloud formation and a change in 
albedo, similar to the effect on synchronously rotating planets
\citep{joshi2003,edson2011,edson2012,yang2013}.

\subsection{Reproducing Milankovitch Cycles}
\label{sec:repmilank}
\begin{figure*}
\begin{centering}
\includegraphics[width=0.8\textwidth]{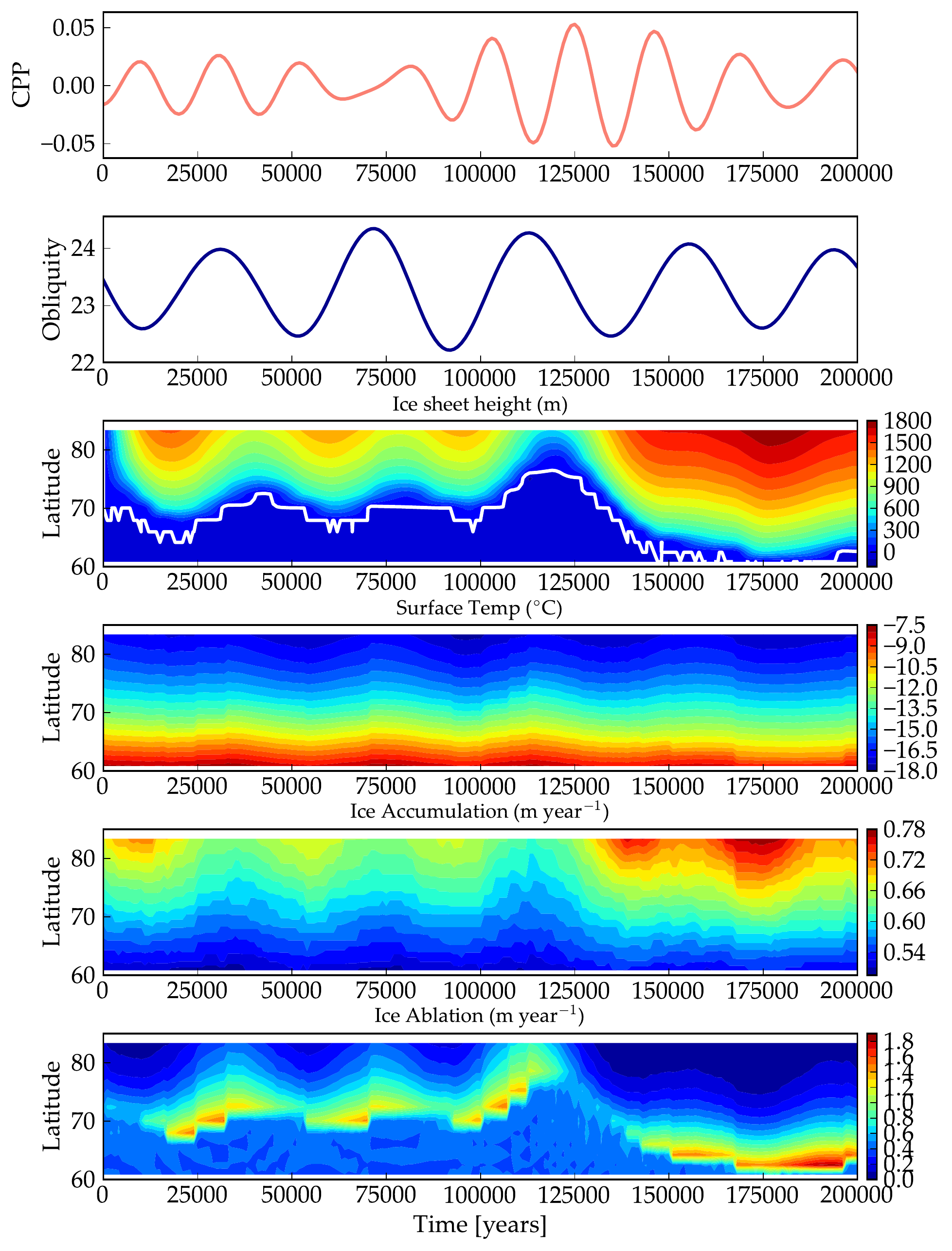}
\caption{\label{fig:huybers} Milankovitch cycles on Earth, in the 
northern hemisphere. The panels are arranged to compare with 
Figure 4 of \cite{huybers2008}. From top to bottom, we have: CPP 
$= e \sin{(\varpi+\psi)}$, obliquity, ice sheet height (m), 
annually averaged surface temperature ($^{\circ}$C), annual ice 
accumulation rate (m yr$^{-1}$), and annual ice ablation rate (m 
yr$^{-1}$). }
\end{centering}
\end{figure*}

For the purpose of this study, we tune the ice deposition rate 
so that the model can reproduce the Earth's ice age cycles at 
$\sim40,000$ years and $\sim100,000$ years over a 10 million 
year simulation. To reproduce the effect of Earth's moon on Earth's obliquity, we 
force the precession rate to be $50.290966''$ year$^{-1}$ 
\citep{laskar1993}. This choice does not perfectly match the dynamics 
of the Earth-moon-sun system, but it is close enough to 
replicate the physics of the ice age cycles. 
The results of this tuning are shown in Figure \ref{fig:huybers} 
(see \cite{huybers2008}, Figure 4, for comparison), for a 
200,000 year window. The ice sheets in the northern hemisphere 
high latitude region grow and retreat as the obliquity, 
eccentricity (not shown), and climate-precession-parameter, or 
CPP ($e\sin{(\varpi+\psi)}$), vary. The ice deposition rate is 
less than that used by \cite{huybers2008} and so the ice 
accumulation per year is slightly smaller. The ice ablation 
occurs primarily at the ice edge (around latitude $60^{\circ}$) 
and is slightly larger than \cite{huybers2008}, but is 
qualitatively similar. 

There are a number of differences between our reproduction of 
Milankovitch cycles and those of \cite{huybers2008}. Most notably, our 
ice sheets tend to persist for longer periods of time, taking up 
to three obliquity cycles to fully retreat. We also require a 
lower ice deposition (snowing) rate than \cite{huybers2008} in 
order to ensure a response from the ice sheets to the orbital 
forcing. We attribute these differences primarily to the 
difference in energy balance models used for the atmosphere.
For example, our model has a 
single-layer atmosphere with a parameterization of the OLR tuned 
to Earth, while \cite{huybers2008} used a multi-layer atmosphere 
with a simple radiative transfer scheme. Further, while the 
model \cite{huybers2008} contained only land, our model has both 
land  and water which cover a fixed fraction of the surface. The 
primary effect of having an ocean in this model is to change the 
effective heat capacity of the surface. This dampens 
the seasonal cycle, and affects the ice sheet growth and 
retreat. Thus, our seasonal cycle is somewhat muted compared to 
theirs, and our ice sheets do not grow and retreat as 
dramatically on orbital time scales. Ultimately, our ice age 
cycles are more similar to the longer late-Pleistocene cycles 
than to $\sim 40,000$ year cycles of the early-Pleistocene. 

Even though we cannot perfectly match the results of 
\cite{huybers2008}, we are comfortable with these results for a 
number of reasons. First, both models make approximations to a 
number of physical processes and thus have numerous parameters 
that have to be tuned to reproduce the desired behavior. Second, 
both models are missing boundary conditions based on the 
continent distribution of the Earth---continental edges can 
limit the equator-ward advance of ice sheets or alter the speed 
of their flow through calving of ice shelves. Finally, because 
the purpose of this study is to understand the response of ice 
sheets and climate to orbital variations, it is enough to merely 
ensure that the ice sheets respond in a way qualitatively 
similar to the Earth's  without being overly sensitive 
(\emph{i.e.}, resulting in ice free or snowball conditions with 
an insolation value of the solar constant, $\sim 1370$ W 
m$^{-2}$, and an OLR prescription similar to Earth's). 

\begin{figure*}
\begin{centering}
\includegraphics[width=\textwidth]{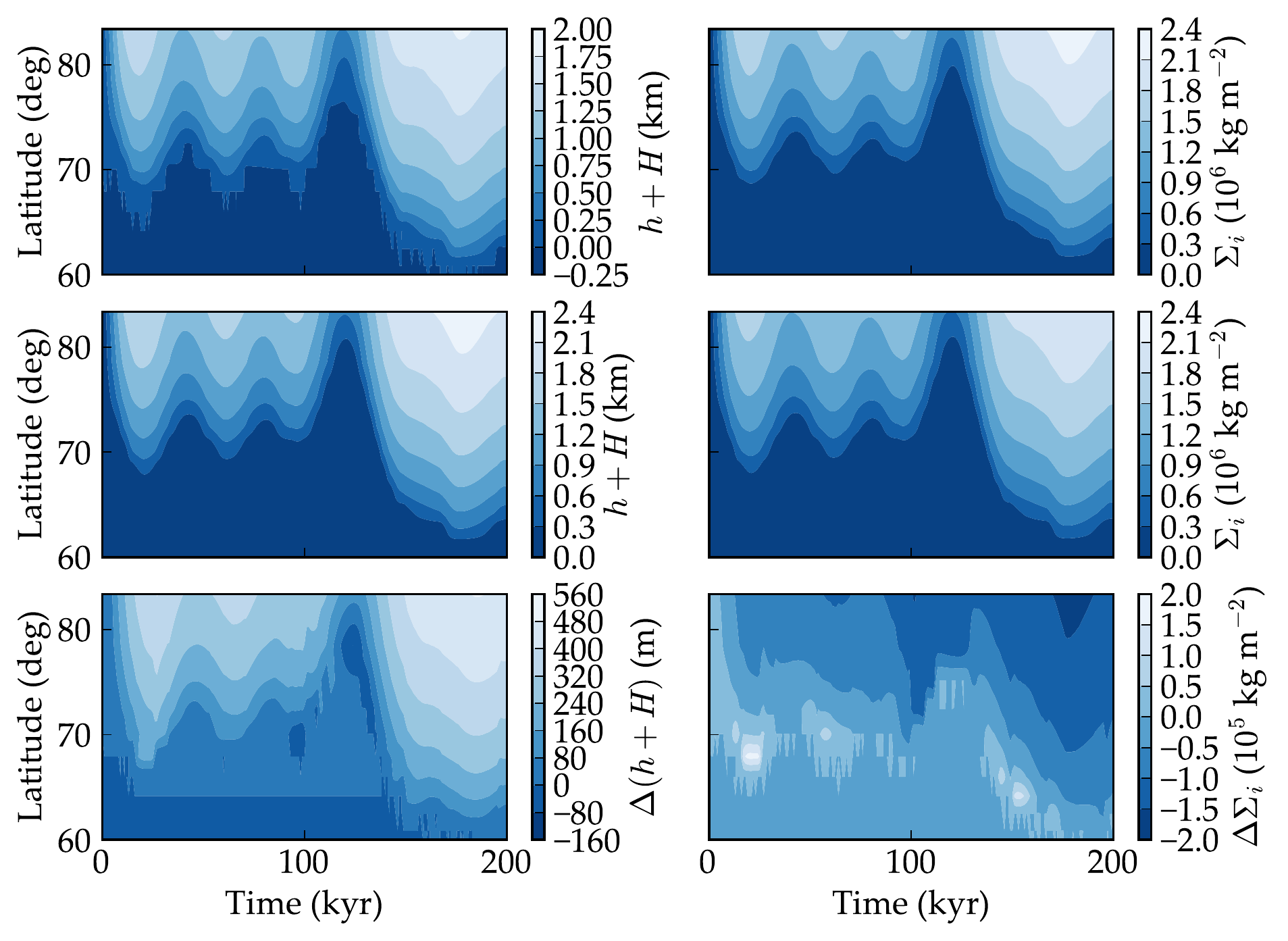}
\caption{\label{fig:icecomp} Ice sheet evolution for Earth with (upper panels) and without (middle panels) isostatic depression and rebound of the bedrock. Also shown is the difference (lower panels). The left panels show the ice sheet height/altitude; the right panels show the surface density of the ice. Without the bedrock model, the ice grows taller (in elevation), but there is less ice overall because the surface does not sink under the weight of the ice.}
\end{centering}
\end{figure*}

To investigate the importance of the bedrock depression/rebound component of the model, we compare this Earth case to one with $\partial H/\partial t$ (Eqn \ref{eqn:brock}) set to zero. Figure \ref{fig:icecomp} shows the ice sheet height, $h+H$, and surface mass density, $\Sigma_i$, with (upper panels) and without the bedrock component (middle panels), and the difference (lower panels). The ice sheets reach higher altitude (by several hundred meters) without bedrock depression, but the ice mass is decreased by $\sim 10^5$ kg m$^{-2}$. The effect of isostasy is thus to confine the ice sheets while allowing them to grow larger. While this subtly increases the thermal inertia, it ultimately makes a minor difference in the prevalence of snowball states in our results (Section \ref{sec:results}).

\section{Results}
\label{sec:results}
\subsection{Static cases}
\begin{figure*}
\includegraphics[width=0.5\textwidth]{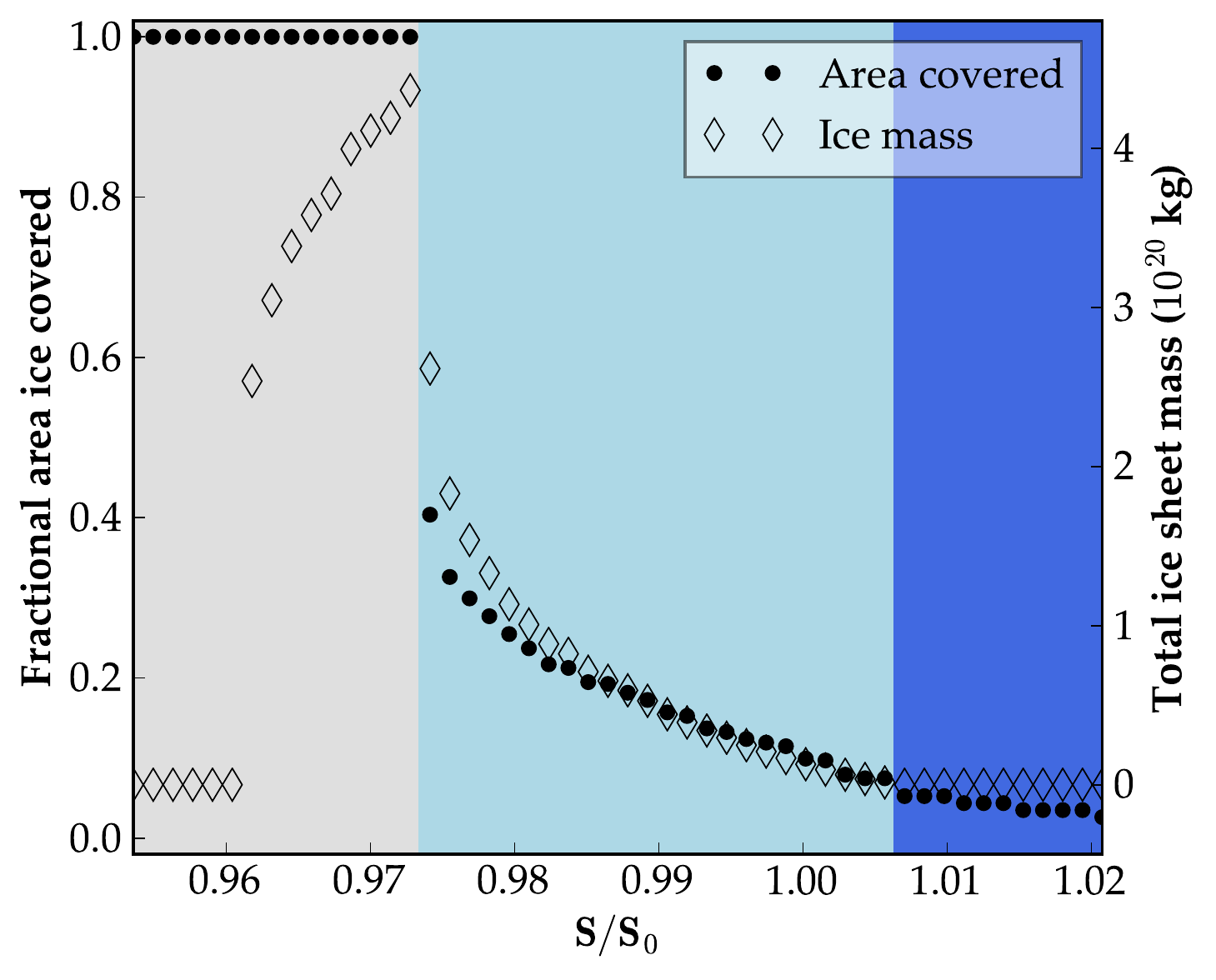}
\includegraphics[width=0.5\textwidth]{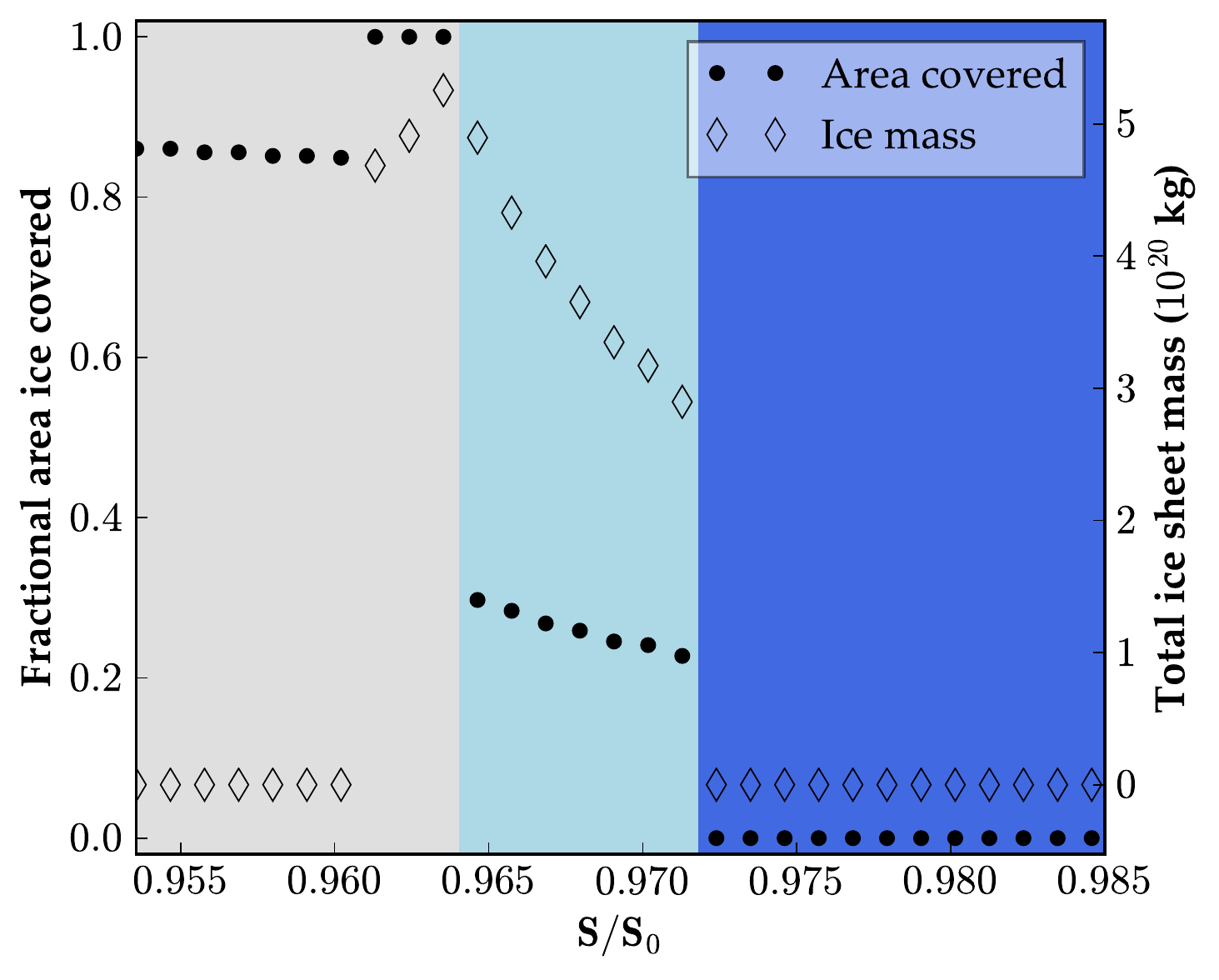}
\caption{\label{fig:trans} The fractional ice cover, $f_{\text{ice}}$, for static
orbital/obliquity conditions as a function of stellar flux, 
$S/S_0$, where $S_0 = 1367.5$ W m$^{-2}$, for $\varepsilon = 
23.5^{\circ}$ (left) and $\varepsilon = 50^{\circ}$ (right). The ice 
covered area includes both land and ocean grid-points. The gray 
shaded area represents snowball states (the ocean surface is 
permanently and completely ice-covered), dark-blue represent 
ice-free (no year-round ice) states, and light-blue is the 
``transition region'', where the ocean is not totally 
ice-covered and ice sheets form on land. For reference, the Antarctic
ice sheet is estimated to be $27 \times 10^6$ km$^{3}$, on the order
of $10^{19}$ kg of ice mass \citep{fretwell2013}.}
\end{figure*}

First, we identify the regimes in which ice sheets are able to 
form. The presence and distribution of permanent ice on land 
will depend on the stellar flux received by the planet and the 
planet's obliquity. In Figure \ref{fig:trans} we show how ice 
covered fraction, $f_{\text{ice}}$ depends on incoming stellar flux at two obliquities 
($\varepsilon = 23.5^{\circ}$ and $\varepsilon = 50^{\circ}$). Note that this 
initial ice coverage in each simulation is determined by the 
initial temperature distribution (Eqn. \ref{eqn:inittemp}), and is 
very different from the final result in most cases. The ice 
coverage includes both land and ocean grid-points. The stellar 
flux is normalized by Earth's value, $S_0 = 1367.5$ W m$^{-2}$. 
No orbital evolution occurs in these simulations, however, the 
spin axis is allowed to precess at a rate set by the stellar 
torque (see Paper I). Two quantities are 
displayed in these plots: the fractional area of the planet that 
is permanently ice covered (\emph{i.e.} ice covered year-round) 
and the total ice mass at the end of the simulation. 

At the lowest stellar flux values, the planet is globally ice 
covered ($f_{\text{ice}}=1$), but the ice sheet mass remains at zero. This is 
because, in our model, precipitation is shut off when the oceans 
are frozen over, and in these coldest cases, the oceans freeze 
over during the spin-up phase of the simulation,
thus no ice accumulates on land. In the $\varepsilon 
= 50^{\circ}$ case, the coldest cases are actually not ice 
covered year round. Since the oceans have frozen before ice 
sheets can grow on land, and the thermal inertia of the land is 
low (compared to the oceans and the ice sheets), the temperature 
over land actually rises above freezing during the summer months. 
Thus, the fact that $f_{\text{ice}}< 1$ is probably a side effect of our 
modeling choices---these cases really are in a snowball state. 
At higher stellar flux values, it takes hundreds to thousands of 
years for the planet to cool into the snowball state, thus ice 
sheets are allowed to grow on land. Because it takes much more 
energy in the model to melt a thick layer of ice (than to simply 
heat the land), these cases remain fully ice covered year-round.

All points within the gray-shaded region entered a snowball 
state in $<200$ kyr, after which all ice sheets appear to be 
stable under static orbital/obliquity conditions. The light-blue 
region corresponds to our ``transition region'', wherein stable 
ice sheets form at some latitudes and persist year-round. In the 
dark-blue region, ice may form seasonally, but no permanent ice 
sheets appear. Note that in the $\varepsilon = 23.5^{\circ}$ cases, the 
ice covered area is not necessarily equal to zero
because the oceans remain frozen at the poles year round, even 
though no ice sheets grow from year to year. 

The higher obliquity case remains clement (not in a snowball 
state) at lower stellar flux, and thus higher semi-major axis, 
than the low obliquity case, consistent with past 
results \citep{spiegel2009,armstrong2014}. The transition region 
is also narrower in this case, and the boundary between the 
transition region and the ice sheet free region (light- and 
dark-blue) is sharper, consistent with
\cite{rose2017}, which demonstrated that ice (as represented by 
$T<0^{\circ}$ C on land or ocean) is less stable on higher 
obliquity planets. Interestingly, even though the obliquity is 
less than $55^{\circ}$ (the approximate value at which the annual insolation 
at the poles begins to exceed that of the equator), the ice 
sheets in the transition region form along the \emph{equator}, 
not the poles. This is a result of the temperature dependence of 
ice ablation---when the atmosphere is warmer, the ice melts 
faster (see Equation (\ref{eqn:ablation})). Even though the 
equatorial latitudes receive more sunlight over the course of an 
orbit, the summers are much more intense at the poles. High 
latitude summers are then much warmer than conditions ever get 
at the equator. So while the snowy season at the poles may be 
colder and longer, the intense summers are more than enough to 
melt the ice accumulated during winters, whereas the melting 
seasons are not hot enough or long enough to fully melt the 
equatorial ice. 

\subsection{Dynamically evolving cases}
Next, we vary the initial eccentricity, inclination, rotation rate, and
obliquity of planet 2 (Earth-mass) in our test system.
Figures \ref{fig:lowoblmidmap}-\ref{fig:highoblmidfastmap} show the
fractional area of the planet that is ice covered for several 
slices of this parameter space at an
incident stellar flux of $S = 1332.27$ W m$^{-2}$, or  $S/S_0 = 
0.974$. This stellar flux puts the planet right at the boundary 
between the snowball state and the transition zone for a planet 
with low eccentricity and $23.5^{\circ}$ obliquity (Figure 
\ref{fig:trans}, left panel), and places the $\varepsilon_0 = 50^{\circ}$ 
simulations in the ice-free regime. 

The obliquity amplitude ($\Delta \varepsilon$) is shown in each panel 
as contours (see Paper I). The blue-white color scale in  
each figure shows the fraction, $f_{\text{ice}}$, of the total area of 
the panel that is permanently ice-covered, where ``permanent'' 
means covered year-round as in the previous section. Thus, some 
cases that have $f_{\text{ice}} = 0$ do have seasonal ice formation. 

The left panels shows the climate conditions assuming a 
static orbit and obliquity fixed at the initial values. Here, inclination
has no direct effect on the insolation or climate, so $f_{\text{ice}}$ 
depends only on the eccentricity  ($S\propto(1-e^2)^{-1/2}$). The planet is
in a snowball state
$f_{\text{ice}}=1$ at $e=0$, but as $e$ is increased, $f_{\text{ice}}$ 
decreases. The stellar torque on the 
equatorial bulge is included and results in a constant axial 
precession rate, but this has minimal impact on the total ice
coverage.

In the middle panels, the orbit and obliquity are also static, 
but they are fixed at the mean values from the 2 Myr simulation. 
The structure of this phase space is very different from that of 
the static initial conditions (upper right). For the cases with 
$\varepsilon_0 = 23.5^{\circ}$ (Figures \ref{fig:lowoblmidmap}, 
\ref{fig:lowoblmidslowmap}, and \ref{fig:lowoblmidfastmap}), using the 
mean properties tends to decrease the portion of phase space 
with $f_{\text{ice}}=1$, however, for the $\varepsilon_0 = 50^{\circ}$ 
cases (Figures \ref{fig:highoblmidmap}, \ref{fig:highoblmidslowmap}, and 
\ref{fig:highoblmidfastmap}), the mean properties produce snowball 
states where none existed before (at the initial values). Hence, 
using the mean orbital/obliquity properties in a climate 
simulation produces very different results from using the 
initial (or, perhaps, observed) properties.

Finally, the right panel in each figure shows $f_{\text{ice}}$
for the full 2 Myr simulation with 
evolving orbits and obliquities. Now, the ice coverage increases 
almost universally, and snowball states are much more frequent 
than under static conditions. There are some configurations 
that had $f_{\text{ice}}=1$ under static conditions but are not 
completely ice covered under evolving conditions (at low 
inclination and low eccentricity, for example), but in general, 
the evolution tends to encourage the snowball instability, 
except at higher $e_0$. Interestingly, there are several blue 
``islands'' (where 
$f_{\text{ice}} < 1$) that are completely surrounded by snowball states 
in the dynamically evolving cases. There is a complex 
interplay between the obliquity and eccentricity that we will 
discuss in more detail in Section \ref{sec:icestab}.

\begin{figure*}
\centering
\includegraphics[width=0.4\textwidth]{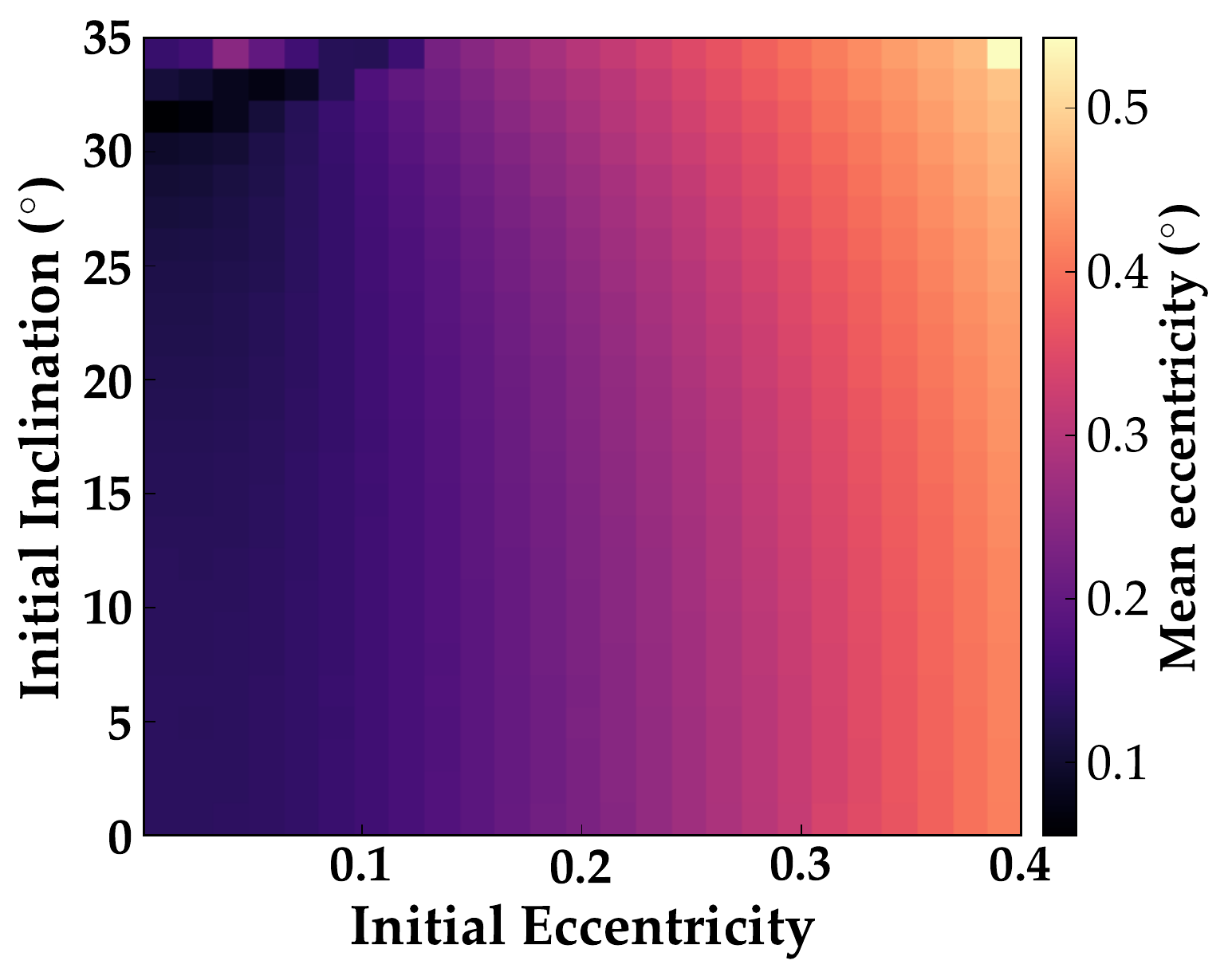}
\caption{\label{fig:orbitmean} Mean eccentricity values as a function of initial inclination and eccentricity. These values are used as input to the climate model for the middle panels of Figures \ref{fig:lowoblmidmap}-\ref{fig:highoblmidfastmap}. There is a single simulation in the upper right corner for which the orbital model fails (the eccentricity exceeds $\sim 0.66$)---we model the system and climate up until the code halts, but this point does not factor heavily into our analysis.}
\end{figure*}

\begin{figure*}
\centering
\includegraphics[width=0.4\textwidth]{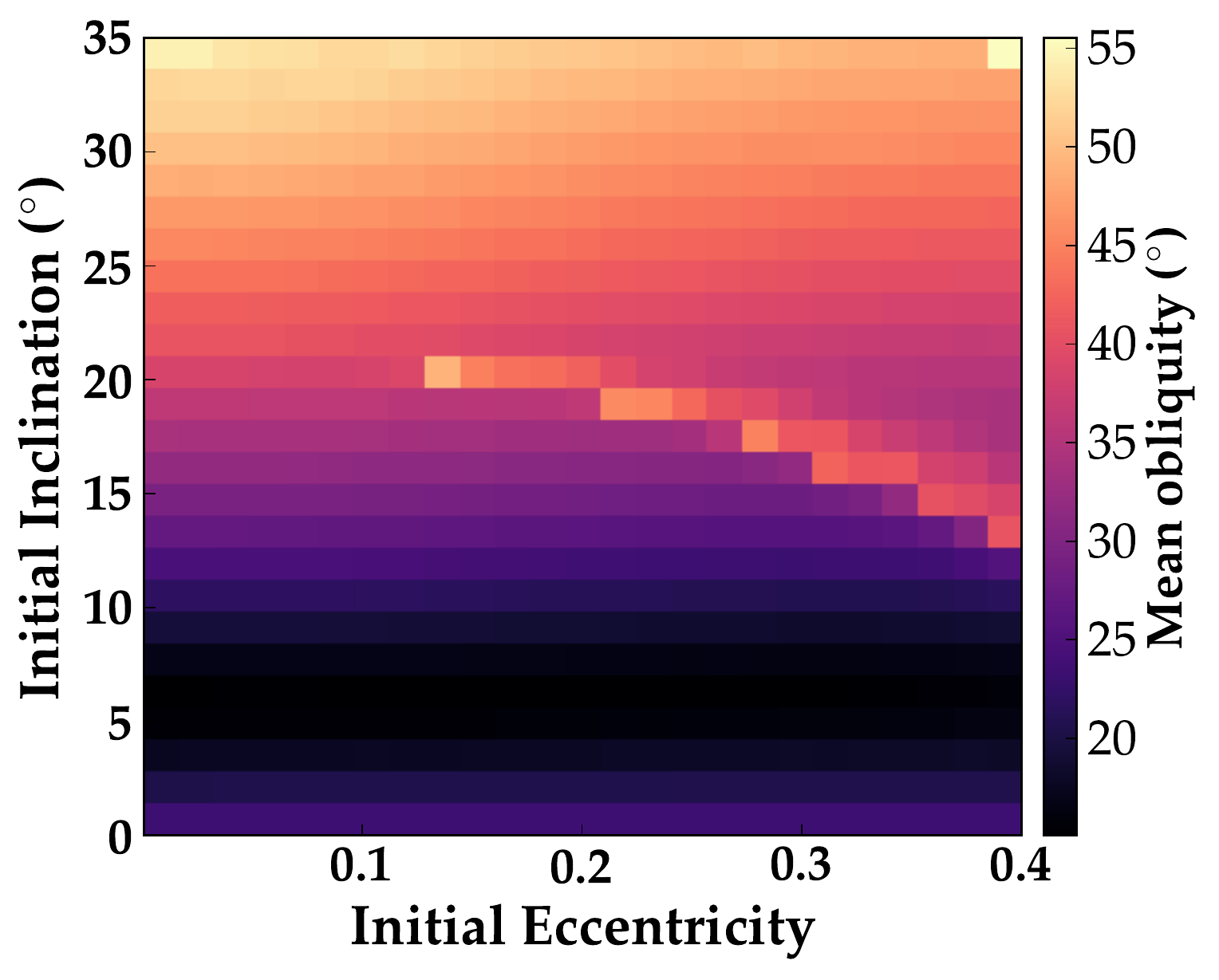}
\includegraphics[width=0.4\textwidth]{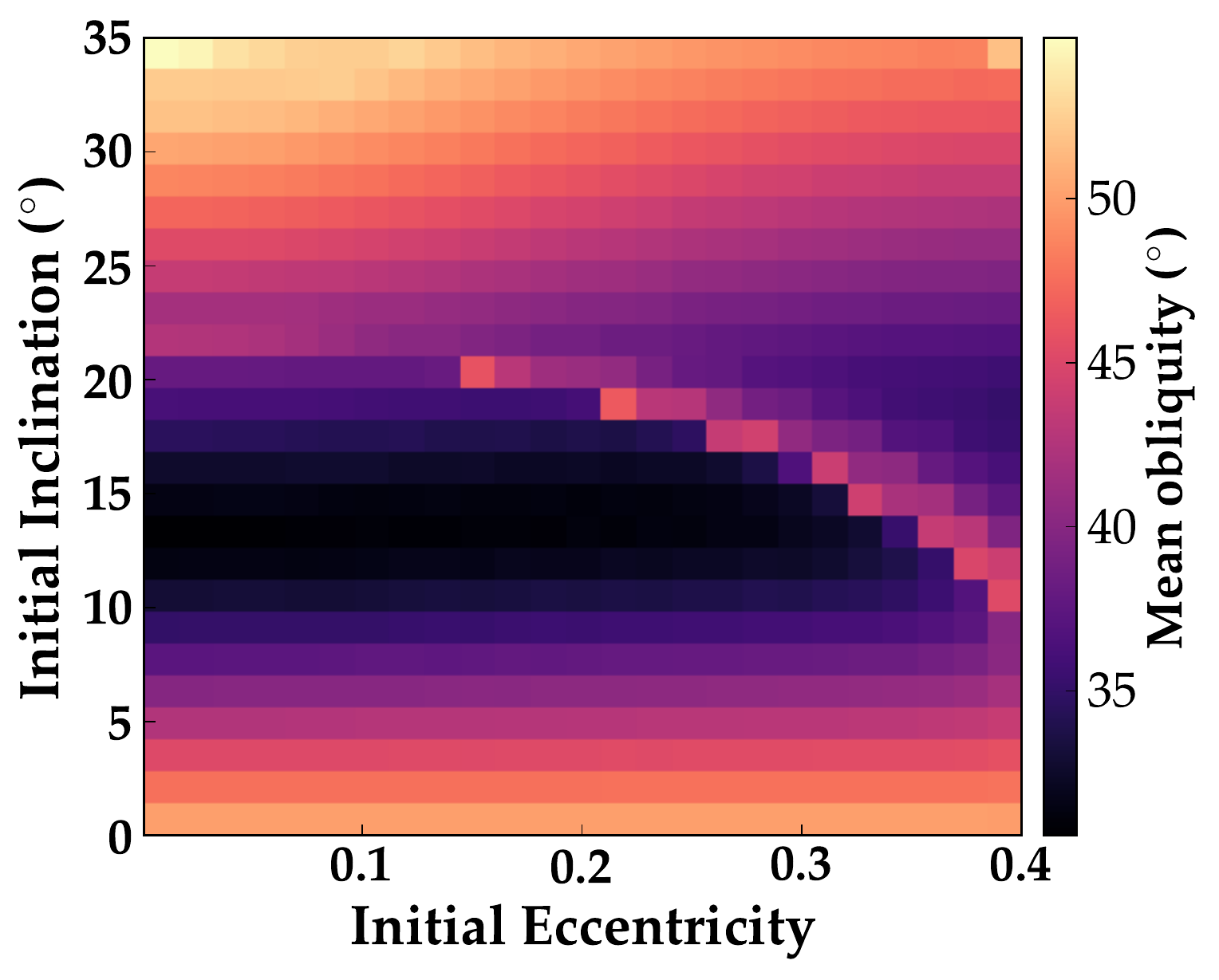}
\caption{\label{fig:oblmean} Mean obliquity values as a function of initial inclination and eccentricity for $P_{rot}=1$ day at $\varepsilon_0 = 23.5^{\circ}$ (left) and $\varepsilon_0=50^{\circ}$. These values are used in the climate model for the middle panels of Figures \ref{fig:lowoblmidmap} and \ref{fig:highoblmidmap}. The high obliquity ``arc'' through the center of each panel is the result of a secular spin-orbit resonance (see Paper I). Corresponding plots for $P_{rot}=1.62$ days and $P_{rot}=0.65$ days (that is, the conditions used in the middle panels of Figures \ref{fig:lowoblmidslowmap}-\ref{fig:highoblmidfastmap}) appear very similar in structure. The range of mean obliquity values is smaller ($15^{\circ}\lesssim\langle \varepsilon \rangle\lesssim50^{\circ}$) for $P_{rot}=1.62$ days, while it is slightly increased ($15^{\circ}\lesssim\langle \varepsilon \rangle\lesssim65^{\circ}$) for $P_{rot}=0.65$.}
\end{figure*}

\begin{figure*}
\includegraphics[width=\textwidth]{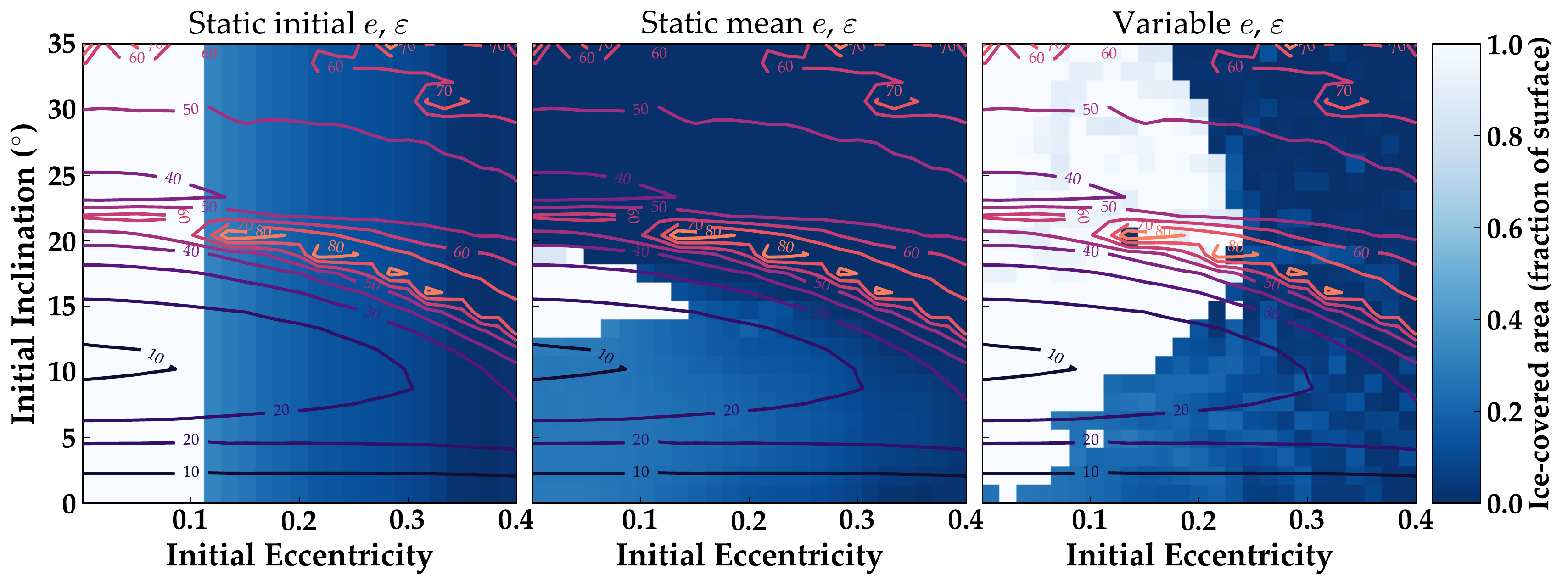}
\caption{\label{fig:lowoblmidmap} Climate states as a function of 
initial eccentricity and inclination, for $P_{rot} = 1$ day and 
initial obliquity $\varepsilon_0 = 23.5^{\circ}$, with a stellar 
constant of $S = 1332.27$ W m$^{-2}$. Each panel shows the fraction 
of the surface area that is 
permanently ice-covered over the final orbit (blue color-scale) 
and contours of $\Delta \varepsilon$ (black lines), under three 
different conditions: left, static orbit and obliquity at 
the initial values; middle, static orbit and obliquity at 
the mean values from the simulation; right, dynamically 
evolving orbit and obliquity.}
\end{figure*}

\begin{figure*}
\includegraphics[width=\textwidth]{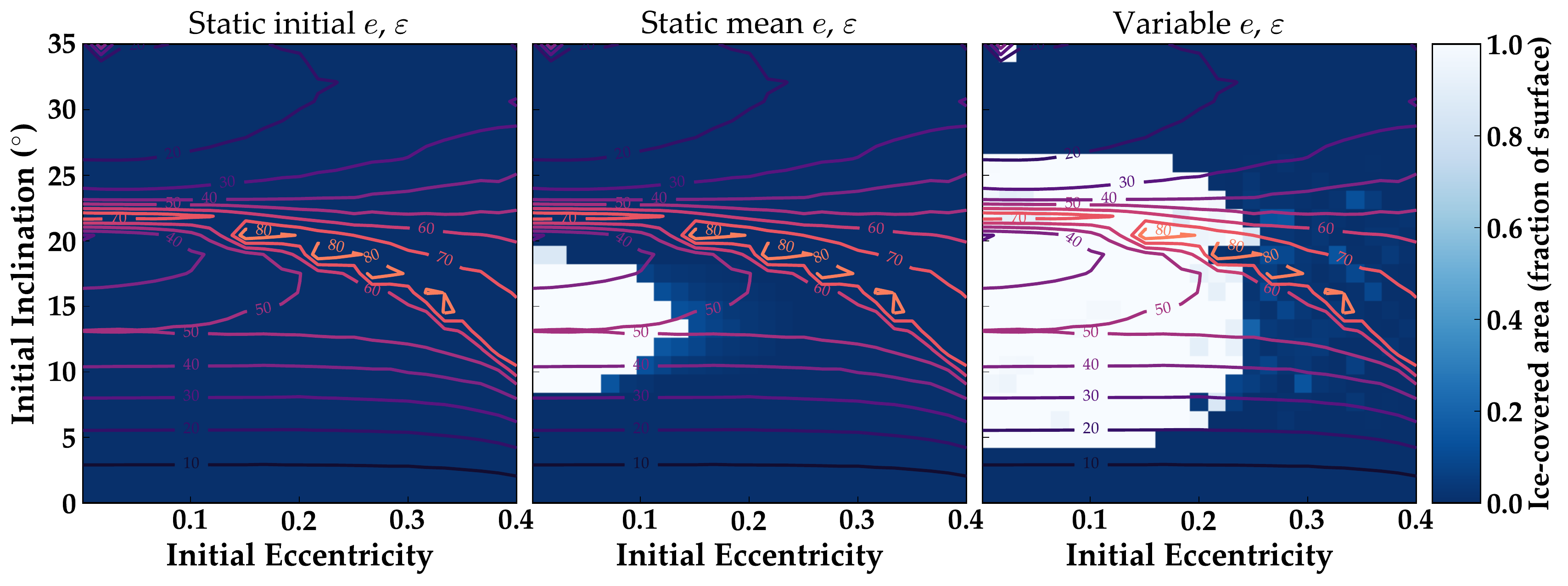}
\caption{\label{fig:highoblmidmap} Same as Figure
\ref{fig:lowoblmidmap} 
but for $P_{rot} = 1$ day and initial obliquity $\varepsilon_0 = 
50^{\circ}$. }
\end{figure*}

\begin{figure*}
\includegraphics[width=\textwidth]{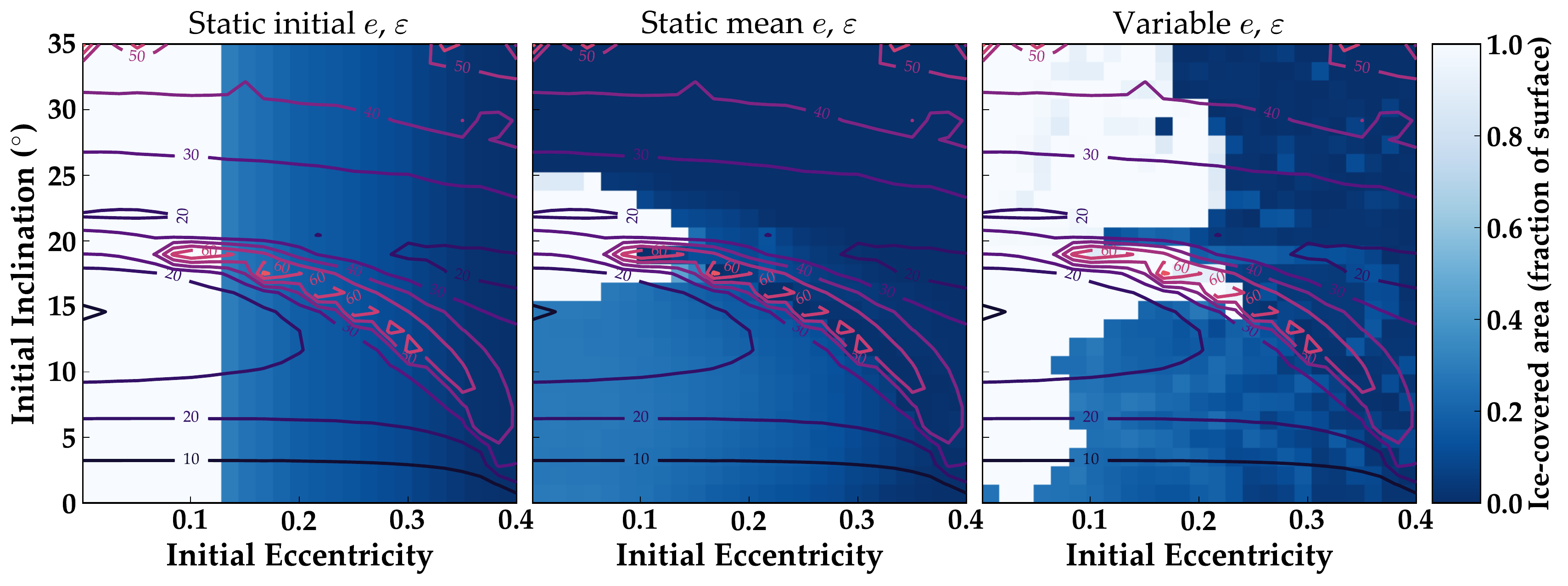}
\caption{\label{fig:lowoblmidslowmap}Same as Figure 
\ref{fig:lowoblmidmap} but for $P_{rot} = 1.62$ day and initial 
obliquity $\varepsilon_0 = 23.5^{\circ}$.}
\end{figure*}

\begin{figure*}
\includegraphics[width=\textwidth]{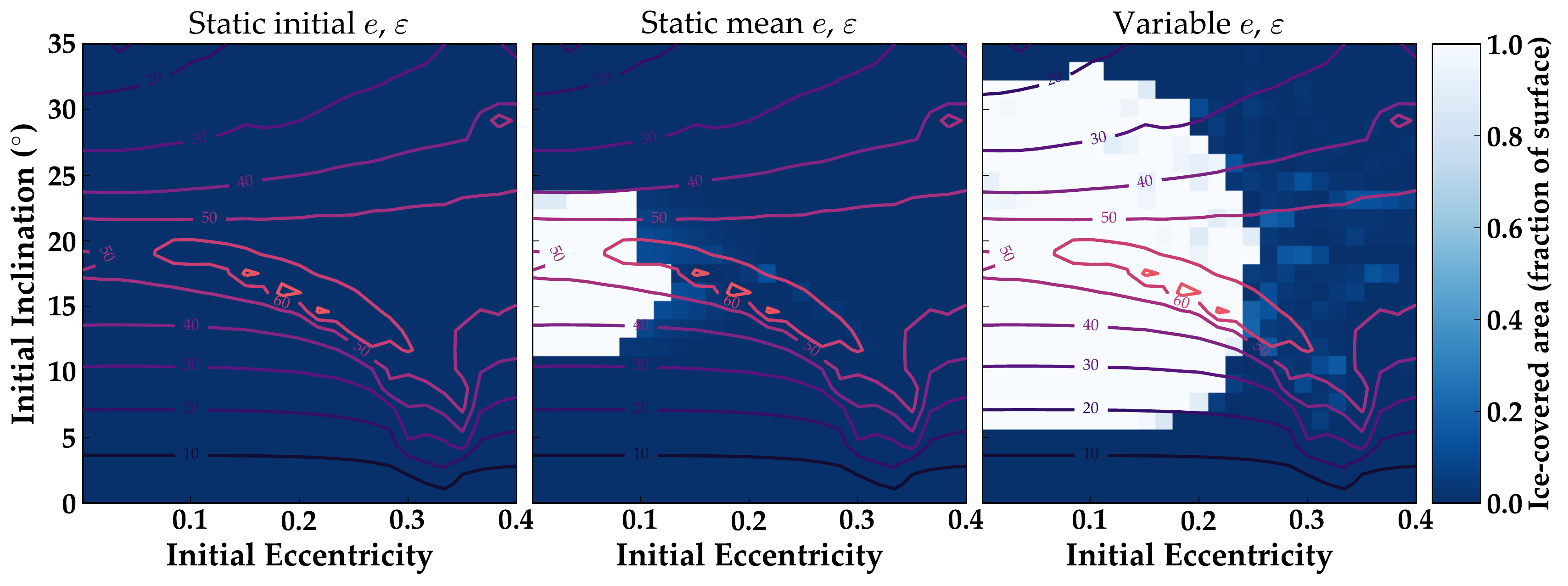}
\caption{\label{fig:highoblmidslowmap}Same as Figure 
\ref{fig:lowoblmidmap} but for $P_{rot} = 1.62$ day and initial 
obliquity $\varepsilon_0 = 50^{\circ}$.}
\end{figure*}

\begin{figure*}
\includegraphics[width=\textwidth]{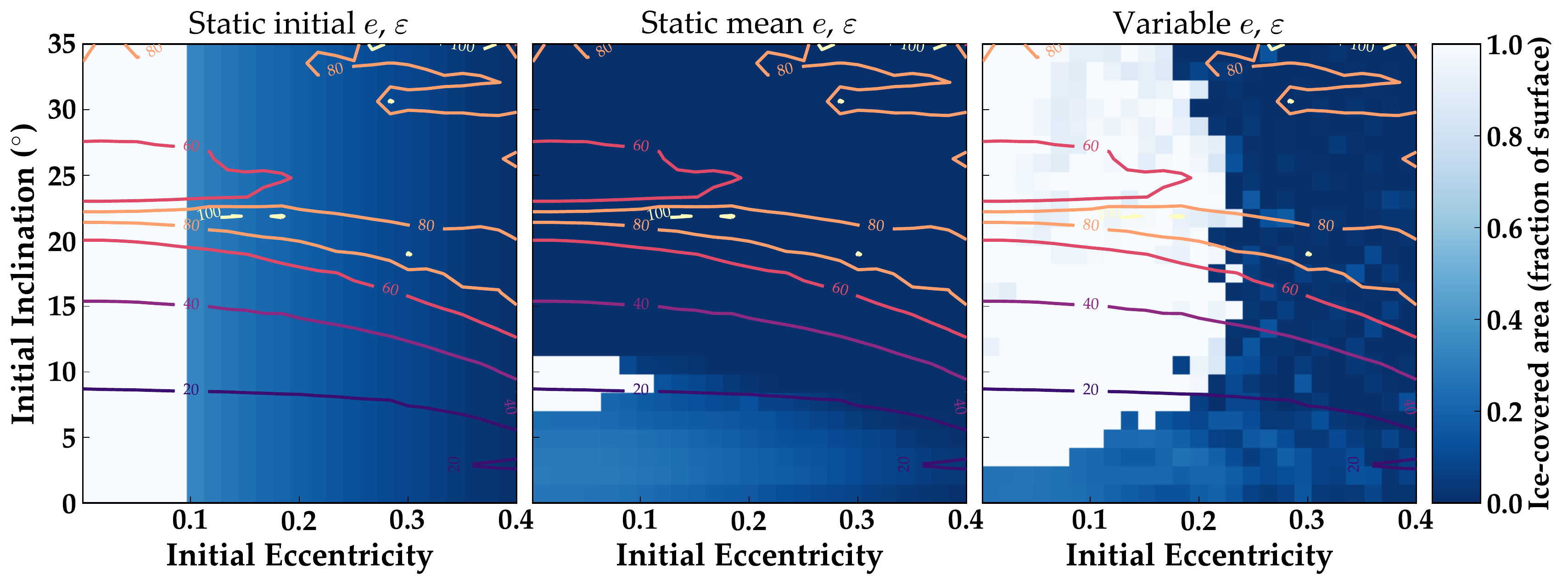}
\caption{\label{fig:lowoblmidfastmap}Same as Figure 
\ref{fig:lowoblmidmap} but for $P_{rot} = 0.65$ day and initial 
obliquity $\varepsilon_0 = 23.5^{\circ}$.}
\end{figure*}

\begin{figure*}
\includegraphics[width=\textwidth]{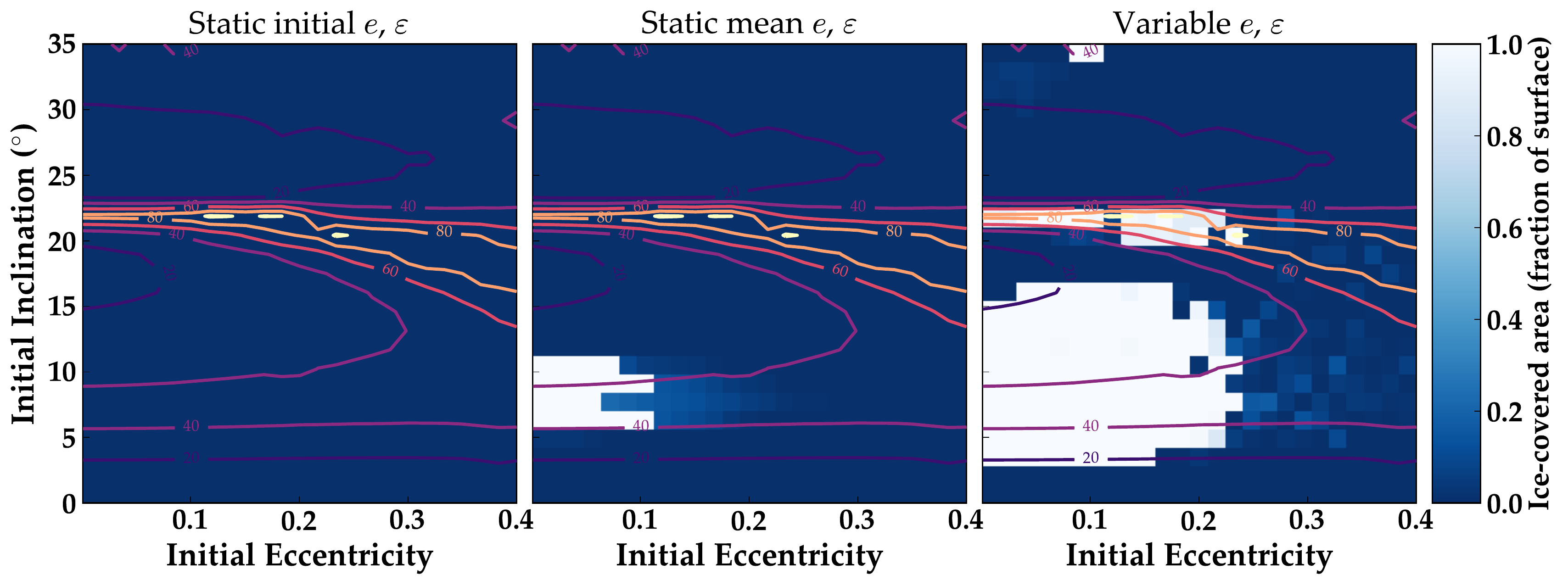}
\caption{\label{fig:highoblmidfastmap}Same as Figure 
\ref{fig:lowoblmidmap} but for $P_{rot} = 0.65$ day and initial 
obliquity $\varepsilon_0 = 50^{\circ}$.}
\end{figure*}

\begin{figure*}
\includegraphics[width=\textwidth]{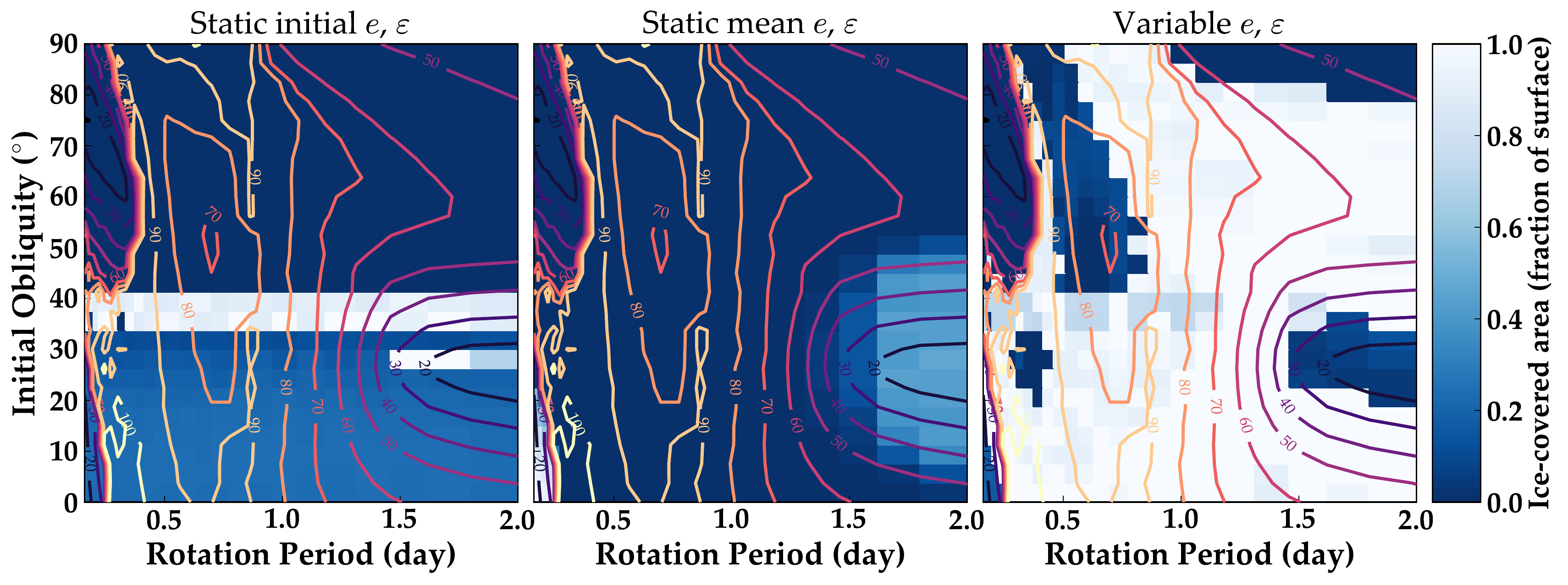}
\caption{\label{fig:oblvsrotmap}Same as Figure 
\ref{fig:lowoblmidmap} but varying $P_{rot}$ and $\varepsilon$ with $e_0=0.2$ and 
$i_0 = 20^{\circ}$. }
\end{figure*}

Figure \ref{fig:oblvsrotmap} illustrates the effects of rotation rate and
initial obliquity. The ice cover is shown in the same style as
Figures \ref{fig:lowoblmidmap}-\ref{fig:highoblmidfastmap}, but with
$e$ and $i$ fixed, and $\varepsilon_0$ and $P_{\text{rot}}$ varied instead.
Under static initial eccentricity and obliquity (left), low obliquity cases form 
some permanent ice, while high obliquity cases form none. From $\varepsilon
\sim 33^{\circ}-40^{\circ}$, the planet enters a snowball state, because
the ice edge is unstable at these obliquities (see Section \ref{sec:icestab}), but
these cases lack the warming effect that comes with even higher obliquity. 
The static 
mean conditions do not enter a snowball state anywhere in this parameter space.
With a variable orbit and obliquity, snowball states occur throughout much of this 
space. Note also that the obliquity variation in some regions is extremely 
large in amplitude and sometimes chaotic (see Paper I).

\begin{figure*}
\includegraphics[width=\textwidth]{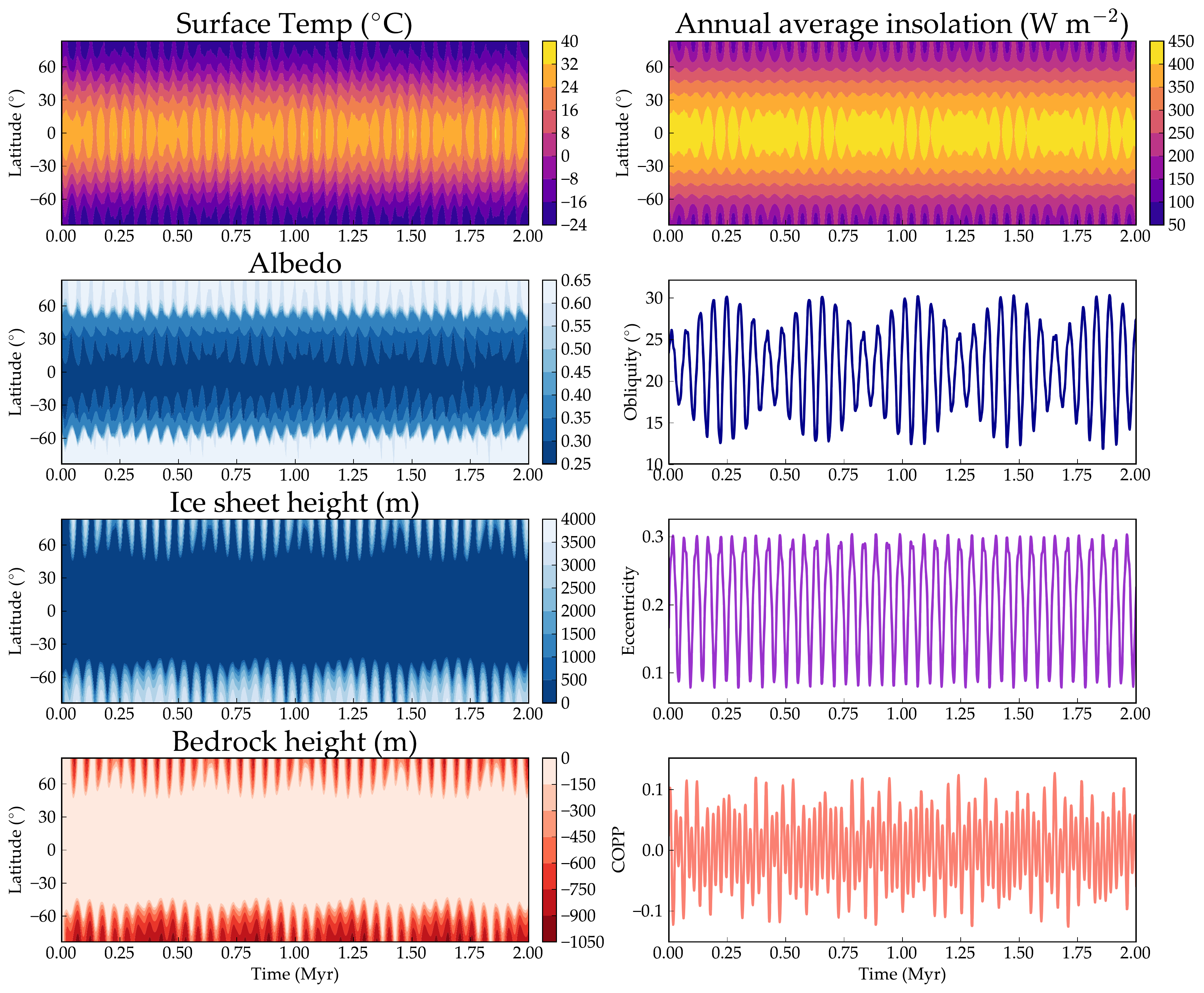}
\caption{\label{fig:caseLowSlow_1} Evolution of climate and orbit 
for a case at initial values: $S = 1332.27$ W m$^{-2}$, $e_0 = 
0.16725$, $i_0 = 14.54^{\circ}$, $\varepsilon_0 = 23.5^{\circ}$, and 
$P_{rot} = 1.62$ day (inside the horizontal blue strip near the 
center of Figure \ref{fig:lowoblmidslowmap}, right panel). The 
climate-obliquity-precession-parameter is defined as COPP $= e 
\sin{\varepsilon} \sin{(\varpi+\psi)}$ and represents the asymmetry 
between the northern and summer hemispheres (see text).}
\end{figure*}

Figure \ref{fig:caseLowSlow_1} shows the climate and orbit evolution 
for a point in the parameter space of Figure 
\ref{fig:lowoblmidslowmap} ($\varepsilon = 23.5^{\circ}$ and $P_{\text{rot}} = 
1.62$ day). In this figure we have the surface temperature, 
planetary albedo, ice sheet height, bedrock 
height, and insolation, all averaged over an orbit or ``year'', 
as a function of latitude and time. Also shown are the three 
parameters that affect the insolation: obliquity, eccentricity, 
and ``climate-obliquity-precession-parameter'' (COPP), which is 
defined as:
\begin{equation}
\text{COPP} = e \sin{\varepsilon} \sin{(\varpi+\psi)},
\label{eqn:coppdef}
\end{equation}where, again, $\varpi+\psi$ represents the 
instantaneous angle between periastron and the planet's position 
at its northern spring equinox. This is essentially the same as 
the commonly used ``climate precession parameter'' or CPP, but 
additionally takes into account the effect of obliquity 
variations (which are neglected in the CPP because Earth's are 
very small). COPP can be thought of as a measurement of the 
asymmetry between the northern and southern hemispheres, and so 
varies with the angle $\varpi+\psi$, modulated by the 
eccentricity and obliquity. When COPP $>0$, the northern 
hemisphere receives more stellar flux than the southern; 
vice-versa for COPP $<0$. 

Despite the climate in Figure \ref{fig:caseLowSlow_1} 
approaching very near to snowball states, the planet 
remains clement throughout this 2 Myr evolution. Ice sheets grow 
and recede at both poles rather dramatically, from almost 
nothing to nearly 4 km in height (in some regions) and back. 
This oscillation is a result of a nearly 200 W m$^{-2}$ swing in the annual 
insolation over $\sim 50,000$ years, due to the combined effects 
of the obliquity and eccentricity variations. The envelope of 
the obliquity oscillation is imprinted on the latitude of 
the ice edge, though the primary driver of growth and retreat is 
the change in eccentricity. 
The ice edge progresses into the mid-latitudes during periods 
when the obliquity oscillation is lowest in amplitude. 

\begin{figure*}
\includegraphics[width=\textwidth]{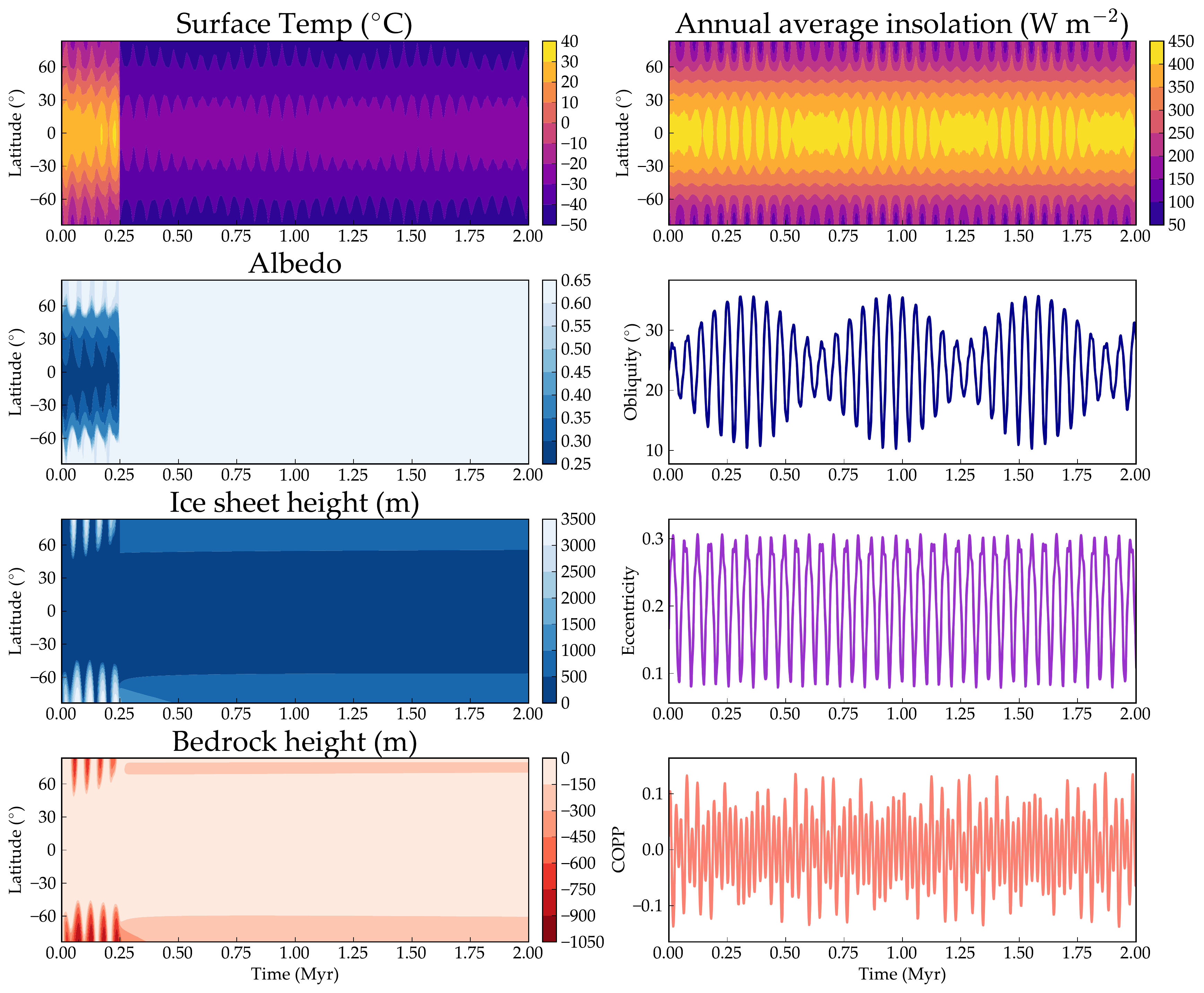}
\caption{\label{fig:caseLowSlow_2} Same as Figure 
\ref{fig:caseLowSlow_1} but for $e_0 = 0.16725$, $i_0 = 
16.04^{\circ}$, $\varepsilon_0 = 23.5^{\circ}$, and $P_{rot} = 1.62$ 
day (slightly lower inclination than the case in that Figure). A 
snowball state occurs at $t\sim750,000$ years---the temperature 
drops globally, the albedo approaches that of ice everywhere, 
and ice sheets no longer grow (precipitation is shut off 
artificially) and instead just gradually flatten.}
\end{figure*}

In Figure \ref{fig:caseLowSlow_2}, we have the same evolution for a 
case immediately adjacent to that in Figure \ref{fig:caseLowSlow_1}. 
The eccentricity and obliquity variations are very similar to 
the previous case, however, the obliquity peaks at a slightly 
higher value ($\sim 35^{\circ}$, compared to $\sim 30^{\circ}$ 
in the previous). The ice sheets grow and retreat in a similar 
fashion until the obliquity approaches its highest value, at 
which point the planet abruptly enters a snowball state. The 
appearance of the large ice cap instability (LICI) is somewhat 
counter to expectation here---as we have shown before (and 
numerous other studies have found), high obliquity tends to 
grant a planet additional warmth at low stellar flux. The 
analytic solution to the annual EBM from \cite{rose2017} 
provides an explanation for how the instability occurs, see Section \ref{sec:icestab}.

\begin{figure*}
\includegraphics[width=\textwidth]{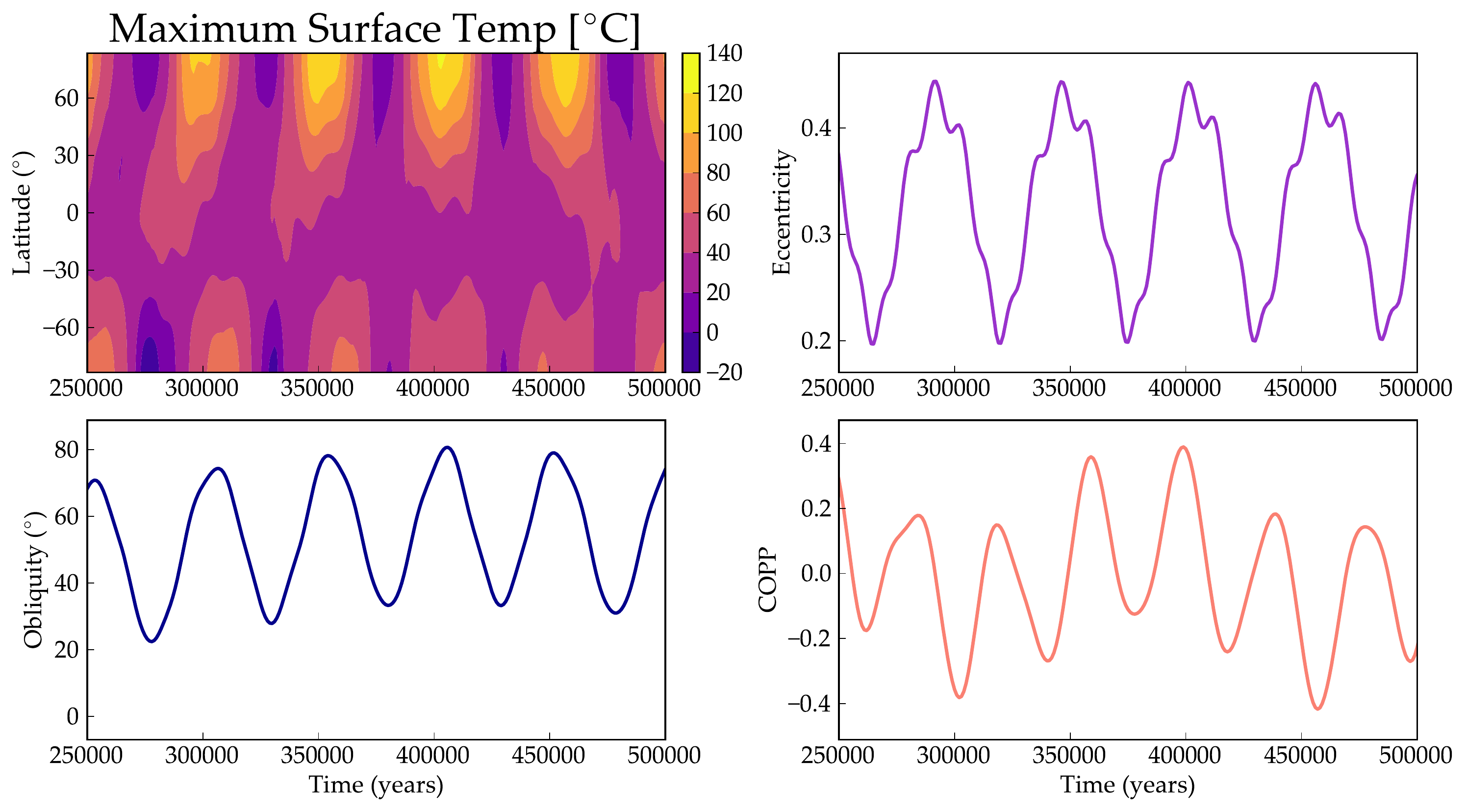}
\caption{\label{fig:hotcasetemp} Evolution of the orbit, obliquity 
and maximum surface temperature for a case with $P_{rot} =1$ day, 
$\varepsilon_0 = 23.5^{\circ}$, $e_0 = 0.3$, and $i_0 = 17.5^{\circ}$ 
over a 250 kyr period. 
The upper left panel is the maximum surface temperature over an orbit 
(averaged over land/ocean); lower left, obliquity; upper right, 
eccentricity; lower right, COPP.
The obliquity reaches large values
because of the secular spin-orbit resonance (see 
Paper I). The highest obliquity times 
correspond to high eccentricity times. As a result, the 
insolation at high latitudes is extremely high during summer and 
the surface temperature exceeds the boiling point of water. This 
effect depends also on the angle $\varpi+\psi$ (the angle between
the equinox and pericenter) and is responsible for the
additional variation in maximum temperature between these warm 
periods.}
\end{figure*}

In addition to snowball states, we also observe some very high
temperatures at high-obliquity, high-eccentricity times. For a 
case with $\varepsilon_0 = 23.5^{\circ}$, $P_{rot} = 1$ day, $e_0 = 
0.3$, and $i_0 = 17.5^{\circ}$, which is inside the secular 
resonance in Figure \ref{fig:lowoblmidmap}, the obliquity reaches 
$\sim 80^{\circ}$ while the eccentricity is $\sim 0.4$. Figure 
\ref{fig:hotcasetemp} shows the orbital/obliquity evolution and the 
resulting average, minimum, and maximum surface temperatures 
(over an orbital period). At the highest obliquity times, the 
north pole of the planet reaches $140^{\circ}$ C. Such strong 
heating should probably result in strong convection, which would 
increase the albedo (due to cloud formation) and cause 
increased horizontal heat flow, but our simple EBM does not 
model such effects (see Section \ref{sec:validcomp}). Thus this temperature
is improbable, except perhaps over dry continental interiors. 
It is beyond the scope of this study to 
comprehensively model this scenario with a GCM, but it is worth 
future investigation in the future. 

\subsection{Examining ice stability}
\label{sec:icestab}

\begin{figure*}
\includegraphics[width=0.5\textwidth]{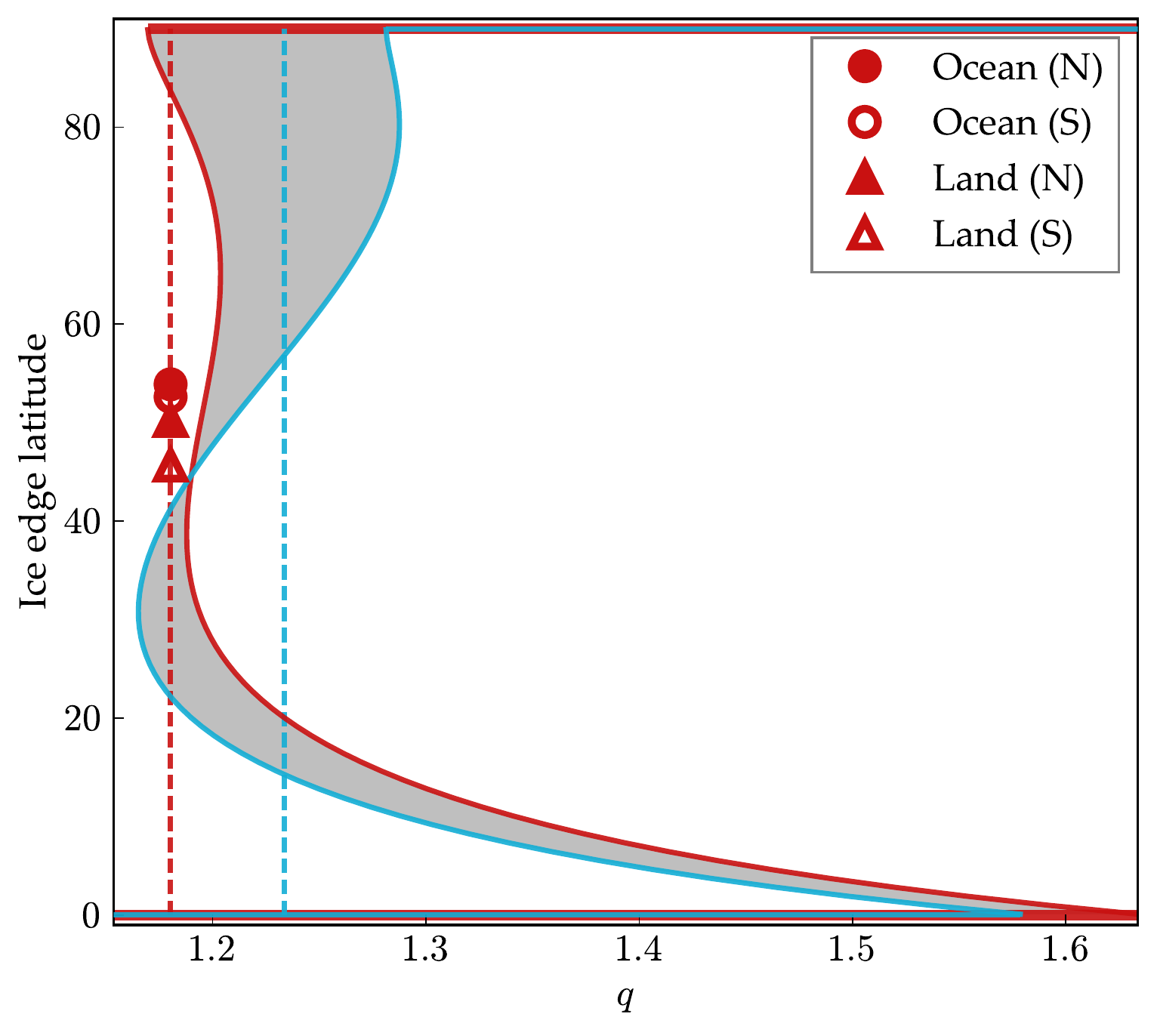}
\includegraphics[width=0.5\textwidth]{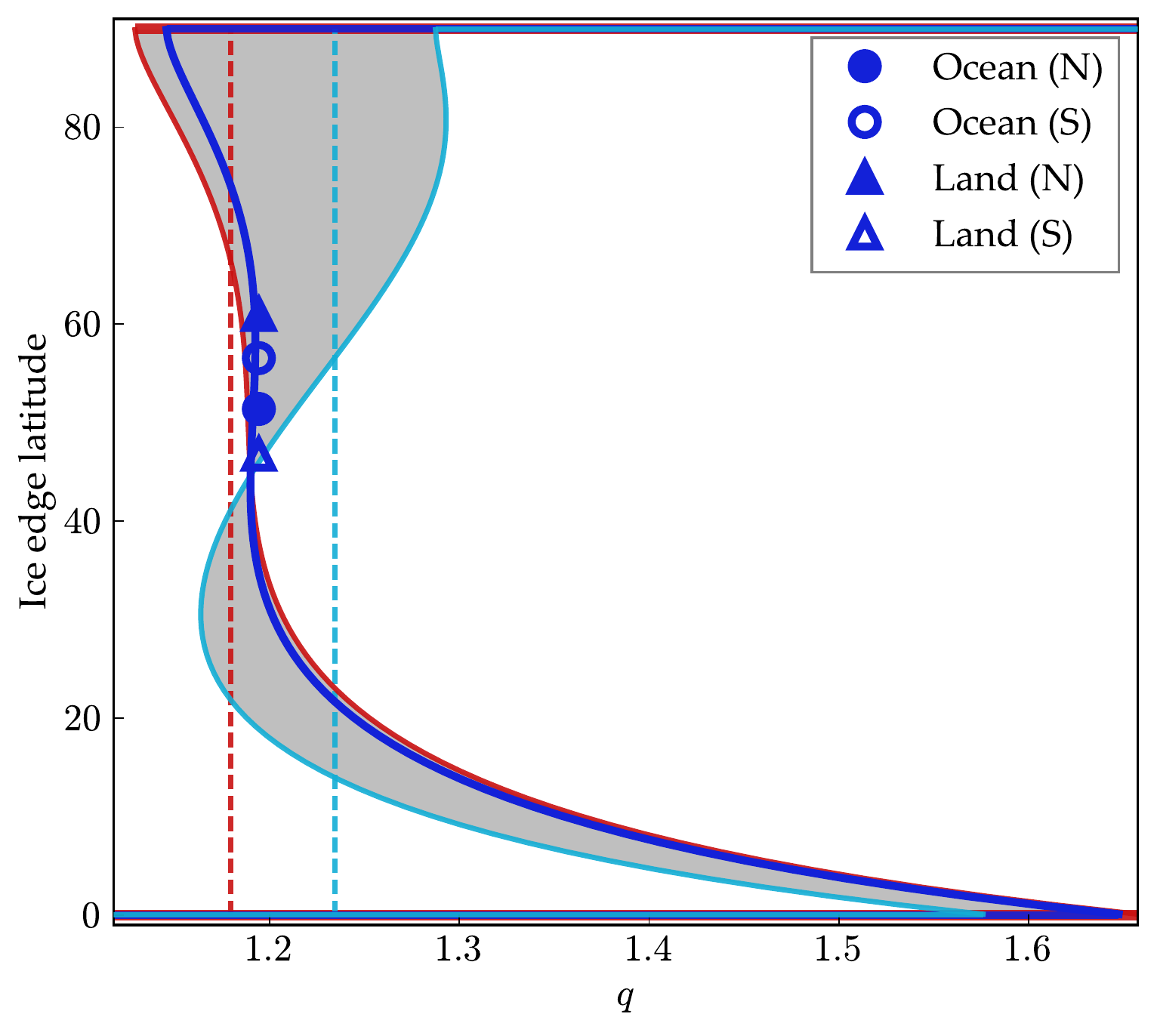}
\caption{\label{fig:caseLowSlow_icestab}Ice edge latitude as a 
function of the parameter $q$ (see Section 
\ref{sec:climatemodel}) from the analytical annual energy 
balance model \citep{rose2017}, for the cases shown in Figures 
\ref{fig:caseLowSlow_1} (left) and \ref{fig:caseLowSlow_2} (right). The 
solution is a function of obliquity: light blue corresponds to 
the minimum obliquity in the simulation, red to the maximum 
obliquity, and the gray-shaded area is the range explored by the 
planet. Vertical dashed lines indicate the value of $q$, which 
is a function of eccentricity, at the corresponding times. In 
the left panel, markers show the ice edge latitude for northern 
and southern land and ocean at the time of maximum obliquity, at 
the coeval value of $q$, which depends on the eccentricity. Triangles and
circles represent land and ocean, respectively, while closed and open
markers represent northern and southern hemispheres, respectively. The 
right panel also shows these ice edge latitudes and the 
analytical solution at 500 years before the planet becomes fully 
glaciated (dark blue).}
\end{figure*}

In the previous section, we saw that the ice caps often become 
unstable as a result of the orbital/obliquity evolution. Though 
we highlighted the snowball instability (or LICI), the SICI can 
also be observed in the rapid retreat of the ice sheets. We can 
use the analytical solution from \cite{rose2017} (Section 
\ref{sec:rosemodel}) to plot the ice edge latitude as a function 
of the dimensionless parameter, $q$ (Figure 
\ref{fig:caseLowSlow_icestab}). As we discussed, the slope of this 
curve indicates whether the equilibrium ice line is stable or 
unstable. 

Figure \ref{fig:caseLowSlow_icestab} shows the ice edge latitude as 
a function of the parameter $q$, from the \cite{rose2017} 
solution, for the two cases discussed above (see Section \ref{sec:rosemodel}).
The dimensionless 
parameter $q$ describes the combined effects of insolation and 
greenhouse warming.

The panels in Figure \ref{fig:caseLowSlow_icestab} show the equilibrium ice edge 
latitude at different obliquities---the light blue line at each 
case's minimum obliquity, and the red line at its highest 
obliquity. The gray shaded area indicates the full range of 
solutions the simulation explores. When the slope of the line is 
positive or zero (as in the upper and lower branches), the ice 
edge is in a stable equilibrium (the annual solution is an 
equilibrium model). When the slope is negative or undefined, the 
ice edge is unstable and gives rise to the small ice cap 
instability (SICI) at the highest latitudes, and the large ice 
cap instability (LICI) at the mid to low latitudes. When the ice 
edge is at 90$^{\circ}$, there is no ice cap; when it is at 
0$^{\circ}$, the planet is in a snowball state. 

The left-hand panel corresponds to the case that does not 
experience the LICI (Figure \ref{fig:caseLowSlow_1}). In this case, 
there is always a stable branch for the ice edge at all 
obliquities. The points shown in the plot are the actual ice 
edge locations from our full seasonal model, for both the land and ocean 
in each hemisphere, at the time of the highest obliquity. The 
vertical dashed lines indicate the average annual value of $q$ 
(which depends on the eccentricity) at each obliquity extreme. 
These points lag the analytic ice edge solution (which 
represents the climate in equilibrium) in time, and are 
dependent on the seasonality and the nature of the ice sheet 
model, and so do not fall directly on the analytical solution at 
most times. Nevertheless, the points stay very near to the 
analytical solution, and give a sense of why the instability 
is avoided. In this case, the instability never occurs 
because the ice edges (land and ocean in each hemisphere) remain 
on a stable branch of the analytical solution at mid-latitudes 
(or retreat to $0^{\circ}$). 

In the right-hand panel, we see the same quantities plotted for 
the second case (Figure \ref{fig:caseLowSlow_2}), which experiences 
the LICI. We can see that at the highest obliquity (red curve), 
there is no stable ice edge between 0$^{\circ}$ and 
90$^{\circ}$. We have additionally plotted the analytical 
solution $\sim 500$ years before the planet has fully entered 
the snowball state. We can see that the ice edges in each 
hemisphere are precariously perched upon a branch of the 
solution where the slope is becoming undefined. At this point, 
the ice must either retreat entirely or expand to the equator. 
Because this occurs near a minimum in global insolation 
(the eccentricity is low), and the ice sheets have high 
thermal inertia, the snowball state is more easily reached. This 
demonstrates the susceptibility of planets with large 
orbital/obliquity variations to the snowball instability. 
Essentially, if planets proceed to a high obliquity and low 
eccentricity state with ice sheets extending to mid-latitudes, 
the ice edge becomes unstable and the entire planet quickly 
freezes. 

For the climate parameters we use here, this instability occurs when the obliquity
reaches $\sim 35^{\circ}$. These climate parameters ($a_0, A, B$, and $D$) are chosen 
to reproduce Earth's atmosphere, however, a planet with different atmospheric 
properties will respond differently to this obliquity oscillation. For some types of 
atmospheres, the instability will occur at a different obliquity, for others, the 
instability may not occur at all 
\citep[for a detailed exploration of the climate parameters, see][]{rose2017}.

\begin{figure*}
\includegraphics[width=0.5\textwidth]{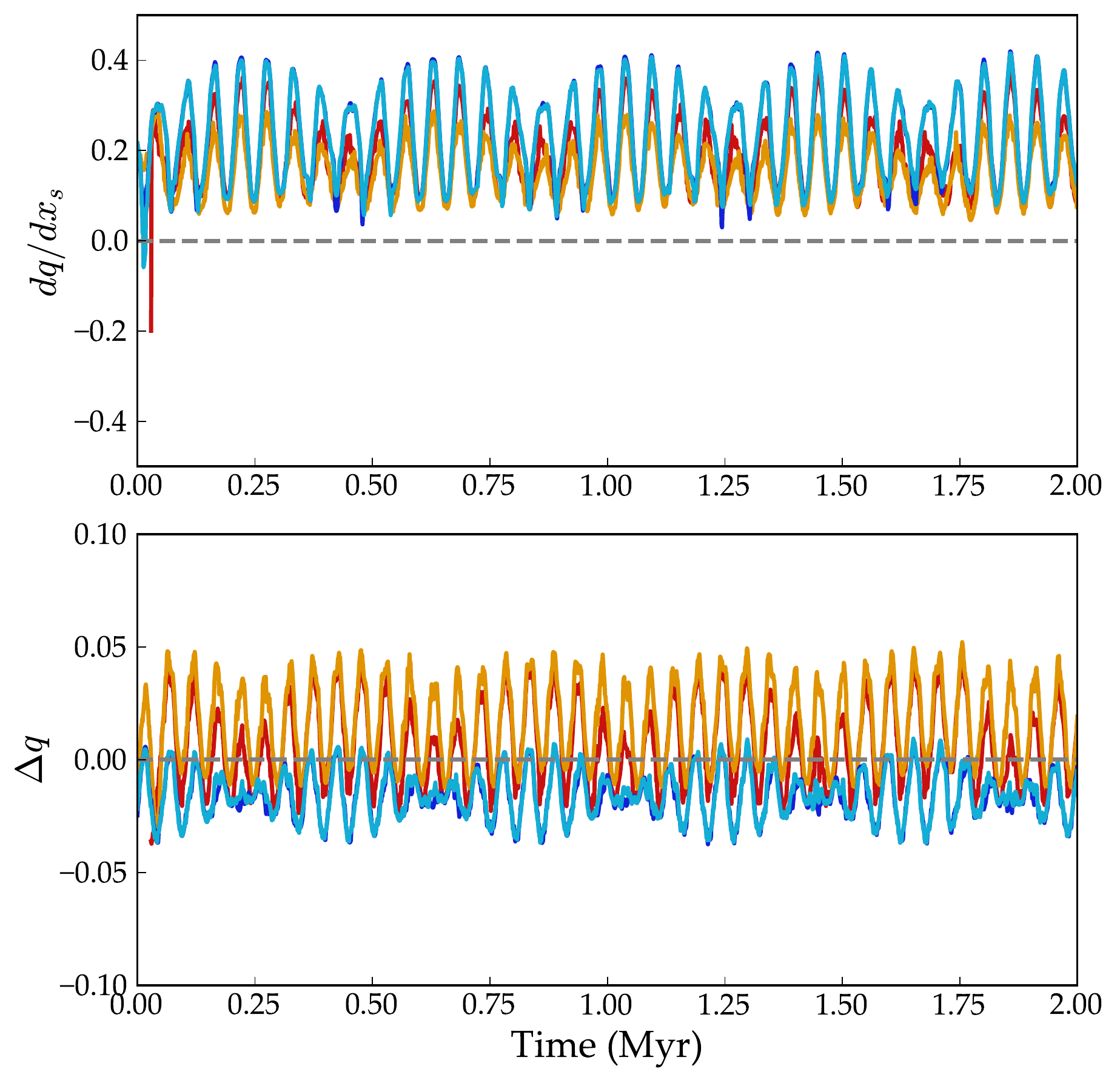}
\includegraphics[width=0.5\textwidth]{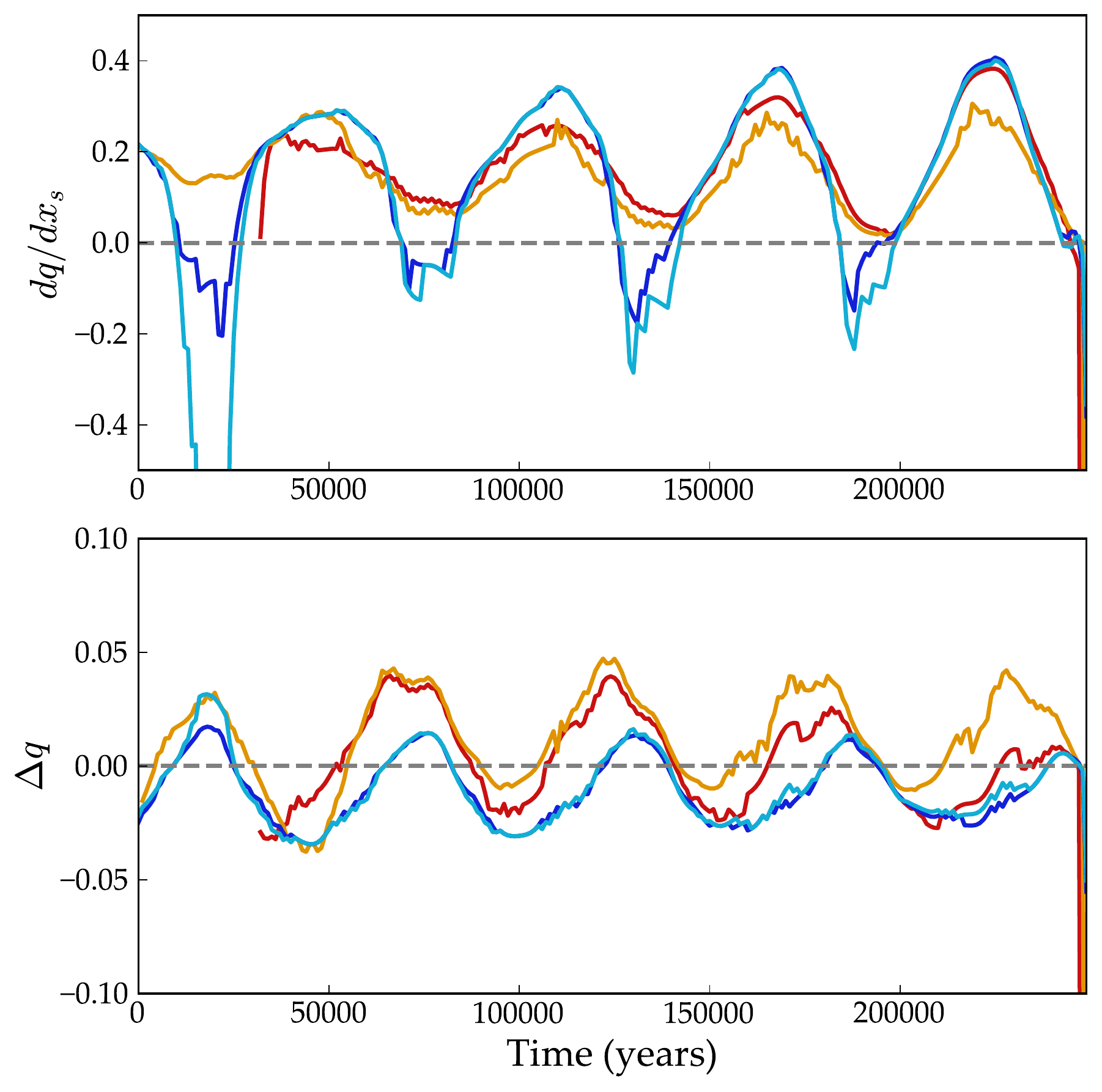}
\caption{\label{fig:icestability1} The quantities $dq/dx_s$ and 
$\Delta q$, which are related to the stability of ice caps in 
the annual EBM (see text), for a case with $P_{rot} =1.62$ day, 
$\varepsilon_0 = 23.5^{\circ}$, $e_0 = 0.167$, and $i_0 = 
14.54^{\circ}$ (left) and $i_0 = 16.04^{\circ}$ (right). The quantities 
$dq/dx_s$ and $\Delta q$ are 
plotted as a function of time for the northern ice sheet on land 
(red), the southern ice sheet (orange), the northern sea ice 
(dark blue), and the southern sea ice (light blue). Negative 
values of $dq/dx_s$ indicate the ice cap is unstable in the 
annual model (but not necessarily in our seasonal model). 
Negative values of $\Delta q$ indicate that the average 
insolation is below that required to maintain the ice edge at 
its current latitude, suggesting that the ice should grow. In the 
left-hand case, the ice-cap is stable over the entire simulation.
In the right-hand case, $dq/dx_s$ periodically dips below zero 
for the ocean in both hemispheres, but the snowball instability isn't
triggered until $dq/dx_s<0$ for land.}
\end{figure*}

Figure \ref{fig:icestability1} shows two parameters that can be used 
to analyze the ice edge stability: $dq/dx_s$ and $\Delta q$, for 
a clement (\emph{i.e.} non-snowball) case with $P_{rot} = 1.62$ 
day and $\varepsilon_0 = 23.5^{\circ}$. Both quantities are calculated 
at the ice edge latitude for northern and southern land and 
ocean, for a total of four ice edges. The ``perturbation'', 
$\Delta q$, is 
\begin{equation}
\Delta q = q_{true} - q_{equil},
\end{equation}
where $q_{\text{true}}$ is the ``true'' value of $q$, calculated from 
the stellar flux and the eccentricity at that instant in time 
and $q_{\text{equil}}$ is calculated from the analytical solution, at 
each ice edge and the current obliquity. Thus, it is when both 
$dq/dx_s$ and $\Delta q$ are negative that we would expect the 
snowball states to occur---this corresponds to the third 
quadrant in the right panel of the figure. Both $dq/dx_s$ and 
$q_{\text{equil}}$ are calculated from the Python package developed in 
\cite{rose2017}, see Section \ref{sec:rosemodel}. 

As described previously, the ice caps will become unstable any 
time $dq/dx_s < 0$. Whether or not the caps collapse to the 
poles or grow to the equator depends on the direction of the 
perturbation, $\Delta q$. Figure 
\ref{fig:icestability1} (left panels) shows a case in which the ice edges are 
truly stable (except in the earliest phase, when the ice sheets 
are growing): $dq/dx_s > 0$ over the entire simulation.

The same quantities are shown in Figure \ref{fig:icestability1} (right panels) for 
an adjacent case which undergoes the snowball instability. In 
this case, $dq/dx_s$ becomes negative several times for the sea 
ice in both hemispheres and $\Delta q$ is negative during some 
of these excursions. The ice edges do not grow immediately to 
the poles, however. This may be due to the fact that the model 
is not in equilibrium, but since the sea ice is treated as a 
thin veneer that melts instantly when $T>-2$, the response time 
of the oceans to changes in insolation should be relatively 
short. \cite{rose2017} shows that the seasonal model does 
deviate from the analytical solution; this is probably the 
reason the instability does not occur during those times. 

Careful inspection of the upper right panel in Figure 
\ref{fig:icestability1} shows that it is actually the northern ice 
sheet (red curve) that leads the way into the snowball state, 
not the sea ice in either hemisphere. It is interesting that 
this happens so quickly after $dq/dx_s$ becomes negative for this 
ice sheet, when the instability did not occur during previous
excursions below zero.
It is possibly a result of hysteresis: one may note 
that $\Delta q$ at the northern ice edge was fairly large and 
positive during the first three eccentricity cycles. During the 
fourth ($\sim 220,000$ years), however, $\Delta q$ barely 
exceeds zero before $dq/dx_s$ becomes negative. In other words, 
the ice sheet receives strong heating during all of the previous 
three eccentricity maxima, but very weak heating during the 
last, which leaves it poised, so to speak, to continue growing the next time
$dq/dx_s< 0$.

\begin{figure*}
\includegraphics[width=0.5\textwidth]{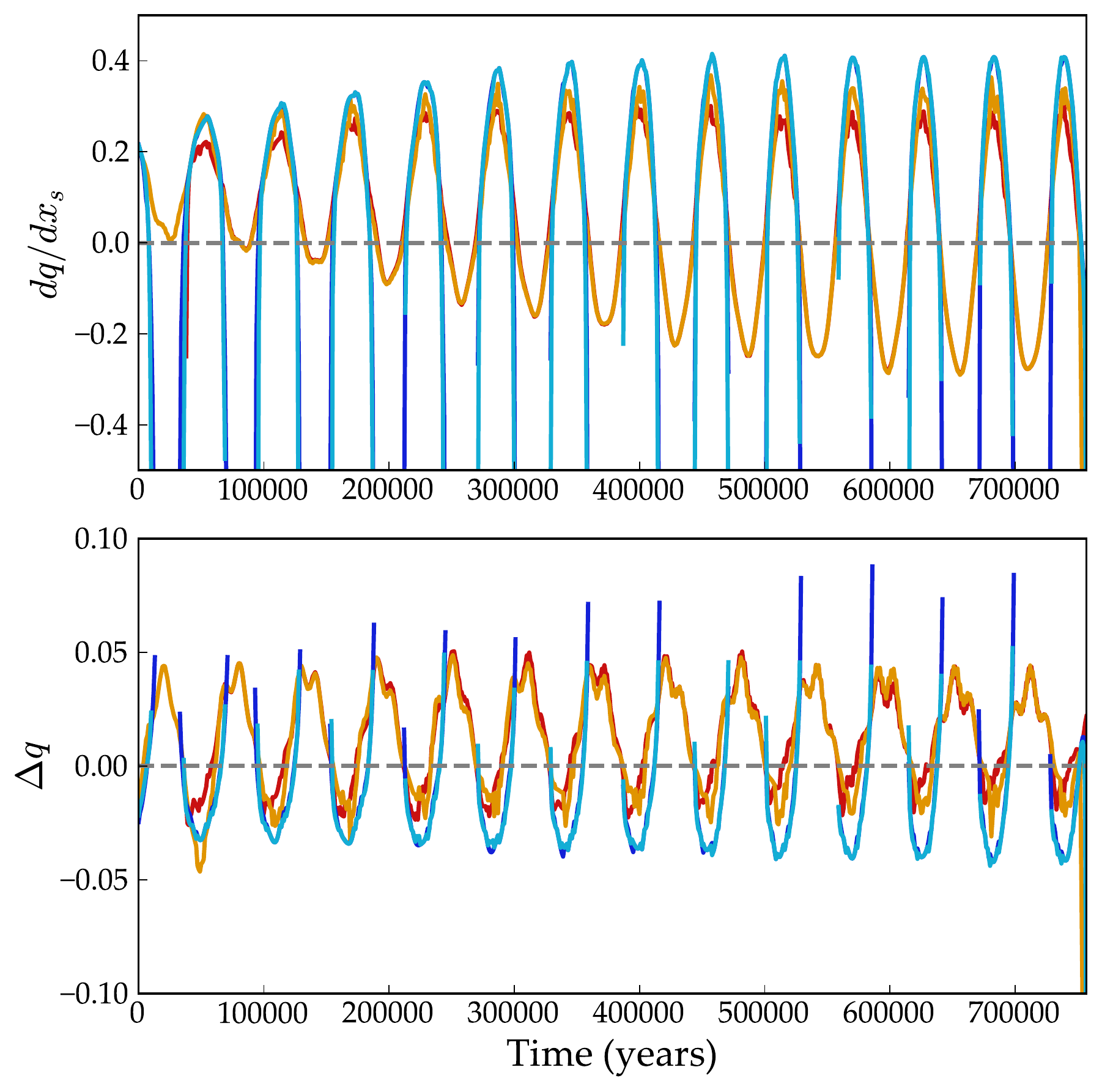}
\caption{\label{fig:icestability2} Same as Figure 
\ref{fig:icestability1}, but for $P_{rot} =1.62$ day, $\varepsilon_0 = 
23.5^{\circ}$, $e_0 = 0.167$, and $i_0 = 18.96^{\circ}$. This 
case enters a snowball state at $\sim750,000$ years. The 
northern and southern sea ice caps melt completely numerous 
times prior to the instability at $\sim760$ kyr---shown as gaps 
in the blue curves. Eccentricity and obliquity are high during 
these times.}
\end{figure*}

The analytical theory does not always provide a simple 
explanation, as it does for the case shown in Figure 
\ref{fig:icestability1}. Figure \ref{fig:icestability2} shows another 
nearby case that undergoes the snowball instability. 
For most of the simulations, whenever $dq/dx_s<0$, $\Delta q$ is positive. At 
these times the sea ice usually disappears entirely (gaps in the 
blue curves left panels). The occurrence of a snowball 
state at $\sim 750$ kyr may be a result of 
hysteresis again---$\Delta q$ does undergo a negative period 
shortly prior to the snowball state, but this period does not 
appear significantly different from the cycles before it. 

\subsection{Relative importance of obliquity, eccentricity and COPP}
With orbital and obliquity cycles as large as our test planet 
here, the periodicity of the ice is plainly visible. It is 
interesting, still, to perform periodogram analysis to 
understand the relative importance of the three insolation 
parameters: obliquity, eccentricity, and COPP. We calculate 
periodograms for each of these variables, for the ice sheet 
heights at $65^{\circ}$ north and south, and for the total 
global ice mass. These are calculated using the periodogram 
function in the \texttt{SciPy} package for Python, with a 
Bartlett window function to produce a clean power spectrum 
\citep{jones2001}.

We first perform a periodogram analysis on a static, but 
eccentric case. Under our ``static'' conditions, the orbit and 
obliquity do not change, but we can still allow the spin axis to 
precess according to Equation (12) in Paper I. This results in a 
sinusoidal variation in COPP. This parameter is 
typically the weakest of the three insolation parameters, so 
this example, which has no variation in $\varepsilon$ or $e$, allows us 
to see its effect more plainly. 
The ice sheets grow and decay in 
response to the planet's precession. The total ice volume's 
strongest peak is at half the period of COPP---this is because 
the northern and southern ice sheets grow and decay at opposing 
times. 

\begin{figure*}
\includegraphics[width=0.5\textwidth]{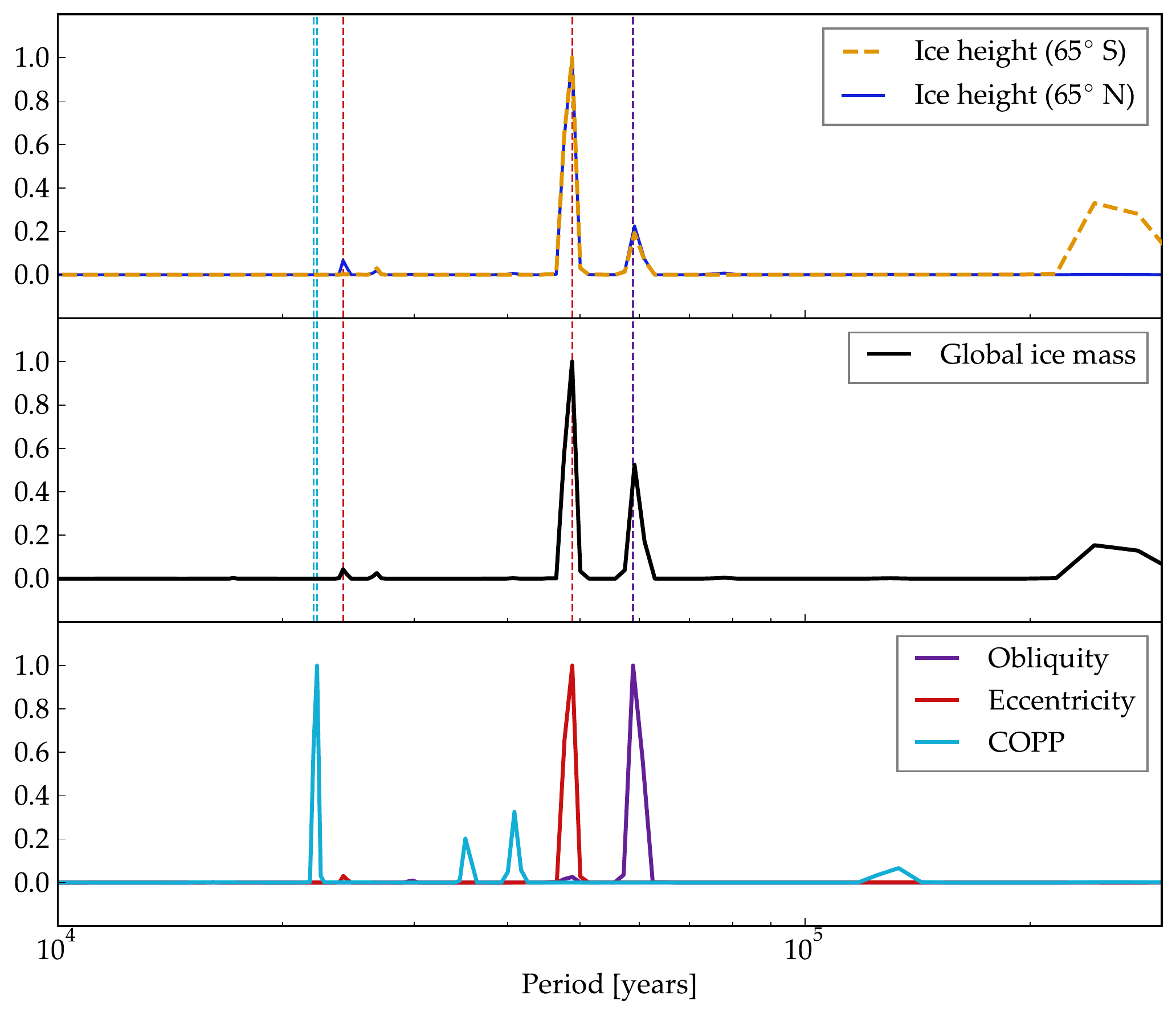}
\includegraphics[width=0.5\textwidth]{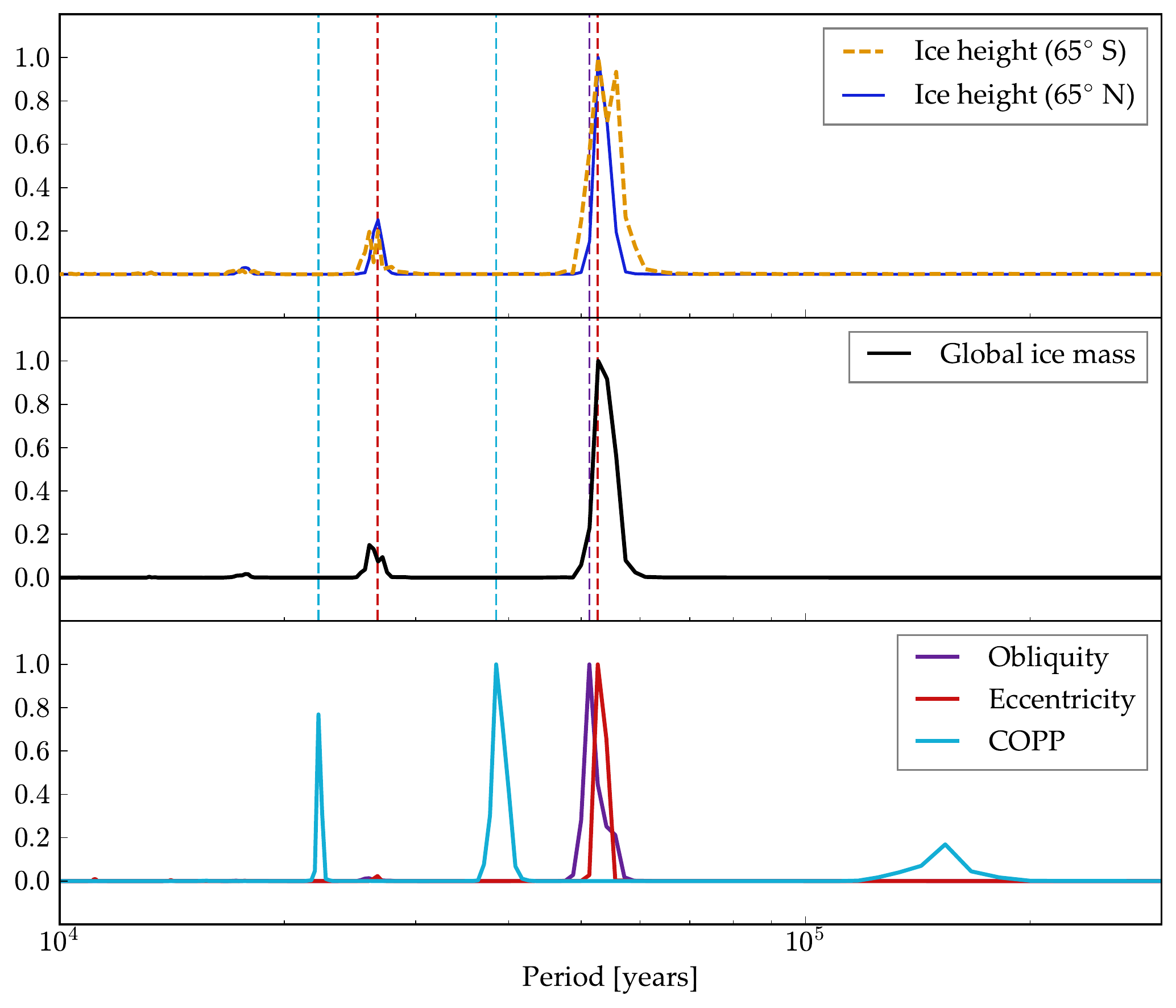}
\caption{\label{fig:fft_slowrot} Normalized power  spectra showing 
the strength at different periods in the ice height (top panel), 
global ice volume (middle panel), and the insolation parameters 
(obliquity, eccentricity, and COPP). Vertical dashed lines in 
the top two panels indicate peaks in the insolation parameters. 
The left panel shows a case with $P_{rot} =1.62$ day, $\varepsilon_0 = 
23.5^{\circ}$, $e_0 = 0.167$, and $i_0 = 11.67^{\circ}$ and the 
right shows a case with  $P_{rot} =1.62$ day, $\varepsilon_0 = 
23.5^{\circ}$, $e_0 = 0.25$, and $i_0 = 16.04^{\circ}$. The ice 
sheets are strongly coupled to the eccentricity and, to lesser 
extent, the obliquity. The case on the right lies within the 
secular spin-orbit resonance, hence the obliquity and 
eccentricity have the same period of oscillation.}
\end{figure*}

Figure \ref{fig:fft_slowrot} shows the periodograms for two cases 
with $P_{\text{rot}} = 1.62$ day and $\varepsilon_0 = 23.5^{\circ}$ that are 
characteristic of the behavior we see over much of this 
parameter space. The left panel shows a case that is outside the 
secular resonance (see Figure \ref{fig:lowoblmidslowmap}) and the 
right shows a case that is \emph{inside} the resonance. 
Outside the resonance, the obliquity and eccentricity have 
distinct peaks, and both can be seen in the ice sheet growth and 
decay. In the secular resonance, the obliquity oscillates with 
almost exactly the same period as the eccentricity (a consequence of
resonance), and the ice sheets follow this period. 
Interestingly, in all of the parameter space we explore, the ice 
mass is dominated by the eccentricity cycle, not the obliquity 
cycle, except in the secular resonance, when the frequencies
are similar and thus difficult to disentangle. 
The periods associated with COPP cannot even be seen in 
the ice sheets on a linear scale. The ice sheets are mostly driven
by the eccentricity, while the obliquity controls their stability
(Section \ref{sec:icestab}). 

\subsection{Importance of ice sheets}
The inclusion of the ice sheet model has important 
consequences. The snowball instability is triggered more easily 
(\emph{i.e.}, at higher $S_{\star}$), because of the extra 
energy required to melt the ice sheets (compared to the energy 
required simply to raise the surface temperature above 
freezing). Thus the climate with ice sheets is 
generally cooler at the same stellar flux than without. Indeed, 
without ice sheets, for our test planet at $\varepsilon = 
23.5^{\circ}$, the snowball state is not reached until $S/S_0 
\approx 0.95$, compared to $S/S_0 \approx 0.975$ with ice sheets
(Figure \ref{fig:trans}). 

\begin{figure*}
\includegraphics[width=\textwidth]{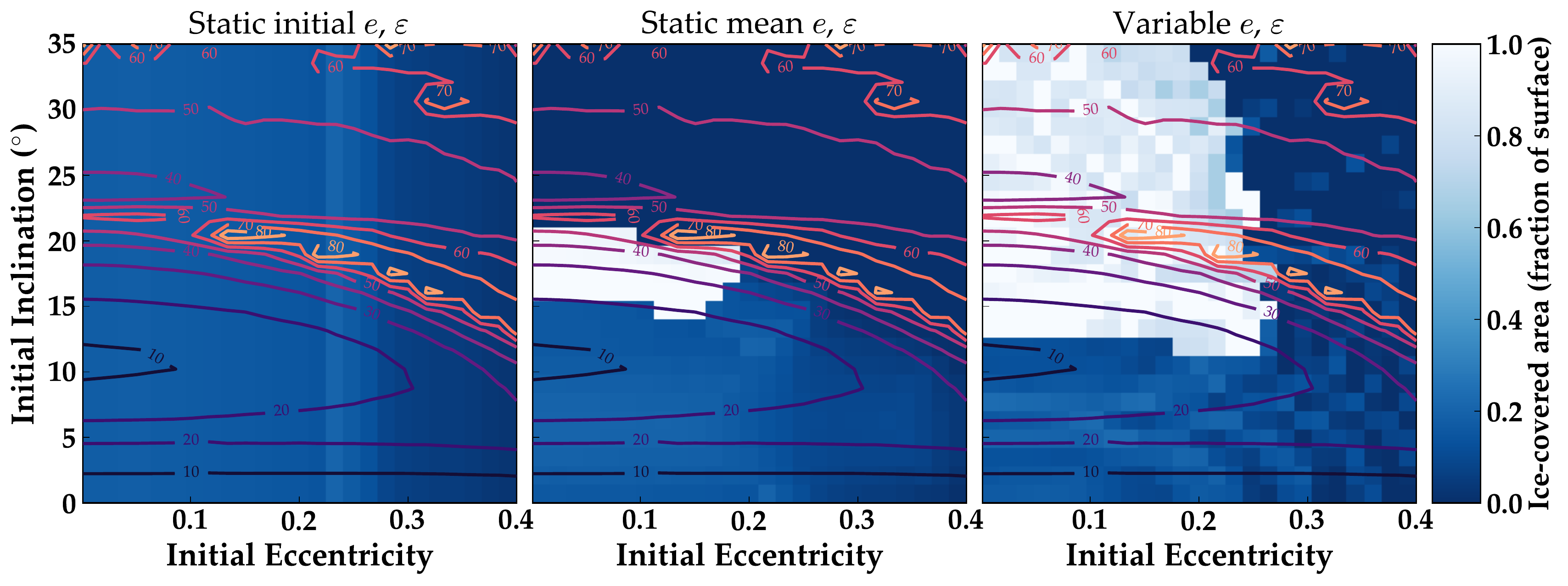}
\caption{\label{fig:areacovnoice} Fractional area of ice coverage at 
$\varepsilon_0 = 23.5^{\circ}$, $P_{rot}=1$ day, with ice sheets 
disabled. On the left are static conditions at the initial 
values; on the right, dynamic orbit and obliquity. Compare to 
Figure \ref{fig:lowoblmidmap}. The stellar flux here is lower than 
in the simulations from Figure \ref{fig:lowoblmidmap}, $S = 1304$ W 
m$^{-2}$. The ice coverage is very different from the cases with 
ice sheets at low inclinations---in the lower left, where 
the obliquity variations are relatively small.}

\end{figure*}

The response to orbital variations is altered as well. Figure 
\ref{fig:areacovnoice} shows the fractional area coverage for 
$\varepsilon_0 = 23.5^{\circ}$, $P_{rot}=1$ day, at $S = 1304$ W 
m$^{-2}$. Without perturbations, at this stellar flux, there are 
no snowball states. At $e \sim 0.25$, the area of ice coverage 
increases slightly, because of increased apoastron distances and 
time spent there, but the ice coverage drops to zero at the 
highest eccentricities. When perturbations are included, the 
area of ice coverage increases in most regions and snowball 
states are reached at $i_0 \gtrsim 12^{\circ}$ and $e_0 \lesssim
0.25$. The change in ice coverage between static and dynamic 
cases is more pronounced here than in the low obliquity cases 
with ice sheets (Figure \ref{fig:lowoblmidmap}). Further, the region
of small obliquity variations (lower left) does not experience snowball
states as often as the cases with ice sheets.   

\subsection{Comparison with \cite{armstrong2014}}
\label{sec:armstrongcomp}
Here, we revisit the 17 test systems from \cite{armstrong2014}. 
Refer to that paper for the physical details of these systems.
We simulate the orbital evolution using 
\texttt{DISTORB} and \texttt{HNBody} and the obliquity
evolution using \texttt{DISTROT}. In cases 1, 2, 5, 6, 7, 13, 14, and 17, 
the combined orbital/obliquity evolution resulting from the secular model
(\texttt{DISTORB}) matches sufficiently well with \cite{armstrong2014}, 
and we couple these directly to the climate model, \texttt{POISE}. In the
rest of the cases,
the eccentricity and/or obliquity evolution (using \texttt{DISTORB})
diverges significantly from
the \cite{armstrong2014} simulations or the semi-major axis evolution 
is large enough that we must use \texttt{HNBody} for the orbital evolution.
Whether we ultimately use \texttt{DISTORB} or \texttt{HNBody}, 
we ensure that the obliquity/orbital evolution 
matches well with \cite{armstrong2014} before running the climate model.

In all cases, we run the climate model with the same parameters and
initial conditions as for our Earth comparison (Section \ref{sec:valid}) 
and the Earth-mass planet in our test system.
For each system, we run three sets of \texttt{POISE} simulations: 
one set with the orbit and obliquity held constant at their initial values,
one set with the orbit and obliquity held constant at their mean values 
(over 1 Myr), and one set with the full orbital and obliquity variations. 

We generate a comparison
with \cite{armstrong2014} by varying the stellar luminosity
and locating the value, $L_{\text{OHZ}}$, at which the transition between warm, 
clement conditions and the
snowball state occurs. The semi-major axis at which the 
outer edge of the habitable zone (OHZ) occurs is then calculated from 
\begin{equation}
    a_{\text{OHZ}} = a_{\oplus}\sqrt{\frac{L_{\odot}}{L_{\text{OHZ}}}}.
\end{equation}
The purpose of this somewhat awkward definition is solely to compare 
directly with \cite{armstrong2014}. We do not vary the \emph{initial}
semi-major axis of the planet ($a_0 = 1$ au in every case) because the 
eccentricity and obliquity evolution would be different at every location.
Varying the stellar luminosity instead gives us a way of isolating
the effects of the dynamical evolution. This definition of 
$a_{\text{OHZ}}$ is also not fully self-consistent because in several cases 
(systems 4, 10, 
and 11), the semi-major axis of the planet varies by $\sim 10 \%$, leading
to a significant change in the stellar flux received by the planet. 
This ultimately leads to a significant decrease ($\sim6-8 \%$) in 
$a_{\text{OHZ}}$ for these three cases. In reality, it is probably more 
accurate to describe this result as an excursion beyond the habitable 
zone due to an increase in semi-major axis $a$, rather than a decrease
in the distance at which the planet enters a snowball state. Such is the 
difficulty in reducing a concept as multi-faceted as orbital evolution
to a single parameter, $a_{\text{OHZ}}$.

\begin{figure*}
\includegraphics[width=\textwidth]{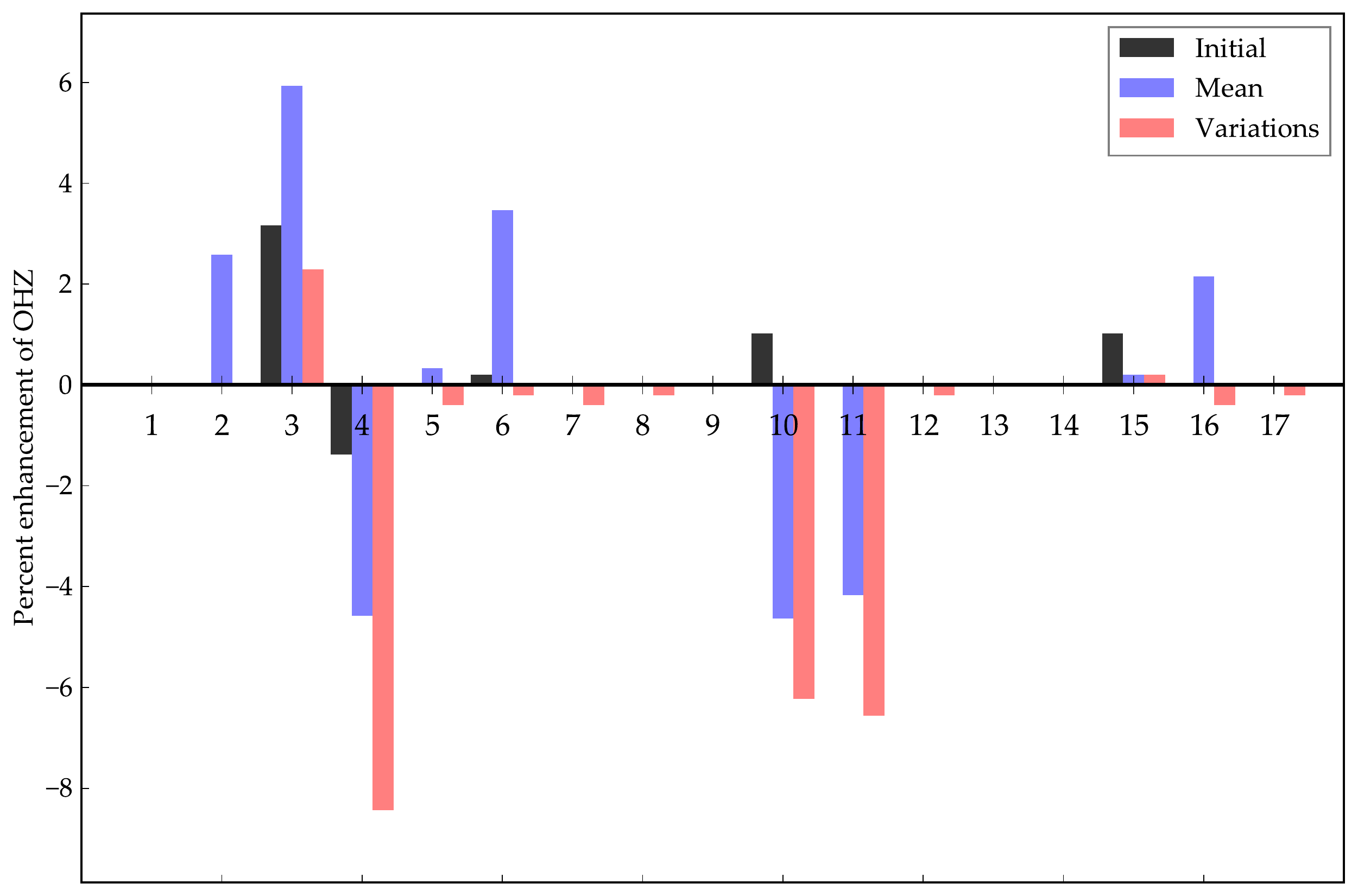}
\caption{\label{fig:ohzcomp} Percent enhancement in the distance to the 
OHZ from the host star for the 17 systems in \cite{armstrong2014}. The 
percent enhancement for each system is measured relative to system 1, 
like in \cite{armstrong2014}. Black bars are for static orbits and obliquity at 
the initial values, blue bars are for static orbits and obliquity at the mean values,
and red bars are for variable orbits and obliquity. In cases 4, 10, and 11, the 
semi-major axis of the Earth-mass planet varies by $\sim10\%$, leading to 
large changes in insolation and subsequent snowball states. In cases 2, 3, 5, 6, 15,
and 16, the large mean obliquity leads to an extension of the habitable zone for 
static mean conditions. In most systems, variable eccentricity and obliquity leads 
decrease in the OHZ distance.}
\end{figure*}

The percent enhancement of the OHZ is then calculated for each system
relative to system 1 and displayed in Figure \ref{fig:ohzcomp} for the static 
initial, static mean, and variable orbit and obliquity (compare to Figure 
11 in \cite{armstrong2014}).
Note also that system 1 has the same $a_{\text{OHZ}}$
for the static initial, static mean, and variable orbit/obliqiuty values, 
so the percent enhancement for each is zero. In most cases, the change
in $a_{\text{OHZ}}$ from system 1 is $\lesssim 1 \%$. The OHZ is enhanced 
under static initial conditions for systems 3, 10, and 15 as a result of the
high initial eccentricity of the planet. In systems 2, 3, 5, 6, 15, and 16, 
the enhancement under static mean conditions is a result of the planet's 
high mean obliquity. Variations enhance the OHZ relative to system 1 only 
in systems 3 and 15, which also saw warmer conditions due to the higher 
initial eccentricity. For the most part, the variations lead to a decrease
in $a_{\text{OHZ}}$. Except in cases where there was no change to the
OHZ, variations always lead to a decrease in the $a_{\text{OHZ}}$ compared to 
static conditions in the same system.

Ultimately, our results are significantly different from \cite{armstrong2014}. 
Compare our Figure \ref{fig:ohzcomp} with their Figure 11. We find that, in 
general, dynamical evolution of the eccentricity and obliquity 
of a HZ planet tends to make the planet more susceptible to 
snowball states than when it has static orbital conditions, 
while \cite{armstrong2014} found that dynamical 
variations tended to inhibit glaciation and snowball states. 
There are two fundamental reasons our results differ from 
that study. 

The first is related to the parameterization of the OLR. The 
stability of the EBM is related to the strength of the longwave (LW)
radiation
feedback and the ice-albedo feedback. The LW radiation feedback is 
negative: a small positive perturbation to the surface 
temperature will cause the OLR to increase, generating more 
cooling and returning the surface to the unperturbed 
temperature. The process also works in the other direction: a 
small negative perturbation to the temperature will cause the 
OLR to decrease, creating additional heating and returning the 
temperature to its previous value. The ice-albedo feedback is 
positive: a small negative perturbation to the surface 
temperature will cause the ice to grow, reflecting more 
radiation to space and causing the surface to cool further. A 
positive perturbation will likewise generate runaway warming, if 
the ice-albedo feedback is the dominant feedback of the model. Of 
course, the real Earth and more sophisticated 3D models have a 
number of other feedback processes that work to alter the 
climate stability, but in a 1D EBM like ours and the model in 
\cite{armstrong2014}, stability is simply a LW competition between the 
radiation feedback and the ice-albedo feedback. 

In this simple formulation, the LW radiation feedback is contained 
within the parameter $B$. A large, positive value of $B$ will 
create a very stable climate, while a smaller value will create 
a less stable climate. For Earth, $B \approx 2.09$ W m$^{-2}$ 
K$^{-1}$ \citep{northcoakley1979}.
A Taylor expansion of the OLR parameterization in 
\cite{spiegel2009}, for example, shows that their model 2 has $B
\approx 2.28$ W m$^{-2}$ K$^{-1}$ at a surface temperature of 
288 K, and so their model should be more stable against snowball 
states when using this formulation than with the OLR from 
\cite{northcoakley1979}. 

The OLR from \cite{armstrong2014} is found by combining their 
Equations (23) and (24) and comparing to the full energy balance 
equation (our Equation \ref{eqn:ebmeq}):
\begin{equation}
I(T) = \frac{\epsilon_s \sigma T_s^4}{1+\tau} - F_{\text{surf}},
\label{eqn:johnolr}
\end{equation}
where $\epsilon_s$ is the emissivity of the atmosphere, $\sigma$ 
is the Stefan-Boltzmann constant, $F_{\text{surf}}$ is a tunable
constant and $\tau$ is a tunable 
parameter used to approximate the greenhouse effect that was 
\emph{not} assumed to be a function of temperature. The authors 
found that setting $\epsilon_s = 1$ and $\tau = 0.095$ 
reproduced Earth and so fixed these values for the rest of the 
study. As stated before, a Taylor expansion of Equation 
\ref{eqn:johnolr} with respect to temperature gives the value of 
$B$:
\begin{equation}
B = \frac{dI}{dT} = \frac{4 \epsilon_s \sigma T_s^3}{1+\tau}.
\end{equation}
Plugging in their constants and a surface temperature of $T_s = 
288$ K, one finds $B = 4.95$ W m$^{-2}$ K$^{-1}$. As far as EBMs 
go, this model is extremely stable against the snowball 
instability. 

The second reason our model differs from \cite{armstrong2014} is 
our inclusion of the horizontal heat transport (however crudely 
it is represented here). A comparison between our energy balance
equation (\ref{eqn:ebmeq}) and that in \cite{armstrong2014} shows
that $D = 0$ in the latter. It can be shown that when $D=0$, the
ice-albedo feedback does not affect adjacent latitudes as it 
should. Conceptually, ice-albedo feedback occurs because, for 
example, when the albedo (and thus temperature) changes in one 
model cell, the temperature gradient between adjacent cells is 
changed. This causes the heat flow between cells to change. The 
feedback works because cooling (or heating) in one cell alters 
heat flow to and from adjacent cells, cooling (or heating) those 
adjacent areas. Without that horizontal heat flow, there is no 
ice-albedo feedback, and no snowball \emph{instability}---that 
is, snowball states can still occur, but only when all latitudes 
in the model \emph{individually} come into radiative equilibrium 
at below freezing temperatures. That occurs at a much lower 
stellar flux than that caused by the instability.

\subsection{Predicting climate states with machine learning}

Results from the statistical analysis and machine learning model are shown in Tables \ref{tab:daveresults1} and \ref{tab:daveresults2}. Correlations are strongest with stellar flux, $S$, and the eccentricity parameters. The MIC values are similar $\sim 0.2-0.3$ across most of the parameters, except for $\varepsilon$'s relationship to $\delta_{\text{snow}}$. Interestingly, $\Delta i$ shows a stronger correlation, $R$, with $\delta_{\text{snow}}$ and $f_{\text{ice}}$ than the obliquity parameters, despite the fact that the inclination has no direct impact on climate. The linear relationships ($R$) between ($f_{\text{ice}}$, $i_0$), ($f_{\text{ice}}$, $\langle i \rangle$), ($\delta_{\text{snow}}$, $i_0$), and ($\delta_{\text{snow}}$, $\varepsilon_0$) are insignificant if a $p-$value of $<0.005$ is desired (see Section \ref{davealgo}).  However, the MIC for these quantities shows a non-linear relationship about as strong as any other parameter. One plausible explanation is that the inclination (especially the variation in inclination) affects both the evolution of the eccentricity and the evolution of the obliquity (see Equations 5,6, 12, and 13 in Paper I), and thus is indirectly coupled to the climate through two variables. 

The stellar flux, $S_{\star}$ (defined here for a circular orbit), is unsurprisingly the most important parameter in determining the final climate parameters, $\delta_{\text{snow}}$ and $f_{\text{ice}}$. The mean eccentricity, $\langle e \rangle$, tends to be the next most important parameter, as expected (see Equation \ref{eqn:annualinsol}). The remaining variables tend to have similar, and relatively small, weighting. About half the time, one could correctly predict the climate state of our test planet with the stellar flux and the mean insolation. However, including all variables, the ML model can predict $\delta_{\text{snow}}$ correctly 97\% of the time, and $f_{\text{ice}}$. For the RF regressor, the accuracy metric is the $R^2$ score, which in this case is $R^2 =0.93$ (the best possible score is 1). The similar weights of the remaining variables illustrates the complexity of the interplay between orbit and climate. Note that feature importances should be interpreted cautiously as correlations between features can skew the features---for example, in the case of two highly-correlated features, one feature can display a high importance ($\xi_i$), while the second displays a low importance. 

\begin{table}
\caption{\textbf{Relative importance of input parameters on $\delta_{\text{snow}}$}}
\centering
\begin{tabular}{lrrrr}
\hline\hline \\ [-1.5ex]
    \multicolumn{1}{c}{Parameter} & \multicolumn{1}{c}{Pearson $R$ ($p$)}  & \multicolumn{1}{c}{MIC} & \multicolumn{1}{c}{$\zeta_{NL}$} & \multicolumn{1}{c}{$\xi_i$}  \\ [0.5ex]
\hline \\ [-1.5ex]
$S_{\star}$ & -0.517486 (0.0) & 0.259659 & -0.008133 & 0.367391 \\
$e_0$ & -0.469633 (0.0) & 0.191850 & -0.028705 &  0.088580\\
$\Delta e$ & -0.281968 (0.0) & 0.181865& 0.102360 & 0.014340 \\
$\langle e \rangle$ & -0.480688 (0.0) & 0.256887 & 0.025826 & 0.227943 \\
$i_0$ & 0.026494 (0.0132) & 0.256149 & 0.255448 & 0.022177\\
$\Delta i$ & -0.318399 (0.0)  & 0.216146 & 0.114768 & 0.024869\\
$\langle i \rangle$ & 0.056757 ($1.08 \times 10^{-7}$)& 0.200756 & 0.197534 & 0.047204\\
$\varepsilon_0$ & -0.026059 (0.01478) & 0.000490 & -0.000189 & 0.015797 \\
$\Delta \varepsilon$ & 0.084789 ($1.95 \times 10^{-15}$) & 0.097013 & 0.089824 &0.094639\\
$\langle \varepsilon \rangle$ & -0.031998 (0.00276) & 0.124936 & 0.123913 & 0.097059\\
\hline 

\end{tabular}

\label{tab:daveresults1}
\end{table}

\begin{table}
\caption{\textbf{Relative importance of input parameters on $f_{\text{ice}}$}}
\centering
\begin{tabular}{lrrrr}
\hline\hline \\ [-1.5ex]
    \multicolumn{1}{c}{Parameter} & \multicolumn{1}{c}{Pearson $R$ ($p$}  & \multicolumn{1}{c}{MIC} & \multicolumn{1}{c}{$\zeta_{NL}$} & \multicolumn{1}{c}{$\xi_i$}  \\ [0.5ex]
\hline \\ [-1.5ex]
$S_{\star}$ & -0.502261 (0.0) & 0.260615 & 0.008349 & 0.396097 \\
$e_0$ & -0.498351 (0.0)  & 0.268657 & 0.020303 & 0.085960\\
$\Delta e$ & -0.322404 (0.0)  & 0.218874 & 0.114929 & 0.012151\\
$\langle e \rangle$ & -0.515085 (0.0) & 0.295807 & 0.030495 & 0.249936\\
$i_0$ & -0.011158 (0.2967) & 0.255632 & 0.255508 & 0.016456\\
$\Delta i$ & -0.361029 (0.0) & 0.216911 & 0.086569 & 0.021697\\
$\langle i \rangle$ &0.020870 (0.0509) & 0.199982 & 0.199546 & 0.036169\\
$\varepsilon_0$ &-0.062202 ($5.77 \times 10^{-9}$) & 0.170839 & 0.166970  & 0.018088\\
$\Delta \varepsilon$ & 0.059806 ($2.16 \times 10^{-8}$)& 0.148690 &  0.145113 & 0.079007\\
$\langle \varepsilon \rangle$ &-0.092422 ($4.61 \times 10^{-18}$) & 0.242192 & 0.233650 & 0.084440\\
\hline 

\end{tabular}
\label{tab:daveresults2}
\end{table}

\begin{figure*}
\includegraphics[width=\textwidth]{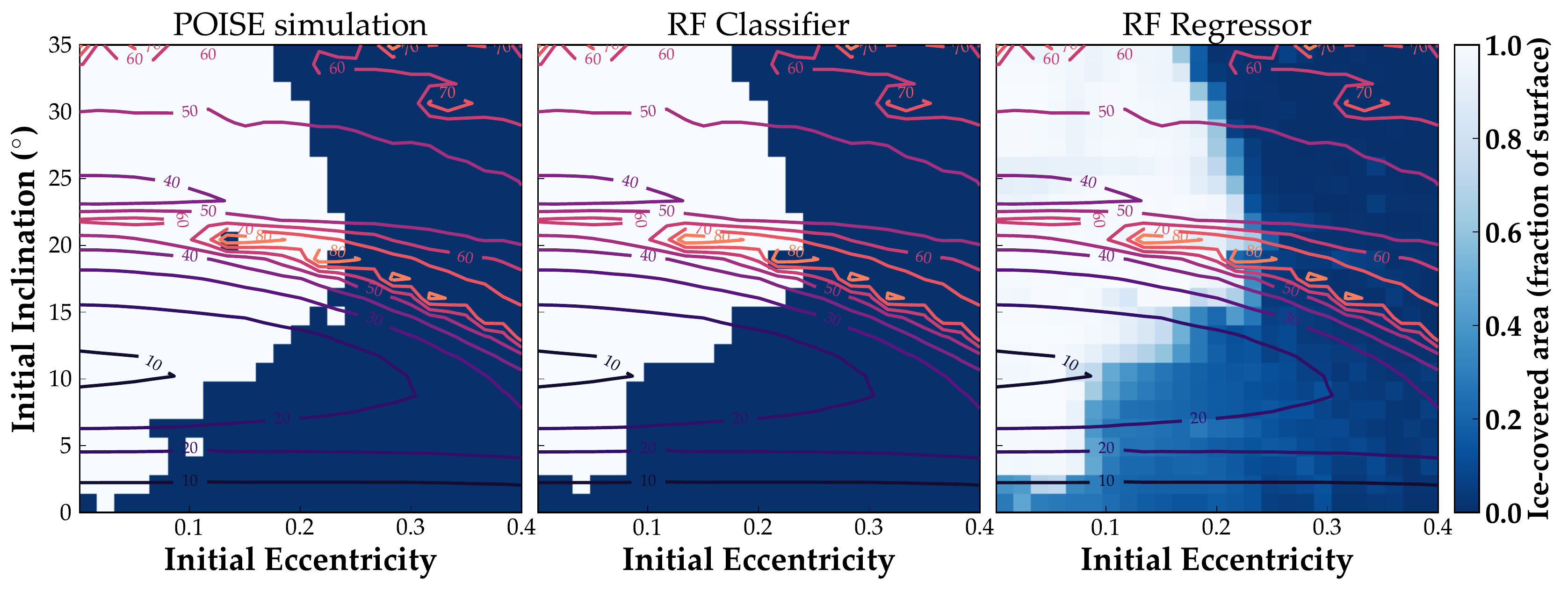}
\caption{\label{mlmodelpredict} Snowball states ($\delta_{\text{snow}}$) for $P_{rot} = 1$ day and initial obliquity $\varepsilon_0 = 23.5^{\circ}$, with a stellar constant of $S = 1332.27$ W m$^{-2}$ from the full orbit/climate simulation (left) and the machine learning algorithm (RF classifier; middle). White regions are simulations that ended in a snowball state; dark blue are those that did not. In the ML case shown here, this slice ($P_{rot} = 1$ and $\varepsilon_0 = 23.5^{\circ}$) of parameter space was excluded from the training set and the algorithm was trained on the remaining data. The right panel shows the fractional ice coverage area for $P_{rot} = 1$ day and initial obliquity $\varepsilon_0 = 23.5^{\circ}$, with a stellar constant of $S = 1332.27$ W m$^{-2}$, as predicted by the random forest regressor. Compare to the right panel in Figure \ref{fig:lowoblmidmap}. }
\end{figure*}


Figure \ref{mlmodelpredict} shows $\delta_{\text{snow}}$ for the full orbit$+$climate simulations (left), compared to the ML algorithm predictions (middle), for one slice of our parameter space. The ML algorithm captures the basic shape of the parameter space, though it does miss a few features such as the blue island at $e\approx0.15$ and $i_0\approx20^{\circ}$. In the case shown, this slice of parameter space ($P_{rot} = 1$ day, $\varepsilon_0 = 23.5^{\circ}$, and $S = 1332.27$ W m$^{-2}$) was excluded from the training set. In the right panel, we show the predicted ice area coverage for $P_{rot} = 1$ day and initial obliquity $\varepsilon_0 = 23.5^{\circ}$ at $S = 1332.27$ W m$^{-2}$. Again, this slice was excluded from the training set. Though the model does slightly better at predicting $\delta_{\text{snow}}$, the algorithm picks out the structure of the original map of $f_{\text{ice}}$ (Figure \ref{fig:lowoblmidmap}).

We conclude that the ML algorithm does very well at predicting the ultimate climate state of this test planet. Though we trained the model on a fixed grid of initial conditions, future studies will probe training sets created with randomized initial conditions. Future analyses will be able to extend the model beyond what is computational feasible via direct integration: when it becomes prohibitive to run a desired number of simulations, we may be able to make do with a fraction of that number when we apply ML. 

\section{Discussion}
We reiterate our primary conclusions here:
\begin{enumerate}
\item In predicting the climate state of a potentially habitable
planet, it is not enough to simply run a climate model with the 
initial conditions (\emph{i.e.} the observed orbit), nor is it 
sufficient to use the averaged quantities. Variations in the 
orbit need to be considered, because of the instability brought 
on by coupled obliquity and eccentricity variations. In 
particular, we note the instability that occurs when the 
planet's obliquity reaches $\sim 35^{\circ}$ during an 
eccentricity minimum, if a large ice cap is present. At this 
obliquity, with the climate parameters we use here, there is no 
stable location for the ice edge; it must either retreat or grow 
uncontrollably. If the incoming stellar flux is decreased 
because the eccentricity is low, the ice will grow to the 
equator. If the eccentricity is sufficiently high at such times, 
the ice caps will collapse entirely. 

\item Coupled orbital and obliquity variations tend to trigger 
the snowball instability. The eccentricity oscillations cause 
the global flux to vary and as a result, the planet can go from 
completely ice free to having large ice caps in a few thousand 
years. If the obliquity remains low enough, the ice caps remain 
stable. When the obliquity is oscillating by a large amount, 
however, the ice latitude can become suddenly unstable. Many 
times, the ice caps are small enough that they disappear 
entirely (the small ice cap instability); other times, the ice 
caps are large enough to trigger the large ice cap instability 
and the planet becomes entirely ice covered. 

\item For eccentricity variations this large ($\Delta e \sim 
0.1-0.3$), the ice ages are primarily controlled by the 
eccentricity, not the obliquity. This is very different from the 
recent Earth, where the insolation variations are dominated by 
the obliquity cycle. Obliquity is important mainly in 
determining the \emph{stability} and \emph{location} of ice 
sheets.

\item The thermal inertia of ice sheets plays an important role. 
The inclusion of ice sheets causes snowball states to be 
triggered at higher incident stellar flux than if a simple 
temperature dependent albedo is used to mimic ice. 
Interestingly, the difference between static and dynamic orbital 
conditions seems to be reduced somewhat by the presence of ice 
sheets. The model is more susceptible to snowball states in 
general, but ice sheets somewhat diminish the response of the 
climate to orbital variations.
\end{enumerate}

In summary, planets undergoing strong orbital forcing are
prone to the snowball or large ice cap instability, and surface 
habitability is therefore compromised. It should be noted, 
however, that Earth potentially went through several snowball 
states during the Proterozoic Eon ($\sim 2.5$ to $0.54$ billion 
years ago), and photosynthetic life persisted during these 
phases \citep{harland1964, kirschvink1992}. One explanation is 
that the surface was not actually completely frozen during such 
time periods---the Earth was in a ``soft'' snowball (or 
``water-belt'') state, with some open ocean in the tropics 
\citep{chandler2000}. An alternative explanation 
is that meltwater ponds
persisted on the surface of the ice, creating a refuge for 
photosynthetic life \citep{hoffman2017}.
Unfortunately, the EBM does not capture 
all the necessary physics to distinguish a soft snowball state 
from a hard snowball state. Therefore, our results are probably 
pessimistic in regard to surface habitability. 

Modeling of Exo-Milankovitch cycles
is difficult because of the timescales 
involved. 3D GCMs can take weeks to converge for static orbital 
conditions and decade long integrations. We have approached the 
problem with a comparatively simple, computationally efficient 
EBM---however, such models lack important phenomena and thus 
must be treated cautiously. As much as possible, we attempt to 
validate our results against a more sophisticated model. In 
terms of average yearly behavior, the EBM does a decent job. The
greatest discrepancies occur in simulations that reach high 
obliquity and have relatively high stellar flux. In these cases, 
the summer insolation at the poles can be intense enough 
(locally) to reach runaway greenhouse temperatures. Undoubtedly, 
there will also be cloud formation, which affects the albedo, as 
observed in GCM simulations of synchronous rotators 
\citep{joshi2003,edson2011,edson2012,yang2013}. 
The difference is that here, the planet is in 
a very different rotation state, which may inhibit the global 
scale redistribution of heat seen in those studies.

The carbon-silicate cycle on a planet like Earth is probably too 
slow to prevent orbitally induced snowball states. Earth's 
carbon-silicate cycle operates on a $\sim 0.5$ Myr time-scale 
\citep{kasting1993,haqqmisra2016}; the planet in this 
configuration can evolve from ice-free to completely ice-covered
in thousands of years. If a planet has significantly higher 
outgassing rate and weathering rates than Earth, there may be 
some hope of preventing the instability through this negative 
feedback. Even with an Earth-like carbon-silicate cycle, 
however, the snowball states could eventually be escaped by building 
atmospheric carbon dioxide pressure. The 
planet may then become extremely warm for an extended period 
until carbon is weathered out of the atmosphere. And, of course, 
the obliquity and eccentricity will continue to vary in the same 
manner as before, perhaps leading to periods of intense polar 
heating. A long term simulation of exo-Milankovitch cycles with 
a carbon cycle would certainly be interesting.

In Paper I, we discussed possibilities for determining whether an 
exoplanet is undergoing Milankovitch cycles. As mentioned there, 
constraining this phenomenon will largely rely on two-dimensional
mapping techniques
\citep{palle2008,cowan2009,kawahara2010,fujii2012,cowan2013,kawahara2016,schwartz2016}. 
A 2-D map of the surface and/or atmosphere of an exoplanet 
will be difficult to generate and will most likely require a large
telescope such as the \emph{Large UltraViolet Optical and InfraRed surveyor}
\citep[\emph{LUVOIR};][]{bolcar2015,dalcanton2015}.

Planets such as we have investigated here,
with large amplitude obliquity and eccentricity cycles, would be 
ideal cases for constraining Milankovitch cycles. Referring to Figures
\ref{fig:lowoblmidmap} - \ref{fig:highoblmidfastmap}, and comparing the left
and right panels in each, we can see that there are regions of parameter
space where we expect the planet to be in a snowball state under 
static obliquity/orbital conditions, but it is clement when these parameters
are allowed to vary. We also see many regions where the planet is warm
under static conditions, but enters a snowball state when variations 
are included. By comparing the climate state under static and dynamic
scenarios with observed 2-D albedo maps, it might be possible to infer that the 
planet is undergoing Milankovitch cycles. This will, of course, depend
heavily on one's trust in the climate models
used and the elimination of alternative explanations.

For the nearer future, the more practical application of the type of
modeling we present here is target prioritization. In scenarios 
where the orbital parameters of a potentially habitable planet and
its companions are well constrained, modeling of dynamical effects on 
climate (such as Milankovitch cycles) may better inform the likelihood
of surface habitability. If there appears to be a high probability of 
snowball states due to such variations, the target will be less 
favorable than another for detecting surface biosignatures. Conversely, 
if one is primarily interested in determining the presence of
Milankovitch cycles, a target in a dynamically ``hot'' system 
will be preferable. Regardless of motivation, our understanding of the
coupling of climate to obliquity and orbital variations will be important 
to the interpretation of \emph{LUVOIR} observations.

\section{Conclusions}
In Paper I, we showed that secular spin-orbit 
resonances can exist even in relatively simple planetary 
systems, and that they can cause very large obliquity 
oscillations. In this paper, we applied a climate model to one
of these systems. We have modeled the climate evolution of a planet with an 
Earth-like atmosphere in response to extreme orbital forcing. 
The large changes in eccentricity and obliquity drive the growth 
and retreat of ice caps, which can extend from the poles to 
$\sim 30^{\circ}$ latitude. These exo-Milankovitch cycles often 
lead to the snowball instability, in which the planet's oceans 
become completely ice covered, as well as the small ice cap 
instability, in which the ice completely disappears. 

We reiterate that planetary systems are extremely complex, and 
in cases like that shown here, the presence of companions can 
affect an Earth-like planet's habitability. It is particularly 
important to understand the eccentricity and obliquity evolution 
in combination, because the stability of ice sheets is 
intimately coupled to the obliquity and the eccentricity affects 
the amount of intercepted stellar energy. At a single stellar 
flux, a planet can be either clement and habitable or completely 
ice-covered, depending on the orbital parameters and the 
planet's recent climate history. This further complicates the 
concept of a static habitable zone 
based on the stellar flux. We have shown that orbital and 
obliquity evolution, and the long time scales of 
ice evolution, should be considered when assessing a 
planet's potential habitability.

 \section{Acknowledgements} 
 This work was supported by the NASA Astrobiology Institute's Virtual 
 Planetary Laboratory under Cooperative Agreement number NNA13AA93A.
 This work was facilitated though the use of advanced computational,
 storage, and networking infrastructure provided by the Hyak 
 supercomputer system at the University of Washington. The results 
 reported herein benefited from the authors' affiliation with the 
 NASA's Nexus for Exoplanet System Science (NExSS) research 
 coordination network sponsored by NASA's Science Mission Directorate.
 Thank you to David Crisp, Andrew Lincowski, Tony Del Genio,
 Ravi Kopparapu, Jacob Haqq-Misra, and Natasha Batalha for helpful discussions,
 and to the anonymous referee, whose feedback resulted in a greatly
 improved manuscript.
\clearpage

\software{Scipy \citep{jones2001}, minepy \citep{albanese2013}, ebm-analytical \citep{rose2017}}

\bibliographystyle{aasjournal}
\bibliography{paper2}
\end{document}